\documentclass[12pt, nofootinbib, notitlepage, prd, aps, tightenlines,a4paper]{revtex4}
\usepackage{listings}
\usepackage{epsfig}
\usepackage{latexsym}
\usepackage[english]{babel}
\usepackage{enumerate}
\usepackage{amsmath, amssymb, graphicx, color}
\usepackage[lofdepth,lotdepth,caption=false]{subfig}
\usepackage[colorlinks,citecolor=blue,linkcolor=blue]{hyperref}

\makeatletter
\renewcommand{\p@subsection}{}
\renewcommand{\p@subsubsection}{}
\makeatother

\graphicspath{{fig-curtis/}}

\newcommand{\del}{\partial}
\newcommand{\beq}{\begin{equation}}
\newcommand{\eeq}{\end{equation}}
\newcommand{\bes}{\begin{equation*}}
\newcommand{\ees}{\end{equation*}}
\newcommand{\bea}{\begin{align}}
\newcommand{\ena}{\end{align}}
\newcommand{\bra}{\langle}
\newcommand{\ket}{\rangle}
\newcommand{\abs}[1]{\left\lvert #1\right\rvert}

\newcommand{\AdS}{\text{AdS}}
\newcommand{\CFT}{\text{CFT}}
\newcommand{\GN}{G_{\text{N}}}

\DeclareMathOperator{\Tr}{\text{Tr}}

\DeclareMathOperator{\Real}{\text{Re}}

\def\({\left (}
\def\){\right )}

\newcommand{\be}{\begin{equation}}
\newcommand{\ee}{\end{equation}}
\newcommand{\ba}{\begin{eqnarray}}
\newcommand{\ea}{\end{eqnarray}}

\newcommand{\eea}{\end{align}}

\makeatletter
\newenvironment{calc}{\allowdisplaybreaks\start@align\@ne\st@rredtrue\m@ne}
{\addtocounter{equation}{1}\tag{\theequation}\endalign}
\makeatother

\begin{document}

\title{Mutual information after a local quench \\ in conformal field theory}
\author{Curtis T. Asplund}
%\date{\today}
\email{curtis@itf.fys.kuleuven.be}

\author{Alice Bernamonti}
\email{alice@itf.fys.kuleuven.be}
\affiliation{Institute for Theoretical Physics, 
	KU Leuven, Celestijnenlaan 200D, B-3001 Leuven, Belgium}

\begin{abstract}

We compute the entanglement entropy and mutual information for two disjoint intervals in two-dimensional conformal field theories as a function of time after a local quench, using the replica trick and boundary conformal field theory. We obtain explicit formulae for the universal contributions, which are leading in the regimes of, for example, close or well-separated intervals of fixed length. The results are largely consistent with the quasiparticle picture, in which entanglement above that present in the ground state is carried by pairs of entangled, freely propagating excitations. We also calculate the mutual information for two disjoint intervals in a proposed holographic local quench, whose holographic energy-momentum tensor matches the conformal field theory one. We find that the holographic mutual information shows qualitative differences from the conformal field theory results and we discuss possible interpretations of this. 

\end{abstract}

\maketitle

\tableofcontents

%%%%%%%%%%%%%%%%%%%%%%%%%%%%%%%%%%%%%%%%%%%%%%%%%%%%%%%%%%%%%%%%%%%%%%%%%%%
%%%%%%%%%%%%%%%%%%%%%%%%%%%%%%%%%%%%%%%%%%%%%%%%%%%%%%%%%%%%%%%%%%%%%%%%%%%

\section{Introduction}
\label{sec:CFTRev}

The entanglement entropy $S_A$
for a single interval $A$, as well as one and two 
point correlation functions of primary operators, after a local quench in a 
(1+1)-dimensional conformal field theory (CFT) were computed in \cite{2007JSMTE..10....4C}.
Certain issues were subsequently found 
with the original analysis in \cite{2007JSMTE..10....4C} and addressed in
\cite{2009JPhA...42X4005C}.
Here we use the same methods to compute the mutual information
\be \label{eq:MI}
I_{A,B} = S_A + S_B - S_{A\cup B}\,,
\ee
between two disjoint intervals 
$A = [u_1, v_1]$ and $B = [u_2, v_2]$.\footnote{
In the following we will often suppress the subscript and just denote 
$I = I_{A,B}$.
}
The computation relies on the replica trick, 
i.e., we first compute the 
mutual R\'enyi information 
\be \label{eq:MIRenyi}
I_{A,B}^{(n)}= S^{(n)}_A + S^{(n)}_B - S^{(n)}_{A\cup B}
\ee
for integer $n>1$, then analytically 
continue in $n$ and take 
\be \label{eq:MIlimit}
I_{A,B} = \lim_{n\to1} I^{(n)}\,.
\ee
Here the $S^{(n)}$ are the R\'enyi entropies, defined by 
\be\label{eq:Renyi}
S^{(n)}_A = \frac{1}{1-n} \log \Tr \rho_A^n\,,
\ee
where $\rho_A$ is the reduced 
density matrix for region $A$, and the entanglement or von Neumann 
entropy is given by 
\beq
\label{eq:vNdef}
	S_A = -\Tr \rho_A \log \rho_A = \lim_{n\rightarrow 1} S^{(n)}_A
	= -\lim_{n\rightarrow 1} \frac{\del}{\del n} \Tr \rho_A^n\,.
\eeq

Computing the entanglement entropy in (1+1)-dimensional quantum field theories 
for more than one interval has been an active subject recently, beginning with 
the initial result in \cite{Calabrese:2004eu}, later recognized to be only 
true in special cases \cite{Caraglio:2008pk,2009PhRvL.102q0602F}.
The R\'enyi and von Neumann entropies for 
an arbitrary number of intervals have been computed in the vacuum of a free fermion 
\cite{2005JSMTE..07..007C,Casini:2008wt,Casini:2009sr,Casini:2009vk} and for two intervals 
for a free (compact) boson 
\cite{2009JSMTE..11..001C, 2011JSMTE..01..021C} (see \cite{Chen:2013kpa} for recent calculations to higher order in the short distance expansion for two intervals in a general CFT).\footnote{The von Neumann entanglement entropy for the 
compact boson theory has only been found in certain regimes, due to the 
difficulty of the analytic continuation of the R\'enyi entropies.}
The apparent discrepancy between these results was explained in 
\cite{Headrick:2012fk}, which also contains further results 
in this vein. 

This can be generalized to certain time-dependent situations via conformal maps.
The first studies along these lines were of global quenches \cite{Calabrese:2005in, Calabrese:2006rx}. 
These are situations in which the system is initially (for $t<0$) prepared in the ground state of a Hamiltonian 
of a gapped field theory. At $t =0$, the Hamiltonian suddenly becomes conformal, and the system is then allowed to 
evolve undisturbed for $t>0$. The strategy of \cite{Calabrese:2005in, Calabrese:2006rx} to compute
correlators after this process is to treat them as path integrals on a bounded 
complex surface in complex spacetime 
(real space times complex time), conformally map this surface to the
complex upper half plane (UHP) and    
compute the correlators in the UHP using boundary conformal field theory (BCFT),
also known as surface critical 
behavior \cite{Cardy:1984bb, DiFrancesco1997, Cardy:2004hm}.
Using conformal symmetry, one can then pull back
this UHP result to compute the correlator of interest, and analytically 
continue to real time.
We review this strategy in detail in Section~\ref{sec:CFT_one_two_EE}. A detailed review of CFT results along these 
lines is \cite{2009JPhA...42X4005C}.

A two dimensional local quench is the process of two semi-infinite lines being joined at 
their endpoints at an instant of time and subsequently evolving as a connected, infinite system.
The first CFT study of this process was \cite{2007JSMTE..10....4C}, later 
elaborated in \cite{2009JPhA...42X4005C}, using a generalization of the same 
computational strategy described above for a global quench. 
These CFT local quench results have been successfully compared 
with semi-analytic and numerical computations in critical spin chains
\cite{2007JSMTE..06....5E, 2007JSMTE..10....4C, 
2008JSMTE..01..023E, 2011JSMTE..10..027D, 2011JSMTE..08..019S}.

\begin{figure}[h]
\centering
\def\svgwidth{8.0cm}
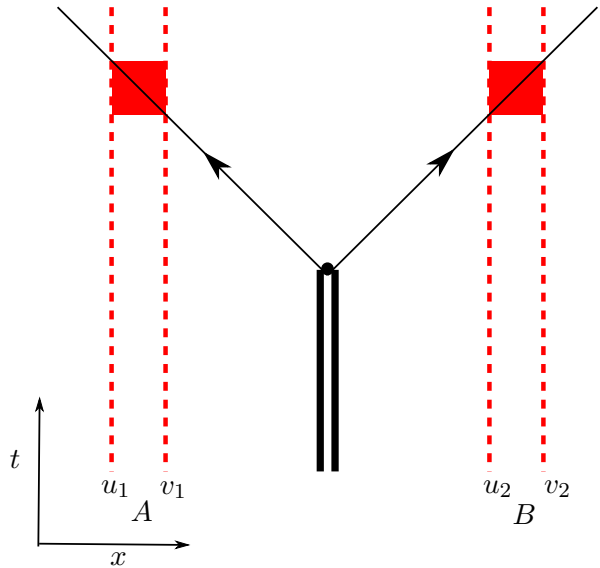
\label{fig:lightcone}
\caption{Illustration of a local quench: two CFTs on half lines are joined 
instantaneously at their boundaries. This generates left and right moving excitations that
propagate along the lightcone of the quench event and are 
entangled with each other. During the time period, in shaded red, that the two 
intervals $A$ and $B$ are intersected 
by this lightcone, the entanglement leads to a mutual information that is larger than the ground state value. 
}
\end{figure}

One of the motivations to study local quenches is that they produce entangled 
particles in a localized region of spacetime 
and some universal aspects of this dynamical process can be analyzed exactly using conformal 
field theory.
Particle production also 
occurs in global quenches but is spread over the entire system in a homogenous 
 \cite{Calabrese:2005in, Calabrese:2006rx} or slightly inhomogenous way \cite{2008JSMTE..11..003S}. 
This particle production can be understood by considering sudden changes to critical spin chain Hamiltonians 
\cite{Calabrese:2005in,2006JSMTE..03..001D,2007JSMTE..06....5E,2007JSMTE..10....4C} 
or from quantum field theory 
subject to changes in the geometry and topology of spacetime \cite{Anderson:1986ww,Manogue:1988}.
The dynamics of the particles goes a long way towards explaining the evolution of entanglement between various subsystems. 
For example, it leads one to expect entanglement 
(above that already present in the ground state) between, and only between, subsystems that 
are separately intersected by the two halves of the lightcone of the quench event, 
see Fig.~\ref{fig:lightcone}. 
This general picture is referred to variously as a 
causality argument, lightcone effects, and the quasiparticle picture \cite{Calabrese:2005in, Calabrese:2006rx}.
The CFT mutual information calculations we perform are, in part, a check of this picture.

Another motivation is that a local quench may 
be realizable experimentally. If so, it might provide a 
way to measure the entanglement (R\'enyi) entropies 
\cite{Klich:2008un,2009PhRvB..80w5412H,2011PhRvL.106o0404C,2012PhRvL.109b0504A}.
It is conceivable, then, that modifications of such setups could be used to 
measure the mutual information between two subsystems, though we do 
not pursue this further here. 
See also \cite{2012arXiv1210.3545S} for possible 
experimental realizations of dynamical topology changes in the context of 
non-relativistic quantum mechanics and applications thereof. 

There has been considerable work on 
relating multiple-region CFT entanglement entropy computations (usually in the limit of large 
central charge) to holographic calculations based on the 
Ryu-Takayanagi holographic entanglement entropy proposal
\cite{Ryu:2006bv} and its generalization to time-dependent 
states \cite{Hubeny:2007xt}. 
For example, there have been studies in the 
ground state \cite{Hubeny:2007re, 2010PhRvD..82l6010H, Hartman:2013mia, Faulkner:2013yia}, 
in thermal (black hole) states \cite{Tonni:2010pv, MolinaVilaplana:2011xt,Fischler:2012uv,2013JHEP...07..081M,Shenker:2013pqa} 
and in some time-dependent situations corresponding to global quenches 
\cite{Balasubramanian:2011at, Allais:2011ys, Hartman:2013qma}.

This leads to the question: what is the holographic dual to a CFT local quench?
Recently, a holographic model of a local quench was proposed in \cite{Nozaki:2013wia}, 
as the backreacted geometry describing a massive particle falling in three-dimensional Poincar\'e AdS space. 
(See also \cite{Roberts:2012aq}, which modelled one feature of a local quench, namely 
a propagating pulse of energy.) 
In \cite{Nozaki:2013wia} the authors compared the entanglement entropy of their model, 
computed using the covariant holographic proposal \cite{Hubeny:2007xt}, with the results for the CFT local 
quench, and found rough agreement for times after the quench. We also show that, for specific parameter choices, 
the expectation value of the holographic energy-momentum tensor exactly matches that of a CFT after a local quench.
In light of our field theory results, it is natural to consider the 
mutual information in the model of \cite{Nozaki:2013wia}, since it provides a more 
refined tool of analysis than the single interval entanglement entropy.  
We compute this below and compare in detail with the CFT results. 
We discuss possible explanations 
and implications of the differences we find between the holographic model and the local quench in the Conclusions in Sec.~\ref{sec:Concl}.

As this work was being completed, \cite{Ugajin:2013xxa} appeared, which contains material that 
overlaps with our Secs.~\ref{sec:Holog_renorm} and \ref{sec:alt_method},  as well as a new model for a holographic local quench. 
As we discuss further in Sec.~\ref{sec:alt_method}, this model has the same behavior, as far as the mutual information, as the model in 
\cite{Nozaki:2013wia} in the regimes that we primarily study. 

In more detail, the structure of the paper is as follows. In Sec.~\ref{sec:CFT_one_two_EE} we review the steps, first worked out in \cite{Calabrese:2005in, 2007JSMTE..10....4C}, in the calculation of the entanglement entropy of one and two intervals in a local quench. We use these results in Sec.~\ref{sec:CFT-MI} to compute the time-evolution of the mutual information between two disjoint intervals. In Sec.~\ref{sec:nrgmom} we work out the expectation value of the energy-momentum tensor after a local quench and show in Sec.~\ref{sec:Holog_renorm} how it can also be obtained through a Lorentzian transformation of the vacuum energy-momentum tensor. 
In the same section we also show that this Lorentzian conformal transformation can be appropriately promoted to a bulk diffeomorphism that maps the holographic dual of a 2d CFT in the vacuum, i.e. empty AdS$_3$, to the presumed holographic dual to a local quench. This geometry was already worked out in \cite{Nozaki:2013wia} and proposed, on the basis of different arguments, as the dual of a CFT local quench. We review the computation of the entanglement entropy in this holographic setup in Sec.~\ref{sec:coord} and work out the holographic mutual information in Sec.~\ref{sec:MI}, extensively comparing with the CFT results. In Sec.~\ref{sec:alt_method}, we present an alternative, equivalent, method to compute the entanglement entropy in the holographic duals of two-dimensional CFTs, first introduced in \cite{Roberts:2012aq} and based on the Lorentzian transformation discussed in Sec.~\ref{sec:Holog_renorm}. We conclude with a discussion of our results in Sec.~\ref{sec:Concl}. The main text is supplemented by two appendices expanding on the regimes of universality of the CFT results (App.~\ref{sec:crossratios}) and the geodesics that compute the holographic entanglement entropy (App.~\ref{sect:geodesics}).         

%%%%%%%%%%%%%%%%%%%%%%%%%%%%%%%%%%%%%%%%%%%%%%%%%%%%%%%%%%%%%%%%%%%%%%%%%%%
%%%%%%%%%%%%%%%%%%%%%%%%%%%%%%%%%%%%%%%%%%%%%%%%%%%%%%%%%%%%%%%%%%%%%%%%%%%

\section{One and two interval entanglement entropies}
\label{sec:CFT_one_two_EE}

We begin by reviewing how to compute the
entanglement entropy after a 
local quench using path integrals, conformal symmetry, and a particular conformal map.
This is described in the first applications of the method \cite{Calabrese:2005in, 2007JSMTE..10....4C}
and further details are explained in a closely related situation in \cite{Asplund:2011cq}. 

%%%%%%%%%%%%%%%%%%%%%%%%%%%%%%%%%%%%%%%%%%%%%%%%%%%%%%%%%%%%%%%%%%%%%%%%%%%

\subsection{Quench background}

First we recall that the density operator 
$\rho(t) = |\psi(t) \ket\bra \psi(t) |$ associated with a given 
time-dependent pure state $|\psi(t)\ket$
can be written in the following form, with respect to an arbitrary 
basis $\{|\phi_i\ket\}$,
\begin{calc}
\label{eq:rho1}
	\bra \phi_2 \vert \rho(t) \vert \phi_1 \ket &= 
		\bra \phi_2 \vert \psi(t) \ket\bra \psi(t) \vert \phi_1 \ket \\
		&= \bra \phi_2 |e^{-iHt} | \psi_0 \ket\bra \psi_0 | e^{iHt} |\phi_1 \ket \\
		&= \bra \psi_0 | e^{iHt} |\phi_1 \ket \bra \phi_2 |e^{-iHt} | \psi_0 \ket\,,
\end{calc}
where $|\psi_0\ket = |\psi(0)\ket$. To regulate 
UV divergences, we modify this matrix element as follows \cite{Calabrese:2005in, 2007JSMTE..10....4C}:
\begin{align}
\label{eq:rho2}
	\bra \phi_2 \vert \rho(t) \vert \phi_1 \ket_\epsilon
	&=  N_\epsilon \bra \psi_0 | e^{iHt -\epsilon H} |\phi_1 \ket 
	\bra \phi_2 |e^{-iHt -\epsilon H } | \psi_0 \ket \\
\label{eq:rho2b}
	&=  N_\epsilon \bra \psi_0 | e^{-H(-it +\epsilon)} |\phi_1 \ket 
	\bra \phi_2 |e^{-H(it + \epsilon) } | \psi_0 \ket \,. %\\
%	&=  \bra \psi_0 | e^{-H(-it +\epsilon)} |\phi_1 \ket 
%	\bra \phi_2 |e^{-H(it + \epsilon) } | \psi_0 \ket \,.
\end{align}
Here $N_\epsilon$ is a normalization factor ensuring that $\Tr \rho(t) = 1$, 
which we write more explicitly below.
If we think of the CFT as arising from the continuum limit of a critical 
lattice system, we can relate $\epsilon$ to the lattice spacing \cite{2007JSMTE..10....4C}, but it 
can also be thought of as an ultraviolet regulator that damps out the high 
energy modes and makes the path integral absolutely convergent.
In the end we will want to consider all quantities asymptotically 
as we remove the cut-off, though we do not strictly take the limit $\epsilon \to 0$ since 
many quantities diverge.
For example, as we will see in Sec.~\ref{sec:nrgmom}, the total energy of the local quench state above the ground state 
is inversely proportional to $\epsilon$.

Following \cite{Calabrese:2005in,2007JSMTE..10....4C}, we represent the product of matrix elements in 
Eq.~\eqref{eq:rho2b} as a path integral analytically continued to imaginary 
time.  
The second factor in Eq.~\eqref{eq:rho2b} 
is interpreted as 
a path integral with boundary conditions that match $|\psi_0\ket$ at 
$\tau = -it - \epsilon$ and $|\phi_2\ket$ at $\tau =0$, while the first matches 
$|\phi_1\ket$ at $\tau =0$ and $|\psi_0\ket$ at $\tau = -it + \epsilon$. 
Together these can be written as a single path integral with a discontinuous 
boundary condition at $\tau = 0$. 
This setup can also be translated so that the discontinuity occurs at an arbitrary imaginary time $\tau$, 
and the $|\psi_0\ket$ boundary conditions occur at imaginary times 
$\pm \epsilon$. 
We denote the Euclidean path integral satisfying these boundary conditions by 
$Z(\{-\epsilon, +\epsilon\}, \tau; \psi_0, \{\phi_1,\phi_2\})$ and 
we first perform the computation in the Euclidean CFT.
We will see that the result is an analytic function of 
$\tau$ in the strip $\Real \tau \in (-\epsilon,+\epsilon)$;
the result for real times is recovered by taking 
$\tau \to it$ at the end of the calculation. 

We therefore have 
\beq
\label{eq:rho3}
	\bra \phi_2 \vert \rho(t) \vert \phi_1 \ket_{\epsilon} =
	 N_{\epsilon} Z(\{-\epsilon,+ \epsilon\}, it; \psi_0, \{\phi_1,\phi_2\})\,.
\eeq
In this notation the normalization factor is given by 
\beq
\label{eq:norm}
N_\epsilon^{-1} = 
\int_X D\phi\ Z(\{-\epsilon,+\epsilon\}, it; \psi_0, \{\phi,\phi\})\,,
\eeq
where $X$ denotes the boundary conditions on the path integral mentioned above.
See \cite{Calabrese:2005in, 2007JSMTE..10....4C,Asplund:2011cq} for further details.

%%%%%%%%%%%%%%%%%%%%%%%%%%%%%%%%%%%%%%%%%%%%%%%%%%%%%%%%%%%%%%%%%%%%%%%%%%%

\subsection{Local quench conformal map} 

\begin{figure}
\subfloat[][]{
\def\svgwidth{9.0cm}
\hspace*{-1.25cm}
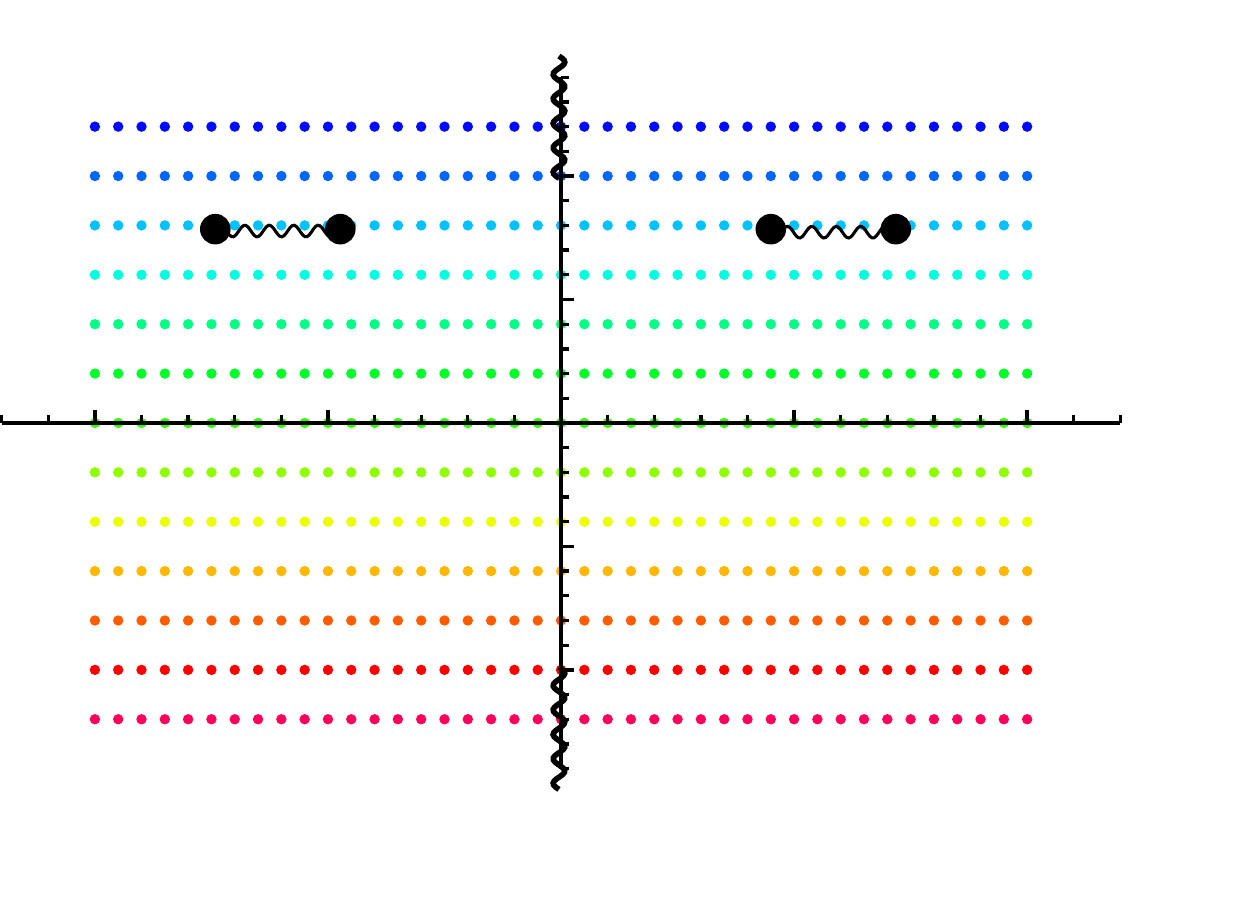
\label{fig:w-points}} %\\
\subfloat[][]{
\def\svgwidth{9.0cm}
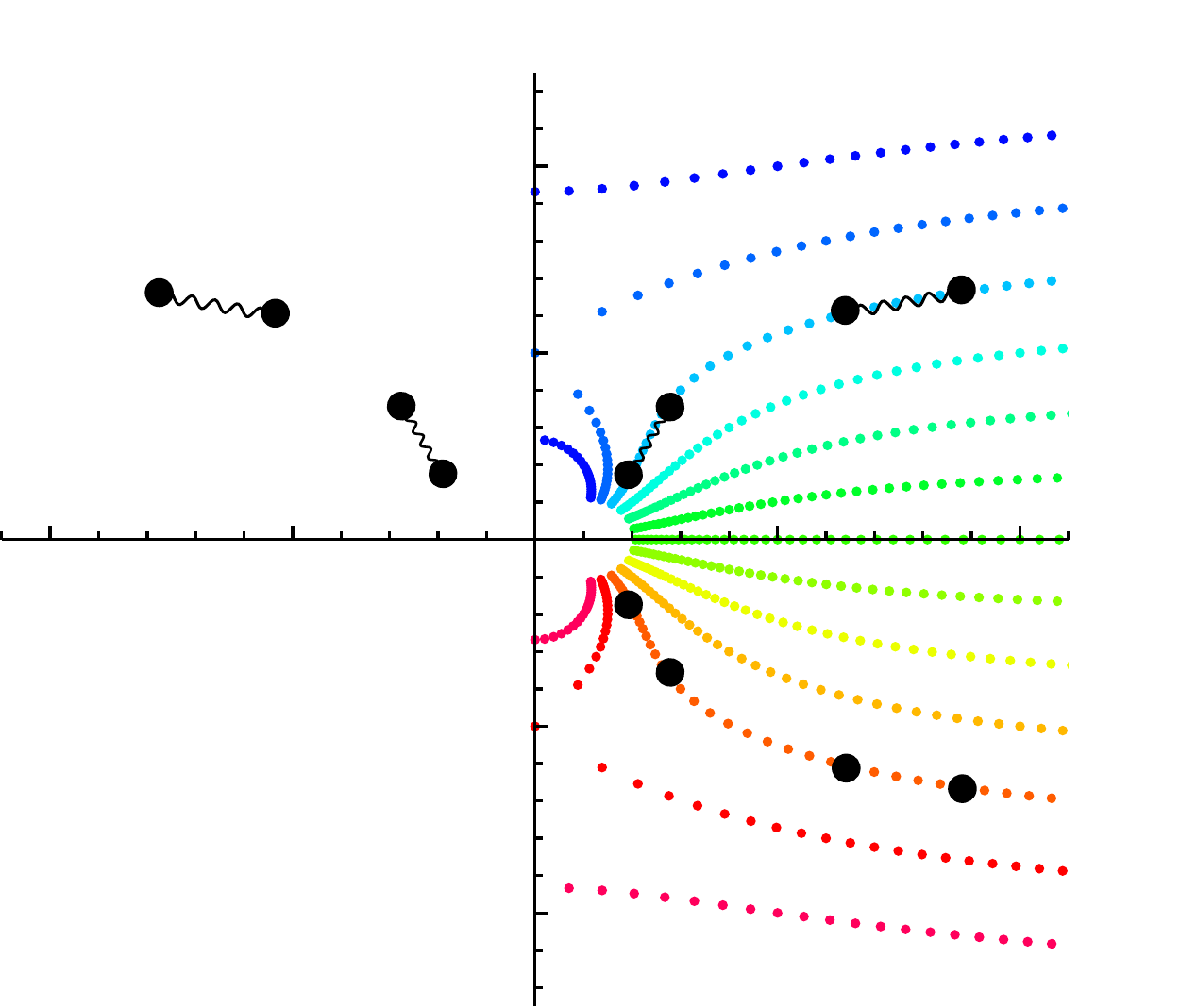
 \label{fig:z-points}}
\caption{Illustration of the mapping \eqref{eq:UHPmap}, with 
$\epsilon =0.1$. The sample of points in the region $[-0.1,0.1] 
\times [-0.12,0.12]$ in the $w$ plane, shown in \ref{fig:w-points}, 
is mapped to the points in the $z$ plane seen in \ref{fig:z-points}. 
The twist operators are inserted at the points $w_1,w_2,w_3$ and $w_4$ and 
the branch cuts they induce are shown with thin curvy lines. The thick curvy 
lines indicate the boundaries of $W$, which are mapped to the imaginary axis in the 
$z$ plane. We also show the image points in the left half of the $z$ plane and the conjugate points in the lower half plane. 
} \label{fig:map}
\end{figure}

So far, everything we have described applies equally well to global or local quenches.
The boundary conditions at $\tau = \pm \epsilon$ corresponding to a local quench, where 
two semi-infinite CFTs are suddenly joined into an infinite CFT, may be implemented 
by an appropriate restriction of the spacetime over which the above path integrals 
are performed. Namely, we consider a Euclidean spacetime $W$ with semi-infinite 
boundaries dividing space into two halves, extending for all $\tau < -\epsilon$
and all $\tau > +\epsilon$, see Fig.~\ref{fig:w-points}. 
We will be imposing conformal boundary conditions on these boundaries 
ensuring, in particular, that neither energy nor momentum can flow past 
them. As we will see, our main results do not depend on any further details of the boundary 
conditions.

Take complex coordinates 
$w = x + i\tau$ for $W$, 
the $w$ plane minus 
two semi-infinite lines on the imaginary axis starting at $\pm i\epsilon$ 
and extending away from the origin to infinity.
We can conformally map this space to the right half plane (RHP), 
$\{z: \Real z > 0\}$, with 
its boundary along the imaginary axis, by the conformal mapping \cite{2007JSMTE..10....4C,2009JPhA...42X4005C}
\beq
\label{eq:UHPmap}
	z = \frac{1}{\epsilon} \left(w + \sqrt{w^2 + \epsilon^2}\right)\,.
\eeq 
Here we are defining the square root to have a branch cut on the negative 
real axis, i.e., it maps the unit circle minus the point $-1$ into the RHP. 
(We choose the RHP instead of the UHP to facilitate comparison with previous 
results \cite{2007JSMTE..10....4C,2009JPhA...42X4005C}.)
The mapping is an example of a Joukowsky transformation.
It is invertible on the RHP and the inverse is given by
\beq
\label{eq:inversemap}
w = \epsilon \, \frac{z^2 - 1}{2z} \,.
\eeq
In the next Section we will see how to compute the entanglement entropy in the RHP and then we will use the conformal transformation \eqref{eq:UHPmap} to obtain the desired results in $W$. 
 
%%%%%%%%%%%%%%%%%%%%%%%%%%%%%%%%%%%%%%%%%%%%%%%%%%%%%%%%%%%%%%%%%%%%%%%%%%%

\subsection{Entanglement entropy calculation}\label{sec:EEcomputation}

Now that we have formulated the problem of computing density matrix elements 
after a local quench, we can turn to the calculation of the entanglement entropy
associated with various spatial intervals.
In the vacuum state of the CFT the various $S^{(n)}$ are proportional to 
the vacuum correlators of $n$-twist operators inserted at points in spacetime 
corresponding to the endpoints of the intervals.
This was first shown in \cite{Calabrese:2004eu} and anticipated in the earlier 
work \cite{Holzhey:1994we}.
We will explain, to some extent, why this is the case below.
The twist operators effectively glue 
$n$ replicas of the spacetime of the CFT into a single 
surface with genus $(N-1)(n-1)$, where $N$ is the number of intervals.
For this reason, this is known as the replica trick for computing the 
entanglement entropy.
For example, $S^{(2)}_{A\cup B}$ is proportional to a four point function of 
twist operators (inserted at the interval endpoints $u_1,v_1,u_2,v_2$) and  
this is proportional to the partition function of the 
CFT on a torus \cite{Lunin:2000yv}.

We will use this same method to compute the R\'enyi entropies
$S^{(n)}_{A\cup B}$ for two disjoint intervals $A = [u_1,v_1]$ and $B = [u_2,v_2]$,
but imposing the boundary conditions described above to obtain results for 
the period after a local quench.
The partial trace over the complement of $A\cup B$, needed to compute 
$\rho^n_{A\cup B}$, and the trace over $\rho^n_{A\cup B}$, 
needed to compute $S^{(n)}_{A\cup B}$, can both be realized by a path integral over 
$n$ replica spacetimes joined together 
along $A$ and $B$. 
This joining procedure can be implemented using twist 
operators, as we mentioned above, and is reviewed in, e.g.,
\cite{2009JPhA...42X4005C}. 

In more detail, denote a configuration of the field in $A, B, C$ 
(where $C$ is the complement of $A\cup B$) by
$\{a, b, c\}$. Then the matrix element of the reduced density operator $\rho_{A\cup B}$ is given by
\begin{calc}
\label{eq:redDO}
	\bra a_2, b_2 | \rho_{A\cup B}(\tau) | a_1, b_1\ket_\epsilon 
	&=  \bra a_2, b_2 | \Tr_C \rho(\tau) | a_1, b_1\ket_\epsilon \\
	&= \int_C Dc\ N_\epsilon Z\(\{-\epsilon,+\epsilon\}, \tau; \psi_0, 
		\{\phi_1 = \{a_1,b_1,c\}, \phi_2 = \{a_2,b_2,c\} \}\) \\
	&= N_\epsilon Z\(\{-\epsilon,+\epsilon\}, \tau; \psi_0, 
		\{\{a_1,b_1\}, \{a_2,b_2\} \}\).
\end{calc}
This indicates that the discontinuous boundary condition is only discontinuous in the region $A\cup B$; in $C$ the configurations
are continuous. Next,
\begin{calc}
	\Tr \rho^n_{A\cup B}(\tau) &= \int  Da Db \int Da_1 Db_1  \cdots \int  Da_{n-1} Db_{n-1} \\
	&\qquad \bra a,b| \rho_{A\cup B}(\tau) |a_1,b_1\ket
	\cdots \bra a_{n-1},b_{n-1}|
	\rho_{A\cup B}(\tau) |a,b\ket \\
	&= \int  Da Db \int Da_1 Db_1  \cdots \int  Da_{n-1} Db_{n-1} \\
	&\qquad  N_\epsilon^n Z\(\{-\epsilon,+\epsilon\}, \tau; \psi_0, 
		\{\{a,b\}, \{a_1,b_1\} \}\)	\\
	&\qquad 	\cdots Z\(\{-\epsilon,+\epsilon\}, \tau; \psi_0, 
		\{\{a_{n-1},b_{n-1}\}, \{a,b\} \}\)\,.
\end{calc}
The result can be put together into a single path integral
over the $n$ replica spacetimes joined together along $A$ and $B$,
which is equivalent to a path integral over $W$ with twist operator insertions at the 
endpoints of $A$ and $B$ \cite{2007JSMTE..10....4C,Calabrese:2004eu,2009JPhA...42X4005C}. 
This latter path integral is a correlation function of four $n$-twist 
operators inserted at locations $w_1 = u_1 + i\tau$, $w_2 = v_1 + i\tau$, 
$w_3 = u_2 + i\tau$ and $w_4 = v_2 + i\tau$, that is,
\be \label{eq:4twist-w}
\Tr \rho^n_{A\cup B}(\tau) = N_\epsilon^{n} \bra \sigma_n(w_1)\tilde{\sigma}_n(w_2)
		\sigma_n(w_3)\tilde{\sigma}_n(w_4)\ket \,.
\ee
The tildes on some of the twist operators indicate that they have opposite 
orientation, which ensures that the various replica spacetimes are connected in 
the correct way. 
This method is difficult to implement in general 
because the genus $(N-1)(n-1)$ spacetime
needed to compute the correlators 
cannot be conformally mapped (uniformized) to the plane (or UHP or RHP). 
The path integral 
for the theory on these surfaces depends, for $(N-1)(n-1) \geq 1$, on the full operator 
content of the theory (not just its central charge), and the same goes for the 
von Neumann and R\'enyi entropies. 
Specifically, for the case of $N$ intervals 
$[u_i,v_i]$, global conformal invariance and the fact that the twist operators 
are conformal primaries implies that for the ground state
\cite{2009JPhA...42X4005C}
\beq
	\label{eq:globalCFT}
	\Tr \rho_I^n = (c_n)^N \left(\frac{\prod_{j<k} (u_k - u_j)(v_k - v_j)}
	{ \prod_{j,k} |v_k - u_j|} \right)^{2 x_n} 
	\mathcal{F}_{n,N}(\{\eta_i\}),
\eeq
where $I =  \bigcup_i [u_i,v_i]$,
\be
x_n = \frac{c}{12} \(n-\frac{1}{n}\)
\ee
is the 
conformal weight of the twist operator \cite{Knizhnik:1987xp, Calabrese:2004eu}, $c_n$ is a non-universal constant 
and $\mathcal{F}_{n,N}$ is a non-universal function of the 
$2N -3$ independent cross-ratios $\{\eta_i\}$ formed by the $2N$ points $\{u_i,v_i\}$. 
The dependence on the full operator content of the CFT
is encoded in the functions $\mathcal{F}_{n,N}$, as exhibited explicitly in the 
short distance expansion in \cite{2011JSMTE..01..021C}.

For local (and global) quenches we must compute $\Tr \rho_I^n$ in a specific boundary CFT (BCFT) 
that has concomitant non-universal constants $\tilde{c}_n$ and functions 
$\tilde{\mathcal{F}}_{n,2N}$ 
(the number of intervals is now effectively $2N$ since we have to include image points when 
performing computations, as we explain below). We use tildes to distinguish quantities in BCFTs from 
those in ordinary CFTs. The two are analogous, 
but those in BCFTs depend on the boundary conditions of the theory, or equivalently on the 
bulk to boundary operators \cite{Cardy1991274}. 

These non-universal quantities are known for some 
specific CFTs and BCFTs, e.g.,
$\mathcal{F}_{n,N} = 1$ for the 
massless fermion theory considered in 
\cite{2005JSMTE..07..007C,Casini:2008wt,Casini:2009sr,
Casini:2009vk},
$\mathcal{F}_{n,2}$ is known for integers $n\geq 1$ 
for the compactified boson theory \cite{2009JSMTE..11..001C, 2011JSMTE..01..021C}
(and also $\mathcal{F}_{2,N}$ for $N \geq 1$ at the self-dual radius \cite{Headrick:2012fk}), 
$\tilde{\mathcal{F}}_{n,2} = 1$ (and $\tilde{c}_n = 1$) for a boundary Gaussian theory and
$\tilde{\mathcal{F}}_{n,2} = \sqrt{(1-\eta)^{1/2}+1} \pm  \sqrt{(1-\eta)^{1/2}-1}$ 
for the critical Ising model/free fermion/self-dual boson boundary theory,
where the sign depends on the boundary conditions
\cite{Cardy:1984bb, DiFrancesco1997}. 
The constants $\tilde{c}_n$ and $c_n$  are known for some
 specific theories.
See \cite{Cardy:2007mb} for discussions 
and calculations of the $\tilde{c}_n$ for the Ising and other models and 
\cite{2009JPhA...42X4005C} for further references. 
In general we have
 $\tilde{c}'_1 - c'_1/2 = \log g$, 
where $\tilde{c}'_1 = \lim_{n \rightarrow 1} 
\del_n \tilde{c}_n$ and $\log g$ is the Affleck-Ludwig boundary entropy 
 \cite{PhysRevLett.67.161,Calabrese:2004eu,2007JSMTE..10....4C}. 

Fortunately, there are some regimes where we do not need to know these
non-universal data.
In cases where all cross ratios $\eta_i \to 0,1$ or $\infty$, we have 
$\mathcal{F}_{n,N} \to 1$ and $\tilde{\mathcal{F}}_{n,2N} \to 1$.
This follows from the short and long distance expansions of the 
twist operator correlation functions
\cite{Cardy1991274, Calabrese:2007rg, 2009JPhA...42X4005C, 2011JSMTE..01..021C}.
The physical reason for this is that in the ground state all correlations 
between two disjoint regions should decay at large separations. 
This allows one to compute universal results for correlators and entanglement 
entropies in specific regimes.
It was applied to CFTs following a global quench in 
\cite{2009JPhA...42X4005C}, since for times, 
interval lengths and separations large compared to both the UV regulator $\epsilon$ and the initial inverse mass gap,
the cross-ratios go to $0,1$ or $\infty$. 

In the case of the mutual information
after a local quench we will need to consider the cross ratios 
formed by the endpoints of one or two intervals and associated image points, 
after the action of the appropriate conformal map. 
We are particularly interested in the regimes where the results are universal, 
i.e., independent of $\tilde{\mathcal{F}}_{n,2}$ or 
$\tilde{\mathcal{F}}_{n,4}$ and so applicable to CFTs in general.    
For example, in the regime of two intervals whose length is much 
smaller than their separation, the relevant cross ratios are indeed all 
approximately $0,1$ or $\infty$, and hence the leading contribution is universal. 
We study this case in detail below.
We analyze the relevant cross-ratios in detail in 
Appendix~\ref{sec:crossratios}; for now 
it suffices to bear in mind that the results we present are typically 
universal in regimes with a large separation between the length scales of the problem. 

%%%%%%%%%%%%%%%%%%%%%

We now address the first step in the computation of 
$S^{(n)}_{A\cup B}$ in Eq.~\eqref{eq:Renyi}, namely the computation of the correlation function of four $n$-twist 
operators $\sigma_n$ in the RHP
\be \label{eq:4twist}
\bra \sigma_n(z_1)\tilde{\sigma}_n(z_2)
		\sigma_n(z_3)\tilde{\sigma}_n(z_4)\ket \,.
\ee
Using the conformal map in Eq.~\eqref{eq:UHPmap}, we can then transform 
Eq.~\eqref{eq:4twist} to obtain 
$\Tr \rho^n_{A\cup B}$ in $W$, via Eq.~\eqref{eq:4twist-w}.
Since the $\sigma_n$ are conformal primaries with weight  $x_n$ \cite{Knizhnik:1987xp, Calabrese:2004eu},
global conformal invariance restricts the form of this four-point function 
to the following function of 
distances between operator insertion points $z_1, \dots,z_4$ and 
image points $z_5,\dots,z_8$ \cite{Calabrese:2004eu}:
\begin{calc}
	\label{eq:corr}
	\bra \sigma_n(z_1)\tilde{\sigma}_n(z_2)
		\sigma_n(z_3)\tilde{\sigma}_n(z_4)\ket &= (\tilde{c}_n)^4 \left(\frac{D_u D_v}{D_{uv}}\right)^{x_n} 
	\tilde{\mathcal{F}}_{n,4}(\{\eta_i\})\,.
\end{calc}
Here 
\begin{calc}
\label{eq:Du}
	D_u &= z_{75} z_{73} z_{71} z_{53} z_{51} z_{3 1} =
	z_{\bar 2 \bar 4} \tilde z_{\bar 2 3} \tilde z_{\bar 2 1} \tilde z_{\bar 4 3} \tilde z_{\bar 4 1} z_{3 1}\\
%\label{eq:Dv}
	D_v &=  z_{8 6} z_{8 4}  z_{8 2}  z_{6 4}  z_{6 2} z_{4 2}
	=  z_{\bar 1 \bar 3} \tilde z_{\bar 1 4} \tilde z_{\bar 1 2} \tilde z_{\bar 3 4} \tilde z_{\bar 3 2} z_{4 2}\\
%\label{eq:Duv}
	D_{uv} &= (z_{2 1} z_{2 3}  z_{2 5} z_{2 7}) 
			(z_{4 1} z_{4 3}  z_{4 5} z_{4 7}) 
			( z_{6 1}  z_{6 3} z_{6 5} z_{6 7}) 
			( z_{8 1} z_{8 3} z_{8 5} z_{8 7}) \\
	&= (z_{2 1} z_{2 3} \tilde z_{2 \bar 4} \tilde z_{2 \bar 2}) 
			(z_{4 1} z_{4 3} \tilde z_{4 \bar 4} \tilde z_{4 \bar 2}) 
			(\tilde z_{\bar 3 1} \tilde z_{\bar 3 3} z_{\bar 3 \bar 4} z_{\bar 3 \bar 2}) 
			(\tilde z_{\bar 1 1} \tilde z_{\bar 1 3} z_{\bar 1 \bar 4} z_{\bar 1 \bar 2}) \,,
\end{calc}
and we have used $z_5=  -\bar z_4 , z_6 =-\bar z_3  ,z_7 = -\bar z_2,z_8 =-\bar z_1$ to express 
the quantities in the second equalities in Eq.~\eqref{eq:Du} in terms of points
in the right half plane alone, 
and defined $z_{i\bar{j}} = \abs{z_i - \bar{z}_j}$, $\tilde z_{i\bar{j}} = \abs{z_i + \bar{z}_j}$, etc.
This is an application of Eq.~\eqref{eq:globalCFT}, modified to the 
case of a boundary CFT as in \cite{Calabrese:2004eu}, and 
corresponds to an eight point function of twist operators inserted at 
$z_1,\ldots,z_4$ and at the image points $z_5,\ldots,z_8$
(see Fig.~\ref{fig:map}).
Here we are using 
the method of images, which allows one to compute 
correlation functions of primary operators in the right half plane  
by computing a correlation function in the full complex plane, 
with each operator insertion 
doubled to include an image insertion in the left half plane
\cite{Cardy:1984bb, DiFrancesco1997}.
As we said, $\tilde{c}_n$ is a non-universal constant that depends on the particular 
boundary CFT; we will see it cancels out in the computation of the mutual information.
As mentioned above, we consider regimes where
$\tilde{\mathcal{F}}_{n,4}(\{\eta_i\}) \to 1$,
because of the behavior of the $\eta_i$, which we discuss in detail in Appendix~\ref{sec:crossratios}. Finally, Eq.~\eqref{eq:corr} should be 
understood as an asymptotic result in the limit in which the ratio of 
$\epsilon$ over all time and length scales goes to zero. 
We will be primarily interested in the leading order results in this limit.

We want to compute the correlator in $W$ as in Eq.~\eqref{eq:4twist-w}.
Since the $\sigma_n$ are conformal primaries, 
we can apply the conformal map in Eq.~\eqref{eq:UHPmap} to Eq.~\eqref{eq:corr}, 
and then pull back to get the desired result in $W$:
\bea
\label{eq:wcorr2}
\Tr \rho^n_{A\cup B} 
	=   (\tilde{c}_n)^4  \left(a^4\abs{\frac{dz}{dw}}_{w_1}\abs{\frac{dz}{dw}}_{w_2}
		\abs{\frac{dz}{dw}}_{w_3}\abs{\frac{dz}{dw}}_{w_4}
		\frac{D_u D_v}{D_{uv}}\right)^{x_n}
		\tilde{\mathcal{F}}_{n,4}(\{\eta_i\}),
\end{align}
where we have made explicit the dependence on a UV cutoff length $a$, e.g., 
the lattice spacing, and now the various $D_i$ and $\eta_i$ are written as 
functions of the $w_i$ via Eq.~\eqref{eq:UHPmap}. The quantity $a$ is required on dimensional grounds; we will see 
that it cancels out in the mutual information.

The expressions for the entanglement entropy of the single intervals $A$ and $B$ are similar but simpler, 
in that they involve two point functions in the right half plane,
\be \label{eq:wcorr1int}
		\bra \sigma_n(z_1)\tilde{\sigma}_n(z_2)\ket \,,
\ee
and are given in \cite{2007JSMTE..10....4C}, 
although their expressions need to be modified by the non-universal factors 
$\tilde{\mathcal{F}}_{n,2}(\eta)$ in general. 
The reasoning is the same as above, and we end up with
\bea
\label{eq:wcorr21int}
	\Tr \rho^n_{A} &=  (\tilde{c}_n)^2
		\left(a^2 \abs{\frac{dz}{dw}}_{w_1}\abs{\frac{dz}{dw}}_{w_2}
		\frac{\tilde z_{\bar 2 1} \tilde z_{\bar 1 2}}{z_{2 1} \tilde z_{2 \bar 2} \tilde z_{\bar 1 1} z_{\bar 1 \bar 2}} \right)^{x_n}
		\tilde{\mathcal{F}}_{n,2}(\eta)\,,
\end{align}
where in this case there is only one cross-ratio $\eta$, which 
we discuss in Appendix~\ref{sec:crossratios}.
The expression for $\Tr \rho^n_{B}$ is the same, with $w_1$ and $w_2$
replaced by $w_3$ and $w_4$. 

From Eqs.~\eqref{eq:wcorr2} and \eqref{eq:wcorr21int} we can compute the R\'enyi 
entropies $S_{A \cup B}^{(n)}$ and $S_A^{(n)}$, using Eq.~\eqref{eq:Renyi}, and the entanglement entropy through Eq.~\eqref{eq:vNdef}. 
For the latter we have:
\begin{align}
S_A &= - \frac c 6 \ln \(a^2 \abs{\frac{dz}{dw}}_{w_1}\abs{\frac{dz}{dw}}_{w_2}\frac{\tilde z_{\bar 2 1} \tilde z_{\bar 1 2}}{z_{2 1} \tilde z_{2 \bar 2} \tilde z_{\bar 1 1} z_{\bar 1 \bar 2}}\) +2 \tilde c_1' + \tilde{\cal F}'_{1,2} (\eta) \label{eq:SAsingle}\\
S_{A\cup B}&= - \frac c 6 \ln \( a^4 \abs{\frac{dz}{dw}}_{w_1}\abs{\frac{dz}{dw}}_{w_2}
		\abs{\frac{dz}{dw}}_{w_3}\abs{\frac{dz}{dw}}_{w_4} \frac{D_u D_v}{D_{uv}} \) + 4 \tilde c_1' + \tilde{\cal F}'_{1,4} (\{\eta_i\})  \,, \label{eq:Sdouble}
\end{align}
where $\tilde{\cal F}'_{1,2N}  (\eta) = -\del_n \tilde{\cal F}'_{n,2N} (\eta) |_{n=1}$, $\tilde c_1'  =  -\del_n\tilde c_1 |_{n=1} $ and we have used $\tilde c_1 =  \tilde{\cal F}_{1,N} =1$, which follow from the normalization condition $\Tr  \rho =1$. Analytically continuing $\tau \to it$ gives us the values of the entanglement entropies a time $t$ after the local quench. From these we can get the mutual information in Eq.~\eqref{eq:MI}, as we do in Sec.~\ref{sec:CFT-MI}. 
Alternatively, from the R\'enyi entropies, one can also obtain the R\'enyi information, using Eq.~\eqref{eq:MIRenyi},
which yields the mutual information in the $n\to 1$ limit. 

%%%%%%%%%%%%%%%%%%%%%%%%%%%%%%%%%%%%%%%%%%%%%%%%%%%%%%%%%%%%%%%%%%%%%%%%%%%

\subsection{Explicit expressions in the limit $t, \ell \gg \epsilon$}

%%%%%%%%%%%%%%%%%%%%%%%%%%%%%%%%%%%%%%%%%%%%%%%%%%%%%%%%%%%%%%%%%%%%%%%%%%%

\subsubsection{Single interval}  \label{sec:EE1}

The entanglement entropy for a single interval $(\ell_1, \ell_2)$ following a local quench, Eq.~\eqref{eq:SAsingle}, was obtained in \cite{2007JSMTE..10....4C,2009JPhA...42X4005C} by Calabrese and Cardy and here we briefly summarize their results. We assume that $\ell_2 >0$ and $\ell_2 > |\ell_1|$; all other cases are easily inferred by renaming the interval endpoints. This situation corresponds to twist operator insertions in $W$ at points $w_p = \ell_p +i \tau$ with $p=1,2$, which are mapped to $z_p$, $p=1,2$ through Eq.~\eqref{eq:UHPmap}.  As we explain in detail in Appendix~\ref{sec:crossratios} when discussing the analogous computation of cross ratios, Eq.~\eqref{eq:SAsingle} simplifies significantly after taking $\tau \to it$ in the $t, \ell_1, \ell_2 \gg \epsilon$ limit. We will often refer to the latter as the small $\epsilon$ limit. For example, to leading order in $\epsilon$, we have:
\begin{align}
\epsilon z_1 &= w_1 + \sqrt{w_1^2} = \ell_1 - t + \max [\ell_1,t] -  \min [\ell_1,t] \\
\epsilon \bar z_1 &= \bar w_1 + \sqrt{\bar w_1^2} = \ell_1 + t +\max[\ell_1,t] +  \min[\ell_1,t]
\end{align}
so that
\be
\epsilon z_{1\bar{1}} = \epsilon \abs{z_1 -\bar{z}_1} = 2 |t + \min [\ell_1,t]| = \left\{\begin{array}{lc}
4 t & 0 < t< \ell_1 \\
2(\ell_1 + t) & t> \ell_1
\end{array}\right.\,. 
\ee
In some cases, the leading non-zero term is at a higher order in $\epsilon$. 
The various Taylor expansions were facilitated by the software program \verb"Mathematica".
  
Combining all contributions, to leading non-trivial order in $\epsilon$, for $t < |\ell_1|$ the entanglement entropy reads \cite{2007JSMTE..10....4C,2009JPhA...42X4005C}:
\be \label{eq:Searly}
S(t< |\ell_1|) =  \left\{\begin{array}{lc}
\frac c 6 \ln \frac{4 \ell_2 |\ell_1|}{a^2} + 2 \tilde c_1' +\tilde{\cal F}'_{1,2} (\eta) & \ell_1 <0  \\
\frac c 6 \ln \frac{4 \ell_1 \ell_2 (\ell_2 -\ell_1)^2}{a^2 (\ell_1 +\ell_2)^2} + 2 \tilde c_1' + \tilde{\cal F}'_{1,2}  (\eta)  & \ell_1 >0 \,,
\end{array}\right.
\ee
while for $|\ell_1| < t < \ell_2$:
\be
S(|\ell_1| < t < \ell_2) =  \frac c 6 \ln \frac{4 \ell_2 (\ell_2 -\ell_1) (\ell_2 -t) (t^2 -\ell_1^2)}{\epsilon a^2 (\ell_1+\ell_2)(\ell_2 + t)} + 2 \tilde c_1' + \tilde{\cal F}'_{1,2}  (\eta)
\ee
and for $t>\ell_2$:
\be
S(t>\ell_2) =  \frac c 3 \ln \frac{\ell_2 -\ell_1}{a} + 2 \tilde c_1' + \tilde{\cal F}'_{1,2}  (\eta) \,.
\ee
The cross ratio $\eta$ is given explicitly in Eqs.~\eqref{eq:crosssingle1}-\eqref{eq:crosssingle3} in Appendix~\ref{sec:crossratios}. Notice in particular that $\eta(t > \ell_2) =1$, so that $\tilde{\cal F}'_{1,2}  (\eta)$  vanishes for $t > \ell_2$. 
Moreover, $\tilde{\cal F}'_{1,2} (\eta)$ vanishes for all times if $\ell_2 \gg \ell_1>0$ or for $\ell_1 \gg \ell_2 -\ell_1( >0)$, and in these cases the results above are universal. We plot these universal contributions in Figures~\ref{fig:1-int-ent} and \ref{fig:onesideoverlap} below. 

The early time expressions in Eq.~\eqref{eq:Searly} correspond respectively to the sum of ground state entanglement entropies for two slits $(0, |\ell_1|)$, $(0, \ell_2)$ in a half line and to the ground state entanglement entropy for a slit $(\ell_1, \ell_2)$ in a half line \cite{2007JSMTE..10....4C,2009JPhA...42X4005C}.  

As in \cite{2007JSMTE..10....4C,2009JPhA...42X4005C}, we fix the regulator $\epsilon$ in terms of the non-universal constant $\tilde c_1'$ by demanding that at $t=0$ the two half lines are disentangled, i.e.
\be \label{relationaepsilon}
\frac{\epsilon}{a} = \frac{e^{-6 \tilde c_1' /c}}{2}\,.  
\ee
It is then possible to eliminate the UV cutoff $a$ and the constant $\tilde c_1'$ from the formulae above in favor of $\epsilon$ alone, leading to:
\be \label{eq:EEsingle1}
S(t< |\ell_1|) =  \left\{\begin{array}{lc}
\frac c 6 \ln \frac{\ell_2 |\ell_1|}{\epsilon^2} + \tilde{\cal F}'_{1,2}  (\eta) & \ell_1 <0  \\
\frac c 6 \ln \frac{\ell_1 \ell_2 (\ell_2 -\ell_1)^2}{\epsilon^2 (\ell_1 +\ell_2)^2} + \tilde{\cal F}'_{1,2}  (\eta)  & \ell_1 >0 \,,
\end{array}\right.
\ee
\be\label{eq:EEsingle2}
S(|\ell_1| < t < \ell_2) =  \frac c 6 \ln \frac{\ell_2 (\ell_2 -\ell_1) (\ell_2 -t) (t^2 -\ell_1^2)}{\epsilon^3 (\ell_1+\ell_2)(\ell_2 + t)} +\tilde{\cal F}'_{1,2}  (\eta)
\ee
and
\be\label{eq:EEsingle3}
S(t>\ell_2) =  \frac c 3 \ln \frac{\ell_2 -\ell_1}{2 \epsilon} \,.
\ee
In the case of a symmetric interval $-\ell_1 = \ell_2 \equiv \ell$, the universal contribution to the entanglement entropy reduces to
\be \label{eq:EEsinglesymm}
S= \frac{c}{3} \ln \frac{\ell}{\epsilon}
\ee
for all times. This constant behavior is expected from the quasiparticle 
picture, since at no time during the whole evolution are there quasiparticles inside 
the interval that are entangled with quasiparticles outside.

%%%%%%%%%%%%%%%%%%%%%%%%%%%%%%%%%%%%%%%%%%%%%%%%%%%%%%%%%%%%%%%%%%%%%%%%%%%

\subsubsection{Two intervals} \label{sec:EE2}

\begin{figure}[h]
\centering
\def\svgwidth{8.0cm}
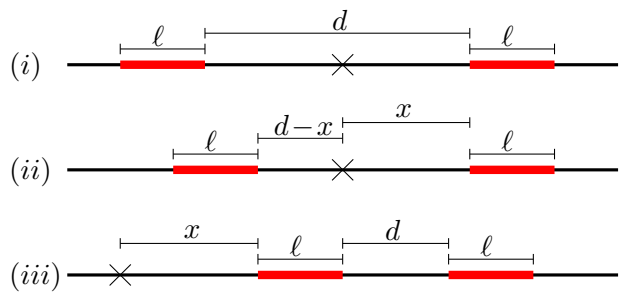
\caption{We illustrate various configurations of intervals and associated 
parameters: $(i)$  
symmetric intervals, $(ii)$ asymmetric intervals and $(iii)$ intervals 
on the same side of the quench. 
}
\label{fig:intervals}
\end{figure}

The formula for the two-interval entanglement entropy in Eq.~\eqref{eq:Sdouble}, 
written in terms of the interval endpoints $u_p$, $v_p$ and time $t$, 
is quite complicated in general, but we may write simpler analytic forms in certain cases.
For example, we can consider the case of symmetric intervals: $A = [-(d/ 2 +\ell), -d/ 2]$, $B=[d/ 2, d/2 +\ell]$ (see Fig.~\ref{fig:intervals} $(i)$), for which the leading order expression in $\epsilon$ is
\beq
\label{eq:SAB-T2}
	S_{A\cup B} (t) = \frac{c}{3} 
	\ln \left( \frac{d l^2 (2l + d)}{a^2 (d+l)^2} \right) + 4 \tilde c_1' + \tilde{\cal F}'_{1,4} (\{\eta_i\}) 
\eeq
for all times. 
The universal part of this expression is equal to the universal contribution to $S_{A\cup B}$ for two intervals of length $\ell$ and 
separation $d$ in the ground state of an infinite two-dimensional CFT, as can be seen 
from Eq.~\eqref{eq:globalCFT}. This constant behavior is expected from the quasiparticle 
picture, as in the symmetric single interval case discussed above.

It is useful to also be slightly more general, and consider equal length, asymmetric intervals 
$A = [-(d-x+\ell),-(d-x)], B= [x, x+\ell]$ (see Fig.~\ref{fig:intervals} $(ii)$). Here $x$ is the amount the intervals have been shifted (to the right) from 
the configuration $A = [-(d+\ell),-d], B= [0,\ell]$. 
This corresponds to insertions of twist operators at $w_1 = -(d-x+\ell) + i\tau, w_2 = -(d-x) + i\tau, 
w_3 = x+i\tau, w_4 =x+\ell+i\tau$ and without loss of generality we have assumed
$0< d-x<x$. 

\begin{figure}[h]
\begin{tabular}{ccc}
\includegraphics[width=0.4\textwidth]{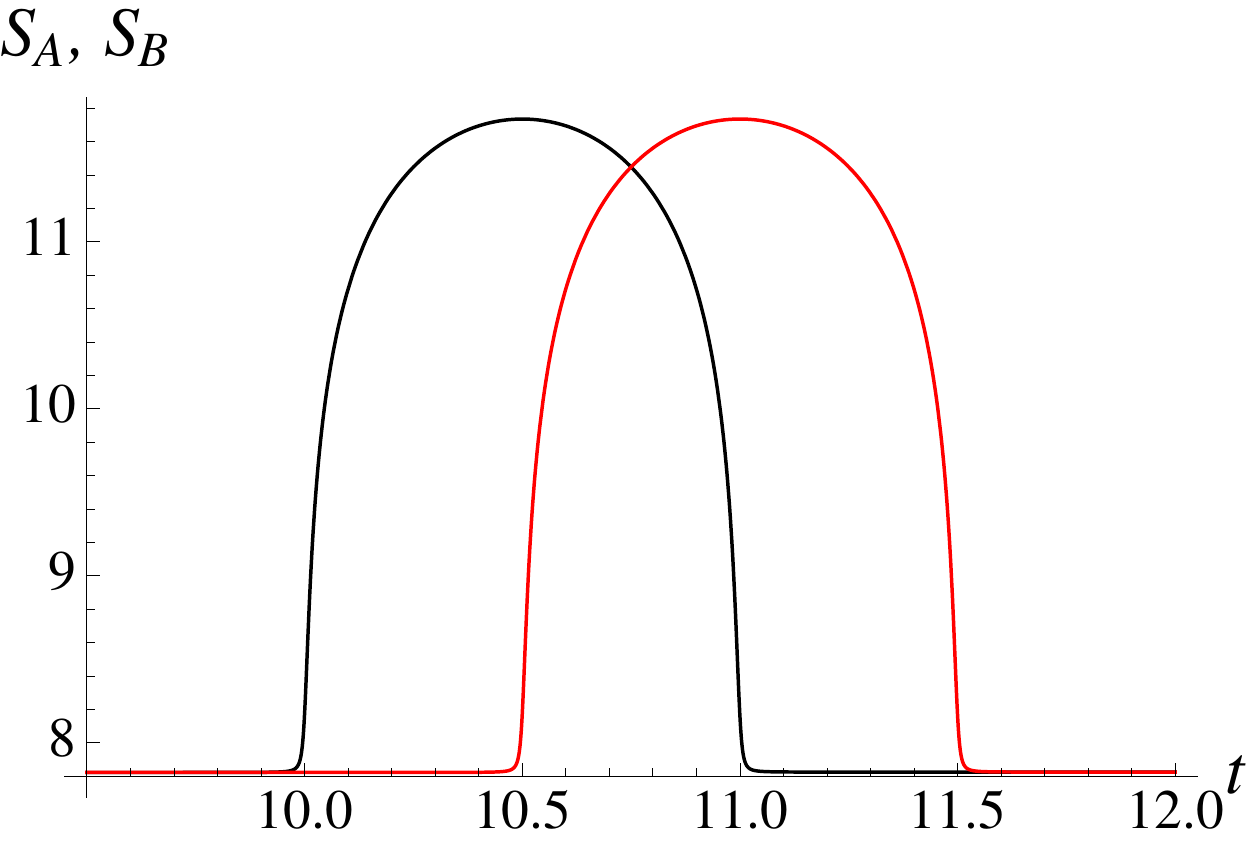} 
&\qquad\qquad\qquad &
\includegraphics[width=0.4\textwidth]{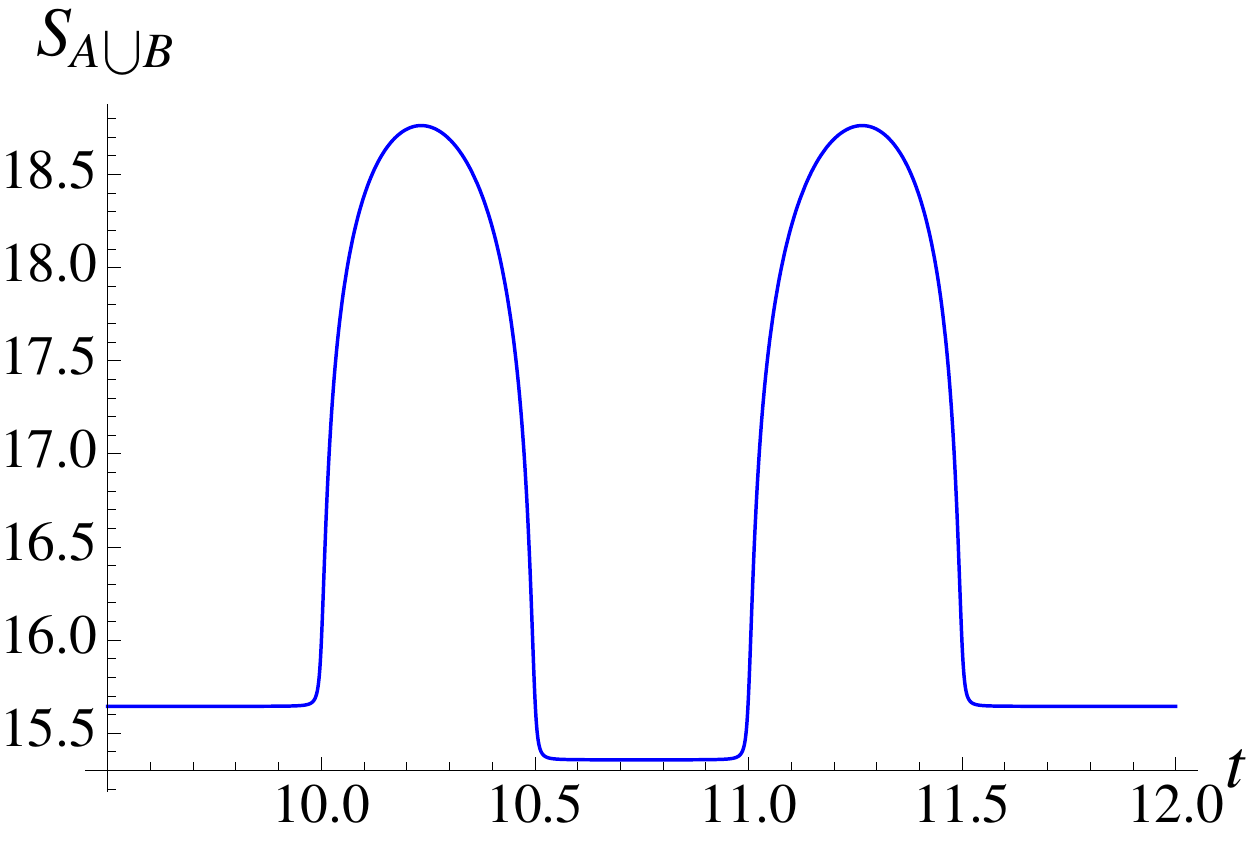}
\\
{\bf (A)} &\qquad \qquad\qquad & {\bf (B)}
\\
\includegraphics[width=0.4\textwidth]{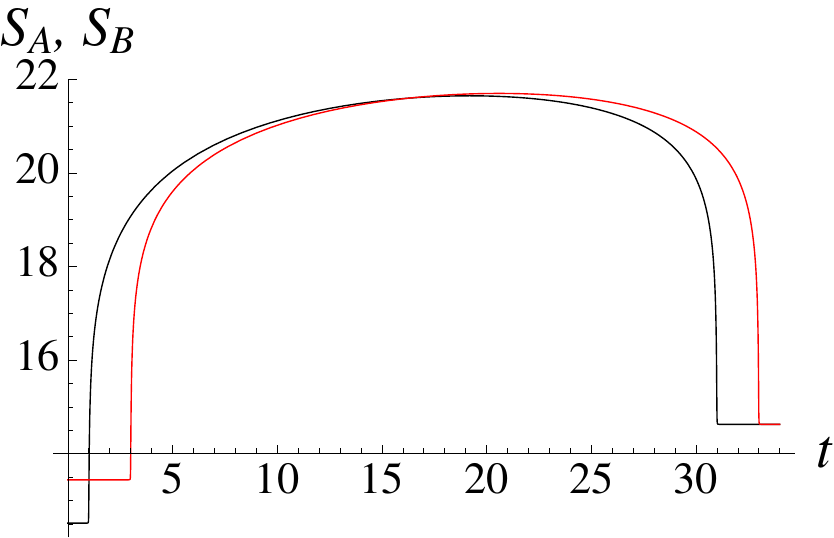} 
&\qquad \qquad\qquad&
\includegraphics[width=0.4\textwidth]{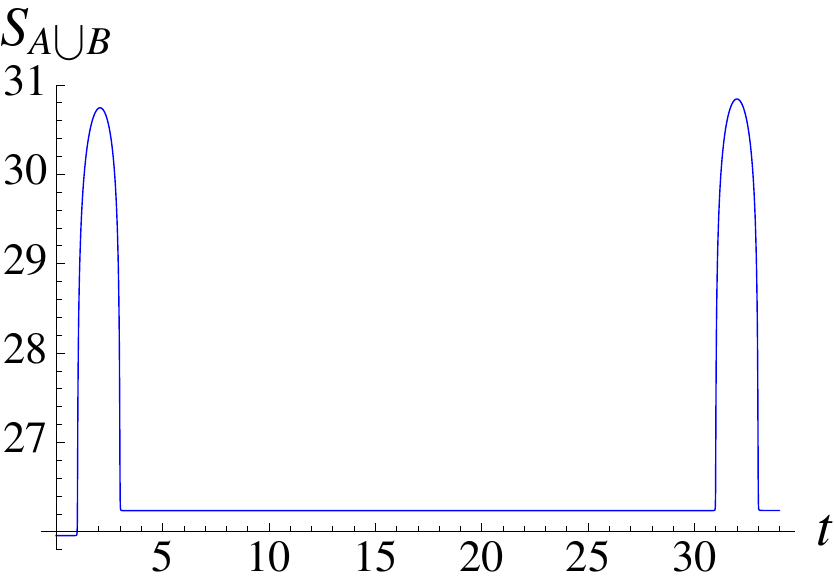}
\\
{\bf (C)} &\qquad\qquad\qquad & {\bf (D)}
\end{tabular}
\caption{Entanglement entropies (rescaled by $c/6$) for one and two asymmetric intervals, in the universal regimes. In the top plots $A = [-11,-10]$ and $B =[10.5,11.5]$, below $A=[-31,-1]$, $B=[3,33]$. Here $\epsilon =0.01$ and $a, \tilde c_1' $ have been eliminated in favor of $\epsilon$ using Eq.~\eqref{relationaepsilon}.
}
\label{fig:1-int-ent}
\end{figure}

In Fig.~\ref{fig:1-int-ent} we show the entanglement entropies $S_A$, $S_B$ and $S_{A\cup B}$ as a function of time. We choose the intervals such that the contributions from the non-universal functions are suppressed.
As derived in Appendix~\ref{sec:crossratios}, this means we consider the two regimes $d/\ell \gg1$ and $d/\ell \ll1$. 
The basic features of these plots can be understood from the quasiparticle picture.
However, as we will see in the next section, the mutual information exhibits some features that go beyond this picture.

It can be seen from Fig.~\ref{fig:1-int-ent} that the time dependence of 
the entanglement entropy of two asymmetric intervals is approximately symmetric about the 
time $t_{\text{mid}} = \frac 1 2 (d + \ell)$, 
and that it is approximately constant outside the interval $[d-x, x+\ell]$. 
There are thus two time intervals with an interesting time dependence: $\Delta t_2 = (d-x,x)$ and $\Delta t_3 = (x, d-x+\ell)$ (when $x< d-x+\ell$). In these time ranges the two-interval entanglement entropy reads respectively
\begin{align}\label{eq:S-T1}
&S_{A\cup B}(t \in \Delta t_2) = \nonumber \\
& \frac c 6 \ln \( \frac{16 x d \ell^3(\ell+d-x)(\ell+x)(\ell-d +2 x)(t^2-(d-x)^2)(x-t) (\ell+d-x-t)(\ell +x+t)}{a^4 \epsilon (2 x-d)(\ell+2 d-2x)(\ell+2 x)^2 (\ell+d) (x+t)(\ell+d-x+t) (\ell+x-t)}\) \nonumber\\
&+ 4 \tilde c_1' + \tilde{\cal F}'_{1,4} (\{\eta_i\})
\end{align}
and 
\beq
\label{eq:S-T2}
S_{A\cup B} (t \in \Delta t_3)  =\frac c 6  \ln \left( \frac{ 4 d^2 \ell^2 (\ell+d-x)(\ell+x)(\ell^2-(d-2x)^2)} {a^4 (\ell+2 d -2x)(\ell+2x)(\ell+d)^2}  \right)+ 4 \tilde c_1' + \tilde{\cal F}'_{1,4} (\{\eta_i\})\,. 
\eeq
In the interval $\Delta t_3 = (x, d-x + \ell)$ the entropy is 
 constant, as also shown in Fig. \ref{fig:1-int-ent}.  
At early times, $t \in \Delta t_1 = (0,d-x)$, and late times, $t\in \Delta t_5  =(x+\ell, \infty)$,
$S_{A\cup B}$ is also approximately constant. 
At early times, it is the sum of the entropies for 
two intervals each in the ground state of a semi-infinite BCFT, 
while at late times it is the entropy for 
two intervals both in the ground state of a connected, 
infinite CFT.

\begin{figure}[h]
\begin{tabular}{ccc}
\includegraphics[width=0.4\textwidth]{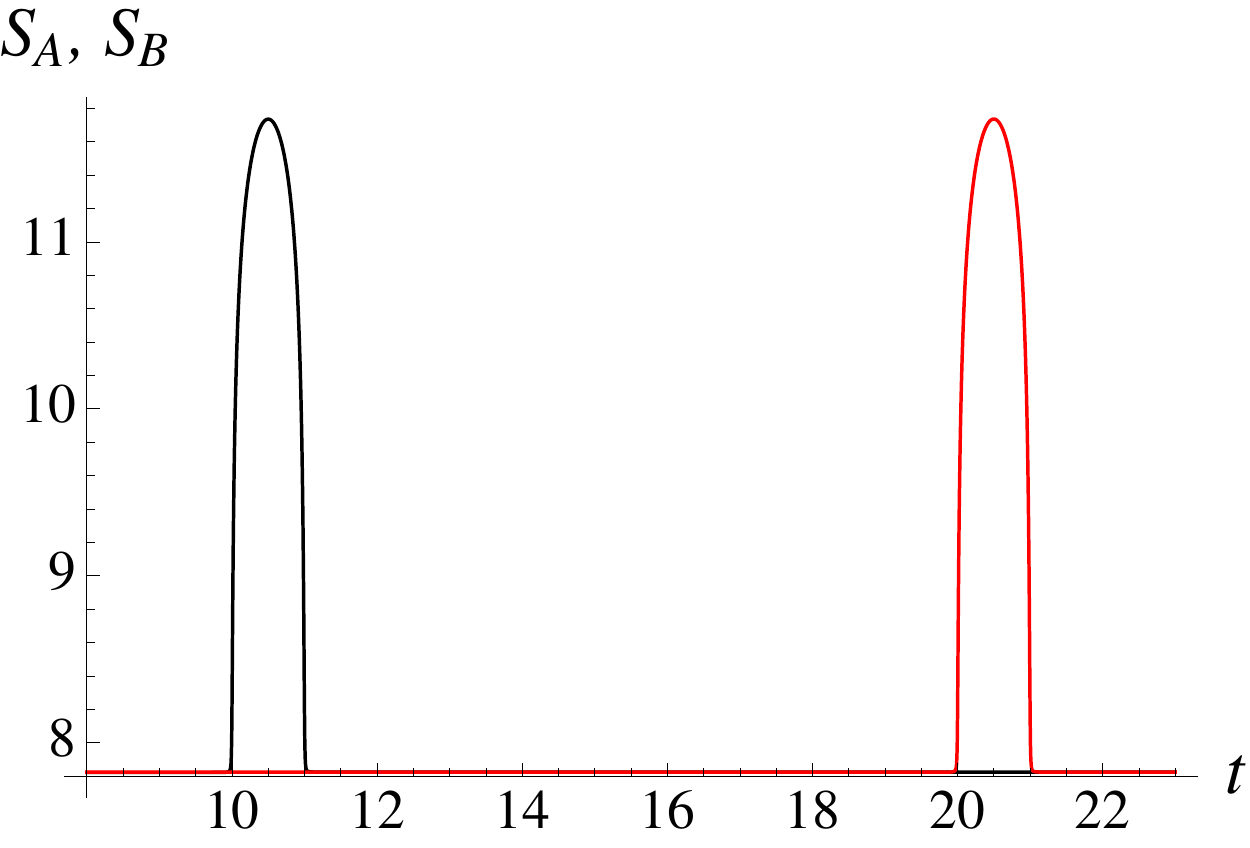}
&\qquad\qquad\qquad &
\includegraphics[width=0.4\textwidth]{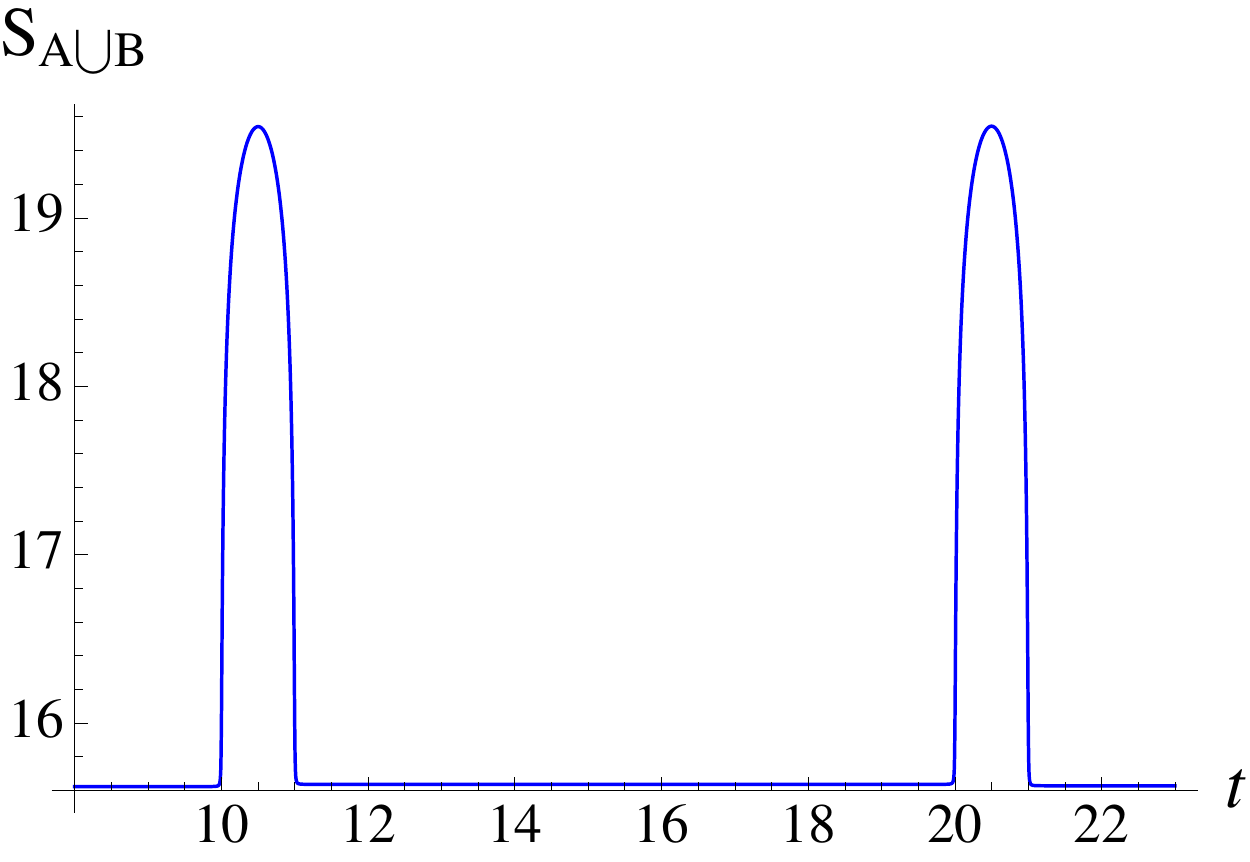}
\\
{\bf (A)}&\qquad\qquad\qquad & {\bf (B)}
\\
\includegraphics[width=0.4\textwidth]{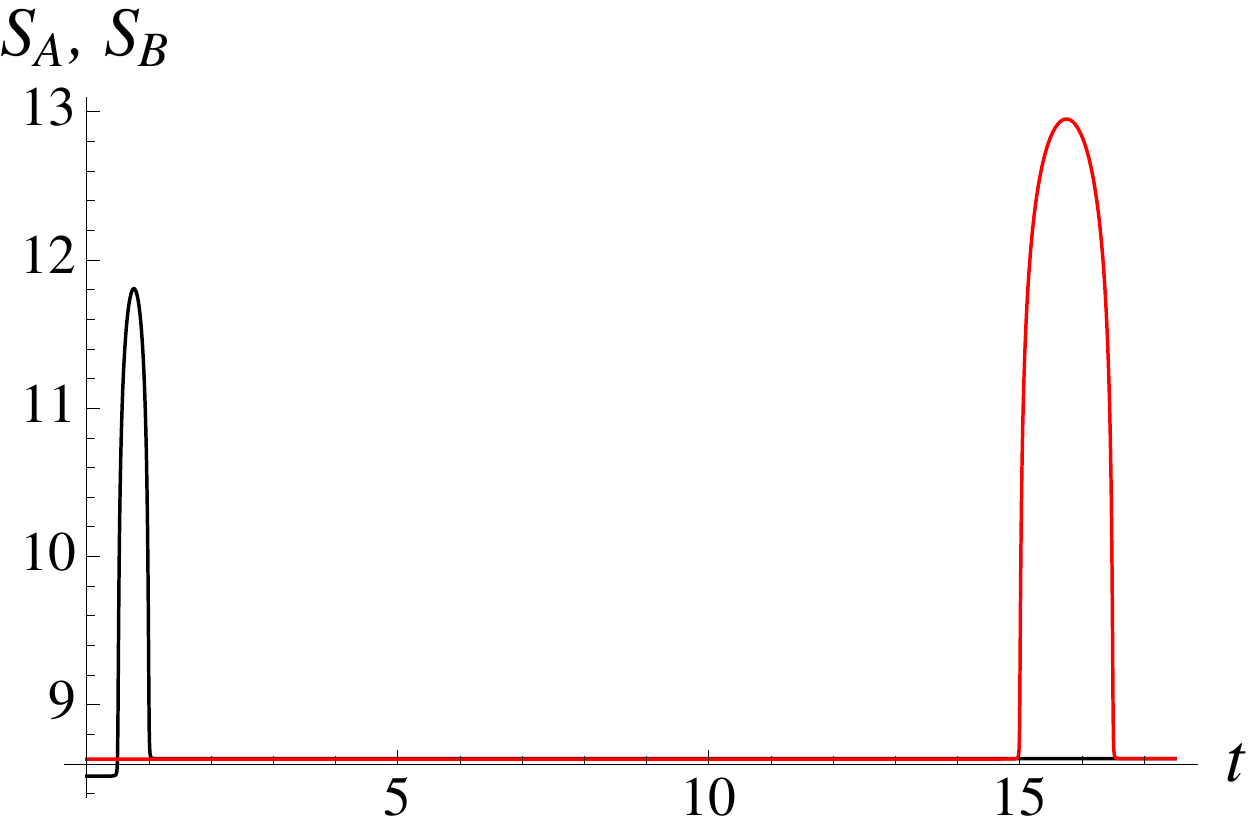} 
&\qquad\qquad\qquad &
\includegraphics[width=0.4\textwidth]{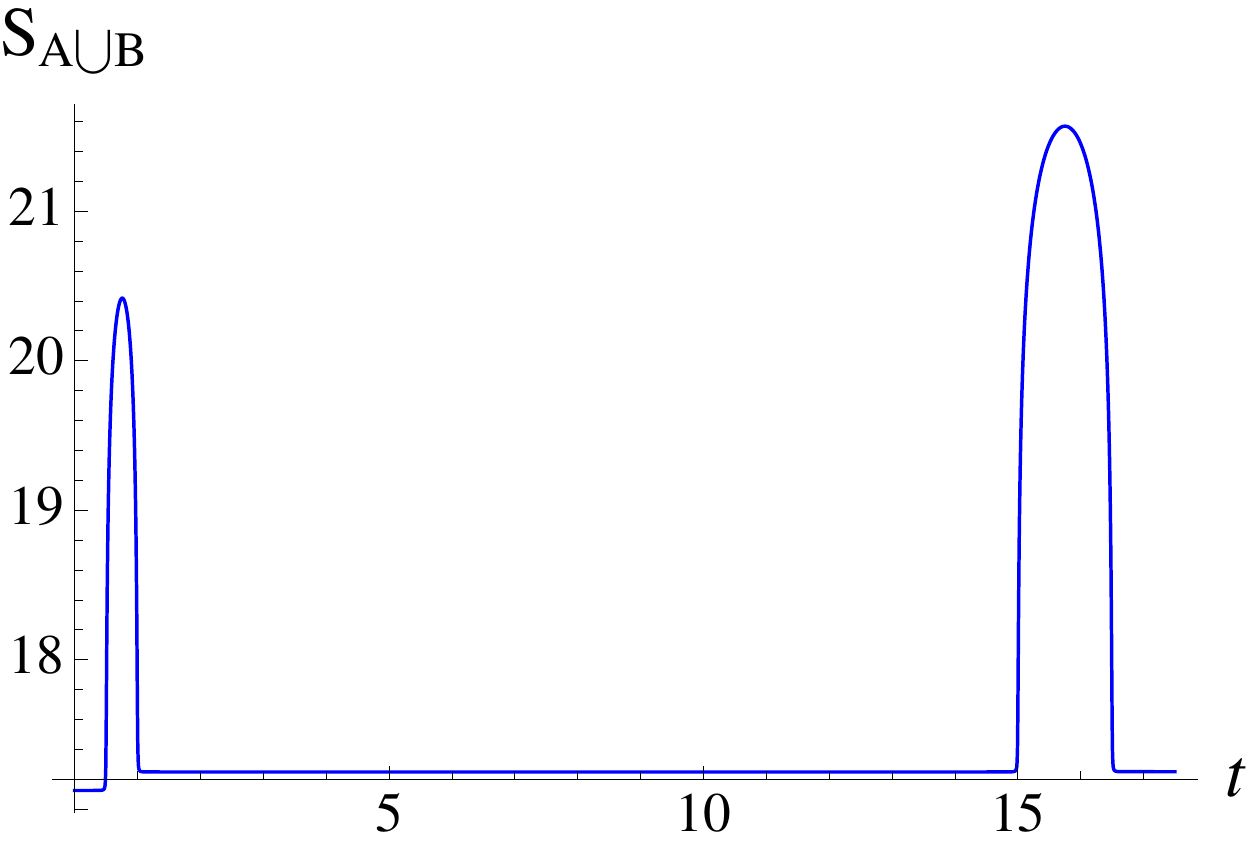}
\\
{\bf (C)} &\qquad\qquad\qquad & {\bf (D)}
\end{tabular}
\caption{Entanglement entropies (rescaled by $c/6$) for one and two intervals on the same side of the quench, or overlapping it, in universal regimes. In the top plots $A = [10,11]$ and $B =[20,21]$, below $A=[-0.5,1]$, $B=[15,16.5]$. Here $\epsilon =0.01$ and $a, \tilde c_1' $ have been eliminated in favor of $\epsilon$ using Eq.~\eqref{relationaepsilon}.
}
\label{fig:onesideoverlap}
\end{figure}

Analogous expressions can be obtained for the case of two intervals lying on the same side of the quench and for the case where one of the two intervals overlaps the defect at the origin, although for compactness we do not report them explicitly here. We plot two such examples in Fig.~\ref{fig:onesideoverlap}. 
%
 
%%%%%%%%%%%%%%%%%%%%%%%%%%%%%%%%%%%%%%%%%%%%%%%%%%%%%%%%%%%%%%%%%%%%%%%%%%%
%%%%%%%%%%%%%%%%%%%%%%%%%%%%%%%%%%%%%%%%%%%%%%%%%%%%%%%%%%%%%%%%%%%%%%%%%%%

\section{Mutual information}
\label{sec:CFT-MI}

From the expressions in Eqs.~\eqref{eq:wcorr21int} and \eqref{eq:wcorr2}, and using Eqs.~\eqref{eq:Renyi} and \eqref{eq:MIRenyi}, one obtains the mutual R\'enyi information:
\begin{align}
\nonumber
	I^{(n)} &= S^{(n)}_A + S^{(n)}_B - S^{(n)}_{A\cup B} \\
	&= -\frac{c(1+n)}{12 n} \ln \left( 
	\frac{ z_{2 3} \tilde z_{2\bar{4}}
	z_{4 1} \tilde z_{4 \bar{2}}
	\tilde z_{\bar 3 1} z_{\bar 3\bar{2}}
	\tilde z_{\bar 1 3} z_{\bar 1 \bar{4}}}
	{ z_{\bar 2 \bar 4}\tilde z_{\bar{2} 3} \tilde z_{\bar{4}1 }z_{3 1}
	z_{\bar{1} \bar 3} \tilde z_{\bar{1} 4} \tilde z_{\bar{3} 2} z_{42}}  \right) + \frac{1}{1-n} \ln \frac{ \tilde{\mathcal{F}}_{n,2}(\eta_A) \tilde{\mathcal{F}}_{n,2}(\eta_B)}{\tilde{\mathcal{F}}_{n,4}(\{\eta_i\})} \,,
\end{align}
where notice that all the Jacobian terms and non-universal constants $a,\tilde c_n$ cancel out.  
Taking the limit $n\to 1$, or directly from the entanglement entropy expressions in Eqs.~\eqref{eq:SAsingle} and \eqref{eq:Sdouble}, one obtains the mutual information 
\begin{align}\label{eq:MIz}
	I &= -\frac{c}{6} 
	 \ln \left( 
	\frac{ z_{2 3} \tilde z_{2\bar{4}}
	z_{4 1} \tilde z_{4 \bar{2}}
	\tilde z_{\bar 3 1} z_{\bar 3\bar{2}}
	\tilde z_{\bar 1 3} z_{\bar 1 \bar{4}}}
	{ z_{\bar 2 \bar 4}\tilde z_{\bar{2} 3} \tilde z_{\bar{4}1 }z_{3 1}
	z_{\bar{1} \bar 3} \tilde z_{\bar{1} 4} \tilde z_{\bar{3} 2} z_{42}}  \right)
	+ \tilde{\mathcal{F}}'_{1,2}(\eta_A) + \tilde{\mathcal{F}}'_{1,2}(\eta_B)- \tilde{\mathcal{F}}'_{1,4}(\{\eta_i\})  \\
	&=  -\frac{c}{6} \ln\left\vert \eta_{8361}
	\eta_{8541} \eta_{7632} \eta_{7452}
	\right\vert
	+ \tilde{\mathcal{F}}'_{1,2}(\eta_A) + \tilde{\mathcal{F}}'_{1,2}(\eta_B)- \tilde{\mathcal{F}}'_{1,4}(\{\eta_i\})  \,.
\end{align}

We now take the analytic continuation to real time and consider specific cases to obtain more insight 
into the evolution of the mutual information after a local quench. 
For simplicity, from here onwards we neglect the non-universal functions in Eq.~\eqref{eq:MIz} and so the results we present below hold only in the universal regimes. 

%%%%%%%%%%%%%%%%%%%%%%%%%%%%%%%%%%%%%%%%%%%%%%%%%%%%%%%%%%%%%%%%%%%%%%%%%%%

\subsection{Symmetric intervals}

\begin{figure}[h]
\begin{center}
\includegraphics[width=0.5\textwidth]{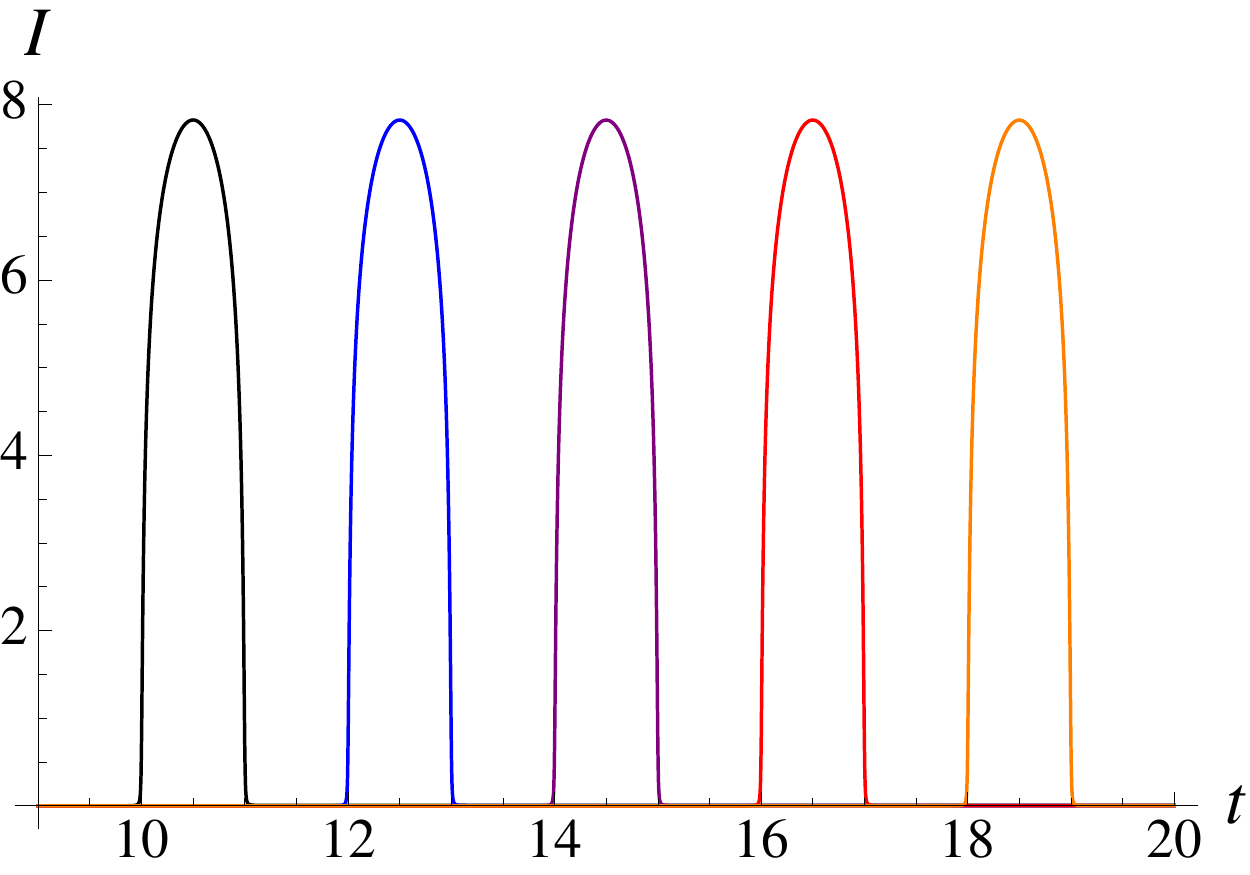}
\end{center}
\caption{Time evolution of the mutual information (rescaled by $c/6$) for symmetric intervals of length $\ell=1$ at distances
$d=20,24,28,32,36$ from left to right. Here we have set $\epsilon = 0.01$.
}
 \label{fig:MI_trans}
\end{figure}

We first consider the case of symmetric intervals $A =[-( d/ 2+\ell),- d/ 2]$, $B= [d/ 2,d/ 2+ \ell]$, see Fig.~\ref{fig:intervals} $(i)$.
In accordance with the quasiparticle picture, the mutual information is always
non-zero only in the time range $t \in [d/2, d/2+\ell]$, where its graph has the hump-shaped profile shown in Fig.~\ref{fig:MI_trans}. 
The latter, to leading order in $\epsilon$ and for large enough $d/\ell$, is independent of the separation between the two intervals, as can be observed in Fig.~\ref{fig:MI_trans}. 

As for the entanglement entropy, we can obtain explicit analytic results in the limit of small $\epsilon$ as compared to all other length and times scales. 
To leading non-trivial order in $\epsilon$, we have: 
\be\label{eq:MI_sym}
I  = \left\{\begin{array}{lc}
0  & t< \frac d 2 \\
\frac c 3 \ln  \frac{2(d+\ell) \(\frac d 2+ \ell- t\)\(t^2-\frac{d^2}{4}\)}{ \epsilon \, d \, \ell \(\frac d 2+\ell+t\)} & \frac d 2 < t< \frac d 2 +\ell \\
\frac c 3 \ln \frac{(d+\ell)^2}{d(d+2\ell)} & t> \frac d 2 +\ell 
\end{array}\right.
\ee
where recall that all non-universal factors are negligible in the two limits $d/\ell \gg 1$ or $d/\ell \ll 1$. 

The time dependence of the universal contribution to $I$ for the symmetric intervals case (the humps 
in  Fig.~\ref{fig:MI_trans}) is identical to that of the entanglement entropy
for a single interval after a local quench in Eq.~\eqref{eq:EEsingle2}. This is because in that time interval the universal part of $S_{A\cup B}$ is  
constant to leading order in $\epsilon$, as discussed in Sec.~\ref{sec:EE2}.
The approximate location of the maximum is $t_{\text{mid}} = (d+\ell)/2$, at which the mutual information for well separated intervals with $d/\ell \gg1$ takes the value 
\beq
\label{eq:smallr}
I(t_{\text{mid}}) \approx \frac c 3 \ln \frac{\ell}{2 \epsilon}\,,
\eeq
which is indeed $d$ independent as anticipated above. This creation of long range entanglement is an important physical effect 
that we discuss further in the Conclusions in Sec.~\ref{sec:Concl}.

Note also that the mutual information grows like $\log\frac{1}{\epsilon}$ at intermediate times, as the ratios $\epsilon/ \ell, \epsilon /d , \epsilon /t$ go to zero, while there is no such growth at early and late times. Recall that $\epsilon$ is a UV regulator, 
so that the mutual information during the time interval $(d/2, d/2 +\ell)$ will be dominant as one includes 
higher energy scales.

%%%%%%%%%%%%%%%%%%%%%%%%%%%%%%%%%%%%%%%%%%%%%%%%%%%%%%%%%%%%%%%%%%%%%%%%%%%

\subsection{Equal length asymmetric intervals}
\label{sec:form-MI}

\begin{figure}[h]
\begin{center}
\includegraphics[width=0.5\textwidth]{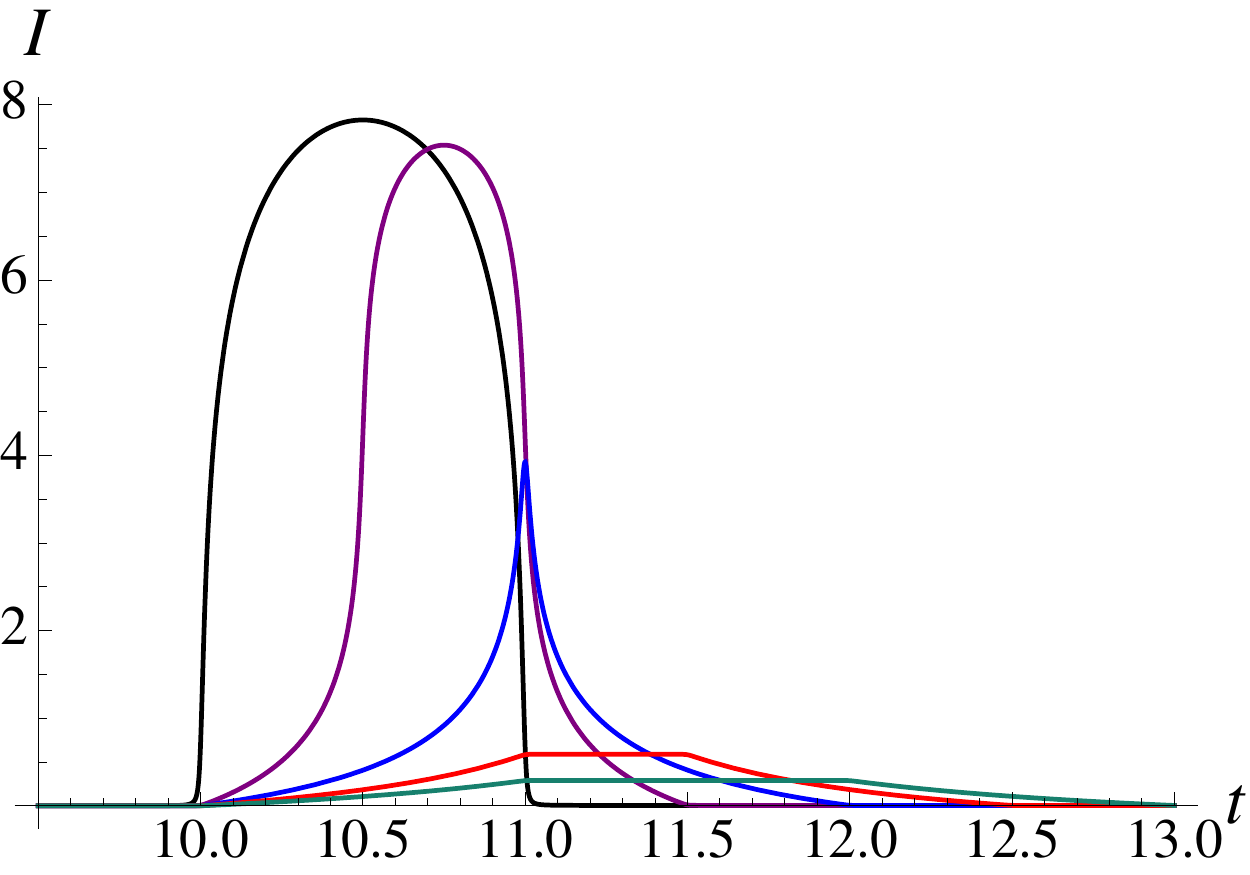}
\end{center}
\caption{Mutual information $I_{[-11,-10],[10 + j/2, 11 + j/2]}$ (rescaled by $c/6$) 
as a function of time for $j=0,1,2,3,4$ (from left to right) with $\epsilon = 0.01$. 
}
 \label{fig:MI-plot2}
\end{figure}

The case of asymmetric intervals $A = [-(d-x+\ell), -(d-x)]$ $B= [x, x+\ell]$, 
see Fig.~\ref{fig:intervals} $(ii)$, is richer, and we plot some examples in Fig.~\ref{fig:MI-plot2}.
The general behavior is again consistent with the quasiparticle picture and 
can be analyzed further using explicit analytic expressions in the limit of small $\epsilon$. 

We consider first the case $ d-x < x< d-x+\ell$ and as before define the time intervals: $\Delta t_1 = (0, d-x)$, $\Delta t_2 =(d-x, x)$, $\Delta t_3 =(x, d-x+\ell)$,  $\Delta t_4 = (d-x+\ell, x+\ell)$ and $\Delta t_5 = (x+\ell, \infty)$. $I$ is constant during $\Delta t_1$ and $\Delta t_5$ and coincides respectively with the early and late time value that we obtained for a symmetric interval:
\begin{align}
I(t \in \Delta t_1) &=0 \,,\label{eq:MI-Dt1} \\
 I (t \in \Delta t_5) &=\frac c 3 \ln \frac{(d+\ell)^2}{d(d+2 \ell)} \,. \label{eq:MI-Dt5}
\end{align}
For asymmetric intervals at intermediate times, the hump of the symmetric case is instead deformed to
\begin{align}
I(t \in \Delta t_2) &=  \frac c 6 \ln \frac{(2x-d)(d+\ell)(x+t)(\ell+x-t)}{d(\ell-d+2x)(x-t)(\ell+x+t)} \,,\label{eq:MI-T1}  \\
I (t \in \Delta t_3) &=\frac c 6 \ln 
\frac{4(x+\ell) (d+\ell)^2 (d-x+\ell-t)(t^2 -(d-x)^2)(x+\ell-t)(t^2-x^2)}{\epsilon^2 d^2 (\ell+x)(\ell^2 -(d-2x)^2)(d-x+\ell+t)(x+\ell+t)}  \,, \label{eq:MI-T2} \\ 
I (t \in \Delta t_4) & = \frac c 6 \ln \frac{(2x-d)(d+\ell)^3 \(t^2-(d-x)^2\)}{d^2(d+2\ell)(\ell -d+2x)\(t^2 -(d-x+\ell)^2 \)} \,. \label{eq:MI-Dt4} 
\end{align}
The non-universal functions can be neglected in the regimes $d-x, x \ll \ell$ and $d-x, x \gg \ell$ (see Appendix~\ref{sec:crossratios}).

In the case in which $x> d-x+\ell$, corresponding to the last two curves in Fig.~\ref{fig:MI-plot2}, the mutual information instead reaches a constant at intermediate times
\be
I(d-x+\ell < t< x) = \frac{c}{6} \ln \frac{(d+\ell)^2(d-2x)^2}{d(d+2\ell)((d-2x)^2-\ell^2)}\,. 
\ee
This behavior, as well as that shown in Eqs.~\eqref{eq:MI-T1} and 
\eqref{eq:MI-Dt4}, is not captured by the quasiparticle picture. Eqs.~\eqref{eq:MI-T1} and 
\eqref{eq:MI-Dt4} show interpolation between the values for the mutual information in the 
ground states of the half-line BCFTs and the infinite CFT.
Note that the mutual 
information in these time periods is independent of $\epsilon$, and so is, in 
some sense, independent of the strength of the quench.

%%%%%%%%%%%%%%%%%%%%%%%%%%%%%%%%%%%%%%%%%%%%%%%%%%%%%%%%%%%%%%%%%%%%%%%%%%%

\subsection{Intervals on the same side of the quench}
\label{sec:MI_same_side}

It is also interesting to consider a pair of non-overlapping intervals on the same side of the 
quench, $A = [x, x + \ell]$, $B = [x+\ell+d,x +2\ell +d]$, where $x, \ell,d> 0$, 
see Fig.~\ref{fig:intervals} $(iii)$. 
Similarly to what we did above, we consider the time intervals 
$\delta t_1 = (0,x)$, $\delta t_2 = (x,x+ \ell)$, $\delta t_3 = (x + \ell, x+\ell+d)$, $\delta t_4 = (x+\ell+d, x +2 \ell +d)$ and $\delta t_5 = (x +2 \ell + d, \infty)$.

In the time intervals $\delta t_1, \delta t_3$ and $\delta t_5$ the leading order behavior in the small $\epsilon$ expansion is time-independent.
Proceeding as above we obtain
\beq
\label{eq:SST1}
	I (t \in \delta t_1) = \frac c 3 \ln \frac{(\ell+d)^2 (2x +2 \ell+d)^2}{d(2x+\ell+d)(2\ell+d)(2x+3\ell+d)}\,.
\eeq
In this first time interval we effectively have two intervals in the ground state 
of a semi-infinite BCFT, which is time-independent and dependent on 
the separation $d$, lengths $\ell$ and distance from 
the boundary $x$, as expected. In the case where $x = 0$ this takes 
the simple form
\beq
	I  (t \in \delta t_1) \Big|_{x=0} = \frac{c}{3} \ln 
	\frac{(\ell+ d) (2 \ell +d)}{d (3 \ell + d)} \,,
\eeq
while in the limit that the intervals are far from the boundary, 
$x \gg d, \ell$, we have
\beq
	I(t \in \delta t_1) \Big|_{x \gg d, \ell}  = \frac{c}{3} \ln 
	\frac{(\ell+ d)^2}{d (2 \ell+d)}  \,. 
\eeq
The latter matches the value for the mutual information between two intervals in
the ground state of an infinite CFT 
(in the regime where also $\tilde{\mathcal{F}}_{1,4} = 1$). 
We also expect this to be the 
value for the mutual information during $\delta t_5$,
after both intervals have been ``notified" that they are in a 
connected, infinite CFT, and indeed we have
\beq
\label{eq:SST5}
	I (t \in \delta t_5) = \frac{c}{3} \ln \frac{(\ell+d)^2}{d (2 \ell+d)} \,.
\eeq
The interval $\delta t_3$, during which the quasiparticles emitted from the quench 
are passing between $A$ and $B$, is an intermediate time where the mutual information is 
constant, but does not correspond to the ground state value for either an 
infinite or semi-infinite CFT:
\beq
\label{eq:SST3}
	I (t \in \delta t_3) =  \frac c 6 \ln \frac{(\ell+d)^2 (2x +2\ell+d)^2}{d(2x+\ell+d)(2\ell+d) (2x+3\ell+d)} \,. 
\eeq
It approaches half of the infinite CFT value in Eq.~\eqref{eq:SST5} for  
$x \gg \ell, d$.

\begin{figure}[h]
\begin{tabular}{ccc}
\includegraphics[width=0.45\textwidth]{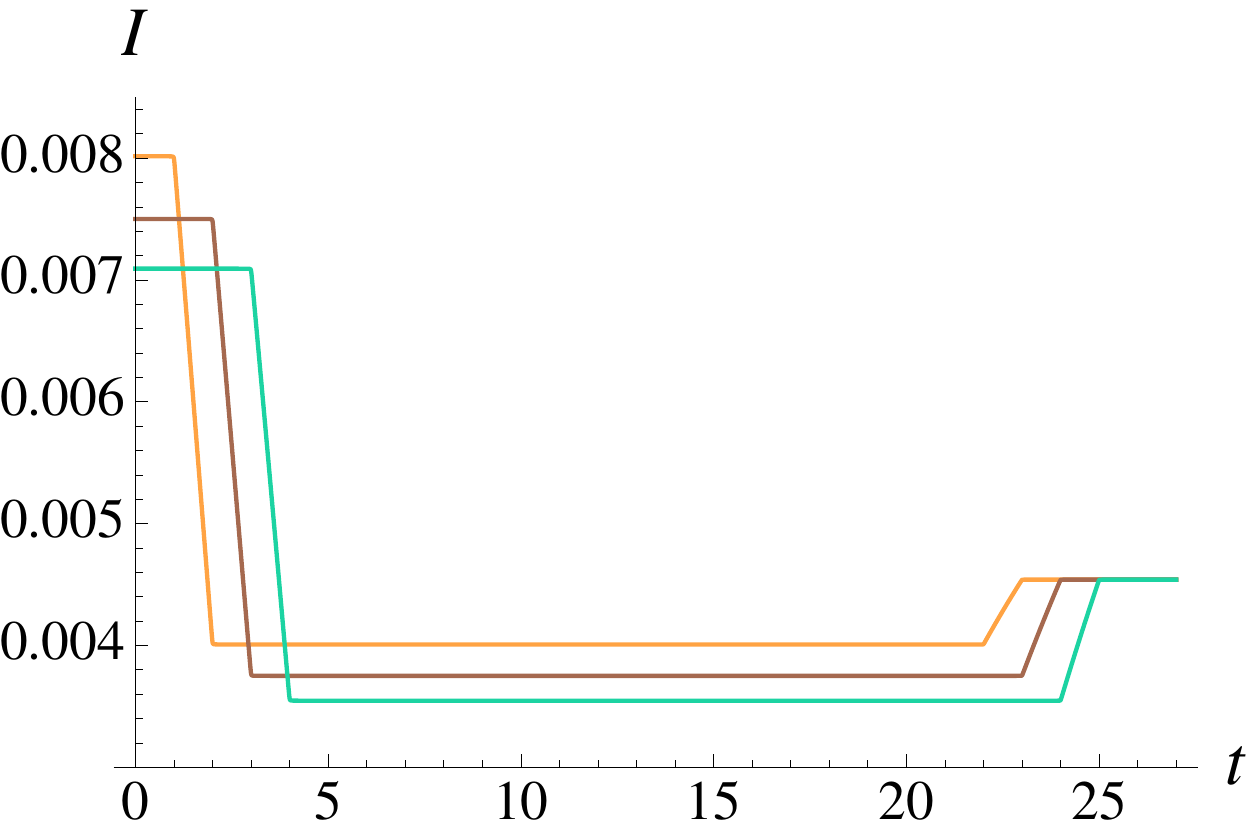}
&\qquad\qquad &
\includegraphics[width=0.45\textwidth]{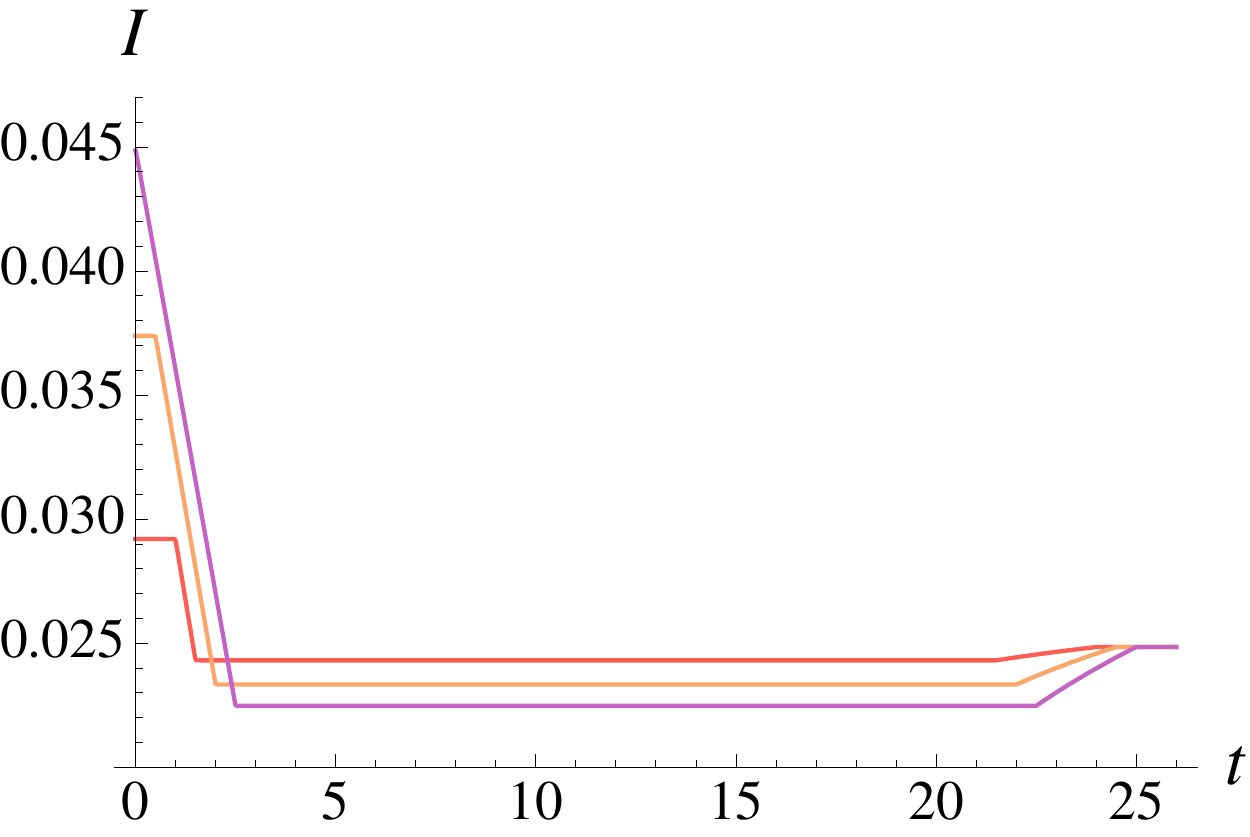}
\\
{\bf (A)} &\qquad \qquad & {\bf (B)}
\end{tabular}
\caption{Evolution of the mutual information (rescaled by $c/6$) in the universal regime for $d=20$ and {\bf (A)} $\ell=1$, $x = 1,2,3$, {\bf (B)} $\ell=2.5$, $x=-1,-0.5,0$ from left to right, and $\epsilon = 0.01$.
}
 \label{fig:MI-sameside-plot1}
\end{figure}

Next we have some time-dependent behavior that transitions between the above 
constant values:
\begin{align}
\label{eq:SST2}
	I (t\in \delta t_2 ) &=  \frac c 6 \ln \frac{(\ell+d)^3 (2x +2\ell+d)^3 (x+\ell+d -t)(x+2\ell+d+t)}{d^2(2x+\ell+d) (2\ell+d) (2x+3\ell+d)^2 (x+\ell+d+t)(x+2\ell+d-t)} \,, \\
\label{eq:SST4}
	I  (t \in \delta t_4)& =  \frac c 6 \ln \frac{(\ell+d)^3(2x+2\ell+d)(t-(x+\ell))(x+\ell+t)}{d^2(2\ell+d)(2x+3\ell+d)(t^2 -x^2)}  \,.
\end{align}
These results for the time evolution of the mutual information for two intervals on the same side of the quench are not corrected by non-universal functions if  $d \gg \ell, x$ or if $x\gg d\gg\ell$, as discussed in App.~\ref{sec:crossratios}. We plot some such examples in Fig.~\ref{fig:MI-sameside-plot1} {\bf (A)}.

Finally, one should also consider the case in which the interval to the left has some overlap with the defect at the origin (i.e. $x<0$). The mutual information is in this case similar to the above, as displayed in Fig.~\ref{fig:MI-sameside-plot1} {\bf (B)}.

%%%%%%%%%%%%%%%%%%%%%%%%%%%%%%%%%%%%%%%%%%%%%%%%%%%%%%%%%%%%%%%%%%%%%%%%%%%
%%%%%%%%%%%%%%%%%%%%%%%%%%%%%%%%%%%%%%%%%%%%%%%%%%%%%%%%%%%%%%%%%%%%%%%%%%%

\section{Energy-momentum tensor}
\label{sec:nrgmom}

One and two-point correlation functions of primary operators after a local 
quench were studied in \cite{2007JSMTE..10....4C}.
As a quasi-primary operator, the energy-momentum tensor has a more complicated 
transformation rule than primary operators. 
Under a general conformal transformation $z(w)$ the 
holomorphic part of the 
energy-momentum tensor transforms by Jacobian factors and 
by the addition of a term proportional 
to the Schwarzian derivative \cite{1984NuPhB.241..333B, DiFrancesco1997}.
The precise form depends on which convention one chooses for the definition 
of the energy-momentum tensor. 
Here we use that of \cite{Balasubramanian:1999re}.
In Lorentzian signature this reads
\be
\label{eq:EM-def}
T^{\mu\nu} \equiv \frac{2}{\sqrt{\abs{g}}} \frac{\delta S}{\delta g_{\mu\nu}}\,,
\ee
which differs in the overall sign from 
\cite{DiFrancesco1997}, but facilitates comparison 
to holographic energy momentum tensors later on. 
We define the Euclidean signature tensor by taking the standard 
tensor transformation rule, see Eq.~\eqref{eq:tens_trans} below, under the change of coordinates 
$\tau = it$. This leads to the following transformation law in complex 
Euclidean coordinates.
\beq
\label{eq:EMtrans}
T_{ww}(w) = \left(\frac{dz}{dw}\right)^2 T_{zz}(z(w)) - \frac{c}{24 \pi}(Sz)(w)\,,
\eeq
where $(Sz)(w)$ is the Schwarzian derivative, given by
\beq \label{eq:Schwarzian}
	(Sz)(w) = \left. \frac{z' z''' - \frac32 (z'')^2}{(z')^2} \right \vert_{w}\,,
\eeq
and $z' = dz/dw$. 
In other words, Eq.~\eqref{eq:EMtrans} gives the pullback of $T_{zz}$ under 
$z(w)$. 
The transformation properties of the anti-holomorphic part $T_{\bar z\bar z}(\bar z)$ are obtained 
by conjugating the above equations.

In the RHP we have $\bra T_{ab}(z) \ket = 0$. 
This may be seen from the boundary version of the conformal Ward identity
\cite{Cardy:2004hm}
\begin{calc}
\bra T_{zz}(z) \phi(w,\bar w)\ket = &{\bigg (}\frac{h}{(z-w)^2} + \frac{1}{z-w}\del_w
	+ \frac{\bar{h}}{(\bar z - \bar{w})^2 } + \frac{1}{\bar{z} + \bar{w}} 
	\del_{\bar{w}} {\bigg )} \bra \phi (w,\bar w)\ket \,,
\end{calc}
by taking $\phi$ to be the identity operator, a primary operator 
with weights $(0,0)$. An analogous equation fixes $\bra T_{\bar z\bar z}(\bar z) \ket =0$.
We can also argue that this is the case as follows.
In the RHP spacetime of a BCFT the global conformal symmetry is reduced 
to the $\text{PSL}(2,\mathbb{R})$ subgroup that preserves the boundary on the 
imaginary axis. Invariance under this group forces the vacuum expectation values 
$\bra T_{zz} (z) \ket$ and $\bra T_{\bar z\bar z}(\bar{z}) \ket$ to be constants. 
The conformal boundary condition, which ensures that no energy or momentum flows across the boundary, 
then implies that these constants are equal on the imaginary axis and so equal everywhere. 
In order to have a finite total energy, this constant has to be zero.

Taking the ground state in the RHP means that there are no operator insertions 
anywhere in the right half plane, and under the action of the map in Eq.~\eqref{eq:UHPmap} this 
translates to no operator insertion in $W$, i.e., we are considering 
the ground states $|\psi_0\ket$ of the disconnected theories as an initial 
condition. 
Taking expectation values of Eq.~\eqref{eq:EMtrans} and applying 
the local quench mapping of Eq.~\eqref{eq:UHPmap},
we obtain:\footnote{The energy-momentum tensor after a local quench was also computed 
in \cite{2011JSMTE..08..019S, Ugajin:2013xxa},
and after a global quench in \cite{Calabrese:2007rg}.}
\beq
\label{eq:EMW}
	\bra T_{ww}(w) \ket = \frac{c}{24\pi} (Sz)(w) =\frac{c}{16\pi} \frac{\epsilon^2}{(w^2 + \epsilon^2)^2}
\eeq
and similarly 
\beq
\label{eq:EMWc}
\bra T_{\bar w\bar w}(\bar w) \ket = \frac{c}{24\pi} (S\bar z)(\bar w) = 
	\frac{c}{16\pi} \frac{\epsilon^2}{(\bar{w}^2 + \epsilon^2)^2} \,.
\eeq
We can transform these to $x = \frac12 (w+ \bar{w})$ and $\tau = \frac{i}{2} (\bar{w} -w)$ 
coordinates using 
\be
\label{eq:tens_trans}
T_{ab}(x,\tau) = \del_a w^c \del_b w^d T_{cd}(x+i\tau,x-i\tau)\,,
\ee
so that 
\begin{align}
T_{\tau\tau}(x,\tau) &= -T_{xx}(x,\tau) = -(T_{ww}(x+i\tau) + T_{\bar w\bar w}(x-i\tau)) \\
T_{\tau x} (x,\tau) &= T_{x \tau}(x, \tau) =  i (T_{ww}(x+i\tau) - T_{\bar w\bar w}(x-i\tau)) \,.
\end{align}
Analytically continuing to real time $\tau \to it$, we have
\begin{align}\label{eq:Ttt}
	\bra T_{tt}(x,t) \ket &= \bra T_{xx}(x,t)\ket = 
	\frac{c}{16\pi} \left(\frac{\epsilon^2}{((x-t)^2 + \epsilon^2)^2} 
		+ \frac{\epsilon^2}{((x+t)^2 + \epsilon^2)^2}\right) \\
	\bra T_{tx}(x,t) \ket &= \bra T_{xt}(x,t)\ket = 
	  -\frac{c}{16\pi} \left(\frac{\epsilon^2}{((x-t)^2 + \epsilon^2)^2} 
		- \frac{\epsilon^2}{((x+t)^2 + \epsilon^2)^2}\right)\,.
\end{align}
We plot the energy density $\bra T_{tt}(x,t) \ket$ in Fig. \ref{fig:Ttt} below. 
\begin{figure}[h]
\begin{center}
\includegraphics[width=0.5\textwidth]{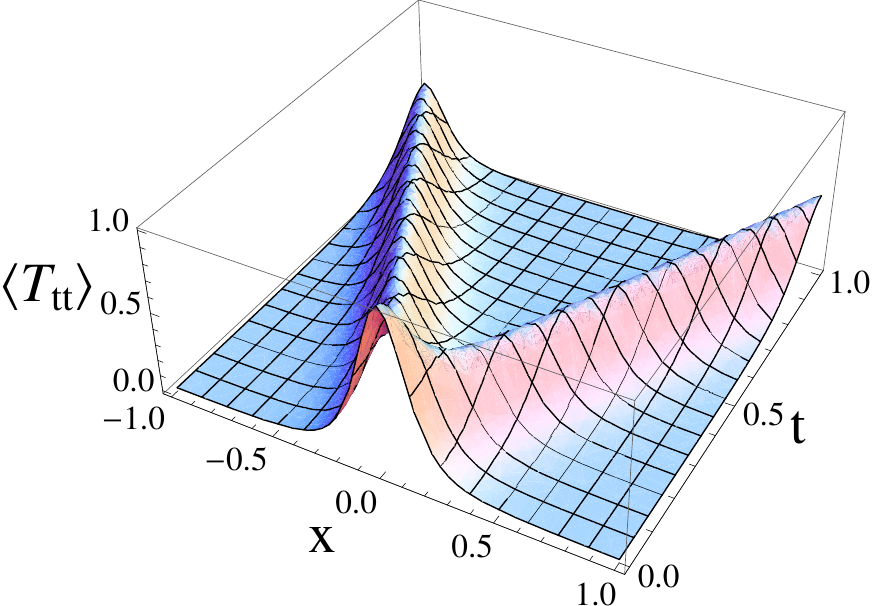}
\end{center}
\caption{The expectation value of the energy density $\bra T_{tt}(x,t) \ket$ given in Eq.~\eqref{eq:Ttt} with $c=1$ and $\epsilon = 0.2$.
}
 \label{fig:Ttt}
\end{figure}

In null coordinates $x_{\pm} = x \pm t$, we have 
\begin{align}
\label{eq:CFT_EM_null}
	\bra T_{\pm\pm}(x_\pm) \ket = \frac{c}{16\pi} \frac{\epsilon^2}{(x_\pm^2 + \epsilon^2)^2}\,,
\end{align}
and $\bra T_{\pm\mp}(x_\pm) \ket = 0$. 
The total energy is given by 
\beq
\label{eq:TotE}
	E_{\rm CFT} = \int_{-\infty}^\infty \bra T_{tt}(x,t)\ket\ dx = \frac{c}{16\epsilon}\,.
\eeq 
Notably, both the total energy and the energy density at $x = t$ diverge as 
$\epsilon \rightarrow 0$. This may be attributed to the instantaneous nature 
of the quench, which may excite modes of arbitrarily high frequency,
as well as to the nature of the quench as a change in the topology and geometry of the 
spacetime of the theory. The resulting UV divergence is anticipated by earlier 
work in quantum gravity \cite{Anderson:1986ww,Manogue:1988}.
Note also the proportionality to the central charge. Topology changes of 
the kind that occur in critical string theories would not suffer from these kinds of divergences 
since these theories have no conformal anomaly, i.e., 
their total central charge is zero.

%%%%%%%%%%%%%%%%%%%%%%%%%%%%%%%%%%%%%%%%%%%%%%%%%%%%%%%%%%%%%%%%%%%%%%%%%%%
%%%%%%%%%%%%%%%%%%%%%%%%%%%%%%%%%%%%%%%%%%%%%%%%%%%%%%%%%%%%%%%%%%%%%%%%%%%

\section{Lorentzian transformations and dual bulk geometries}
\label{sec:Holog_renorm}

Given the form of the energy-momentum tensor in Eq.~\eqref{eq:CFT_EM_null}, one 
could use holographic renormalization to reconstruct a holographic bulk geometry \cite{Balasubramanian:1999re,Balasubramanian:1999jd, deHaro:2000xn, Skenderis:2002wp}. 
We can imagine unitarily evolving the system either forwards 
or backwards in time from the moment just after the quench, 
and so consider the energy-momentum tensor in 
Eq.~\eqref{eq:CFT_EM_null} to hold for all time, not just $t>0$.
This simplifies things and brings 
up several questions, which we address in this section. We come back to the 
case of an actual quench, which is discontinuous at $t=0$ and does not conserve 
energy, in the Conclusions.

In $\AdS_3/\CFT_2$, holographic renormalization is simplified by the restrictions on gravity in three 
dimensions, where one can write down the general asymptotically $\AdS_3$
metric in terms of just two undetermined functions $L_\pm(x_\pm)$ \cite{1999AIPC..484..147B}
\beq
\label{eq:metric_gen}
ds^2 = R^2 \left(\frac{L_+}{2} dx_+^2 + \frac{L_-}{2} dx_-^2 + \left( \frac{1}{z^2} + \frac{z^2}{4}
	 L_+L_-\right) dx_+ dx_- + \frac{dz^2}{z^2} \right),
\eeq
which are related to the non-vanishing components of the boundary energy-momentum tensor by \cite{Balasubramanian:1999re}
\beq \label{eq:TL}
	T_{\pm\pm} = \frac{R}{16\pi G_{\text{N}}} L_\pm \,.
\eeq
Here $z>0$ is the Poincar\'e radial bulk coordinate, the spacetime boundary is at $z \to 0$ and $R$ denotes the AdS radius. 
Hence, through the identification $c = 3R/2G_{\text{N}}$ \cite{Brown:1986nw, Balasubramanian:1999re}, one can simply solve for the functions $L_{\pm}$, and thus the exact metric, given the boundary energy-momentum tensor. In the geometry in Eq.~\eqref{eq:metric_gen} with Eq.~\eqref{eq:TL} given by the local quench energy-momentum tensor \eqref{eq:CFT_EM_null}:
\be \label{eq:Lxpm}
L_\pm = \frac{3}{2} \frac{\epsilon^2}{(x_\pm^2 + \epsilon^2)^2}\,,
\ee
one can then directly compute the holographic entanglement entropy and mutual information by computing geodesic lengths, using the holographic proposal \cite{Ryu:2006bv,Hubeny:2007xt}, and compare to the CFT results. 

However, notice that the energy-momentum tensor we are considering, Eq.~\eqref{eq:CFT_EM_null}, is of the same form as that already found by \cite{Nozaki:2013wia},
\beq
T_{\pm\pm} = \frac{M \alpha^2}{8\pi\GN R (x_\pm^2 + \alpha^2)^2},
\eeq 
obtained from the back-reaction on spacetime of a massive point source falling from near the conformal boundary of $\AdS_3$. This energy-momentum tensor depends explicitly on a mass parameter $M$ as well as a length scale $\alpha$, and we will explain their physical interpretation in Sec.~\ref{sec:coord}. 
To match the energy-momentum tensor in Eq.~\eqref{eq:CFT_EM_null}, we need to identify $\alpha$ with $\epsilon$. We should also match total energies: $E_{\text{CFT}} = E_{\text{AdS}}$, 
where $E_{\text{CFT}}$ was given in Eq.~\eqref{eq:TotE} and $E_{\text{AdS}} = M/(8 R \GN \alpha)$ \cite{Nozaki:2013wia}, leading to 
\be
M = \frac 3 4 R^2\,, 
\ee
again using $c = 3R/2G_{\text{N}}$ \cite{Brown:1986nw, Balasubramanian:1999re}. 

Thus, one can alternatively proceed as in \cite{Nozaki:2013wia}, where the 
bulk metric is obtained by a boost and coordinate change from a BTZ black 
hole solution in global coordinates (The corresponding metric in $\AdS_5$ was first obtained in 
\cite{Horowitz:1999gf}). The fact that both vacuum solutions have the same holographic energy-momentum tensor (on all of 
$\mathbb{R}^{1,1}$) guarantees that we are working with the same metric \cite{1999AIPC..484..147B}.

The special properties of holographic renormalization in three dimensions were exploited by \cite{Roberts:2012aq} to construct the gravity dual to a pulse of energy. \cite{Roberts:2012aq} furthermore gave the explicit expressions needed to obtain the class of metrics in Eq.~\eqref{eq:metric_gen} from large diffeomorphisms of pure $\AdS_3$. Different such solutions 
are related by bulk diffeomorphisms that do not vanish sufficiently rapidly as they approach the conformal boundary. In particular, if one starts with pure $\AdS_3$, i.e. 
$L_\pm = 0$, in coordinates $(y_\pm, u)$ then it was worked out in \cite{Roberts:2012aq}
that acting with the diffeomorphism\footnote{These differ slightly from the 
expressions in \cite{Roberts:2012aq} because there they use coordinates
$x_\pm = \frac{1}{\sqrt{2}}(t\pm x)$, which differ by numerical factors from our conventions $x_\pm = x \pm t$.}
\begin{align}
\label{eq:diffeo}
	y_\pm &= f_\pm(x_\pm) -
		\frac{2 z^2 f'^2_\pm f''_\pm}{4 f'_\pm f'_\mp +  z^2 f''_\pm f''_\mp} \\
\label{eq:diffeo-u}
	u &= z \frac{4 (f'_+ f'_-)^{3/2}}{4 f'_+ f'_- +  z^2 f''_+ f''_-}\,,
\end{align}
one obtains a metric of the form of Eq.~\eqref{eq:metric_gen} with 
\beq
\label{eq:Lpm}
	L_\pm = - (Sf_\pm)(x_\pm).
\eeq 
Here, as in Eq.~\eqref{eq:Schwarzian}, $(Sf)(x)$ denotes the Schwarzian derivative. 
For the local quench energy-momentum tensor, $L_{\pm}$ are given in Eq.~\eqref{eq:Lxpm} and thus the diffeomorphism $f_\pm(x_\pm)$ solves: 
\beq
\label{eq:Sch_null}
	(Sf_\pm)(x_\pm) = -\frac{3}{2} \frac{\epsilon^2}{(x_\pm^2 + \epsilon^2)^2}\,.
\eeq
Notice we have already found one solution of equation \eqref{eq:Sch_null} in Euclidean signature: the conformal map in Eq.~\eqref{eq:UHPmap}.  Thus we here consider the solution 
\beq
\label{eq:null_soln}
	f_{\pm}(x_\pm) = x_\pm + \sqrt{x_\pm^2 + \epsilon^2}\,. 
\eeq 
(Here it is more convenient to drop the factor of $1/\epsilon$, so that $y_\pm$  and $x_\pm$ have the same dimensions).
This solution is not unique: since Eq.~\eqref{eq:Sch_null} is a third order differential equation there is a three parameters family of solutions, but the results we discuss below do not depend on the specific choice we make.

Therefore, performing the bulk diffeomorphism in Eqs.~\eqref{eq:diffeo} and \eqref{eq:diffeo-u}, with $f_\pm$ given in Eq.~\eqref{eq:null_soln}, on pure $\AdS_3$ transforms it to a metric whose holographic energy-momentum tensor matches Eq.~\eqref{eq:CFT_EM_null}. From the boundary point of view, this diffeomorphism is a conformal transformation, under which the energy-momentum tensor transforms as in Eq.~\eqref{eq:EMtrans} 
\beq
\label{eq:EMtens_null}
	T_{\pm \pm}(x_\pm) = \left( \frac{df_\pm}{dx_\pm}\right)^2  
	T^{(y)}_{\pm \pm}(f_\pm(x_\pm)) - \frac{c}{24\pi}(Sf_\pm)(x_\pm)\,.
\eeq
In particular, the Lorentzian conformal transformation in Eq.~\eqref{eq:null_soln} maps the vacuum energy-momentum tensor in $y_\pm$ coordinates, $\bra T^{(y)}_{\mu\nu} \ket = 0$, to an 
energy-momentum tensor satisfying Eq.~\eqref{eq:CFT_EM_null}, taken to hold for all time. 

To recap, we have two distinct ways to obtain the energy momentum tensor in 
Eq.~\eqref{eq:CFT_EM_null}: one from a Euclidean conformal transformation of a (Euclidean) BCFT, 
analytically continued to real time, and one from a Lorentzian conformal transformation applied to
the ground state of a Lorentzian CFT (with no boundary). 
We will refer to them as the Euclidean and Lorentzian CFT states, respectively.
Both can be interpreted as yielding states 
in theories defined on Minkowski space $\mathbb{R}^{1,1}$.
 What, then, is the relation between the two? 

We could analyze these states in a variety of ways. 
We choose to use the entanglement entropy and 
mutual information.  
However, this is somewhat tricky to handle directly for the Lorentzian 
CFT state since, as noted in \cite{Roberts:2012aq}, it is not clear 
how twist operators transform under Lorentzian conformal transformations
(i.e., the twist operators may not be conformal primaries with respect to the Lorentzian 
conformal group), so it is not clear if one can compute
the entanglement entropy via the replica trick. To proceed further, we are thus led to consider the 
holographic entanglement entropy, which is expected to be relevant to all two-dimensional, 
large central charge CFTs \cite{2010PhRvD..82l6010H, Hartman:2013mia}. 
A closely related question was raised in the case of a global quench in \cite{Hartman:2013qma}.
We come back to the relation between the bulk and boundary states, and 
global and local quenches, in the Conclusions, Sec.~\ref{sec:Concl}.

In the following sections we therefore proceed with a detailed analysis of geodesics in the metric of \cite{Nozaki:2013wia}. In \cite{Nozaki:2013wia} the authors computed the single-interval entanglement entropy, but not the mutual information. 
We emphasize that the bulk geometry, described either by Eq.~\eqref{eq:metric_gen} with 
$L_\pm$ given by Eq.~\eqref{eq:Lxpm} or by the 
metric in \cite{Nozaki:2013wia} (reproduced below), is precisely the bulk geometry one obtains from applying 
holographic renormalization to the energy-momentum tensor in Eq.~\eqref{eq:CFT_EM_null}.
From our discussion above on bulk diffeomorphisms and boundary conformal transformations,
we see that it can also be interpreted as the precise bulk dual to the state obtained from 
the vacuum under the Lorentzian conformal transformation in Eq.~\eqref{eq:null_soln}.
This is to be distinguished from the state obtained by the action of the 
Euclidean mapping in Eq.~\eqref{eq:UHPmap} and the analytic continuation 
$\tau \to it$, since, though it has the same energy-momentum tensor, it is different. 
The precise differences will be exhibited, via the mutual information, and discussed below.

%%%%%%%%%%%%%%%%%%%%%%%%%%%%%%%%%%%%%%%%%%%%%%%%%%
%%%%%%%%%%%%%%%%%%%%%%%%%%%%%%%%%%%%%%%%%%%%%%%%%%

\section{Holographic entanglement entropy} \label{sec:coord}

In order to holographically obtain the mutual information between two non-overlapping intervals in the asymptotically $\AdS_3$ model described in \cite{Nozaki:2013wia}, 
one needs to solve for space-like geodesics joining the interval endpoints. Using the covariant generalization of the Ryu-Takayanagi formula \cite{Ryu:2006bv,Hubeny:2007xt}: 
\be\label{eq:RT}
S = \frac{\mathcal L}{4 \GN}\,,
\ee
where $\mathcal L$ denotes the length of such geodesics, one can then compute the corresponding entanglement entropies $S$, and from those the mutual information. We discuss the results of \cite{Nozaki:2013wia} for the entanglement entropy of one interval below and work out the mutual information of disjoint intervals in Sec.~\ref{sec:MI}, extensively comparing to the CFT results.  To this end, and in view of the discussion in the previous section, we will identify the length scales $\alpha =\epsilon$ and set the mass parameter $M= (3/4) R^2$ (see below). 

%%%%%%%%%%%%%%%%%%%%%%%%%%%%%%%%%%%%%%%%%%%%%%%%%%

\subsection{Setup}

The holographic dual of a two-dimensional local quench proposed in \cite{Nozaki:2013wia} consists of a particle of mass $m$ falling in three-dimensional Poincar\'e spacetime with metric
\be \label{eq:Poincare}
ds^2 = \frac{-dt^2 + dz^2 +dx^2}{z^2}\,,
\ee
where $z>0$ and we have set the AdS radius to one. The spacetime conformal boundary is at $z \to 0$ and its location will often be regulated by a IR cutoff $z_\infty \ll 1$. 
The action of a particle with mass $m$ at $x=0$ in the geometry in Eq.~\eqref{eq:Poincare} is \cite{Nozaki:2013wia}
\be
S = -m \int dt \frac{\sqrt{1-\dot z(t)^2}}{z(t)}\,.
\ee
The general trajectory extremizing this action is given by 
\be \label{eq:trajectory}
x(t) =0\,, \qquad z(t) = \sqrt{(t-t_0)^2 +\alpha^2}\,,
\ee
where $t_0$, $\alpha$  are constants and time translation invariance can be used to set $t_0 =0$. The particle energy reads
\be \label{eq:energy}
E= \frac{m}{\alpha}\,,
\ee
and we will often be interested in the limit in which the ratio of $\alpha$ over all time and length scales of the problem goes to zero. If the ratio $m/\alpha$ (in units of the AdS radius) is kept finite in this limit, the energy in Eq.~\eqref{eq:energy} also remains finite. 
The trajectories for different $\alpha$ are related to each other by boost transformations. 
Such boosts are isometries of pure AdS space, but once we include back-reaction this is 
no longer the case and we get distinct metrics for distinct $\alpha$. 

The particle back-reaction on spacetime can be easily computed by observing that the change of coordinates\footnote{
Throughout the holographic computations, we denote by $\tau$ the global time coordinate. This should not be confused with the Euclidean time of the CFT sections. 
}
\begin{align}
z &= \frac{\alpha}{\sqrt{1+r^2} \cos \tau + r \cos \theta} \label{eq:z} \\
t &= \frac{\alpha \sqrt{1+r^2} \sin \tau}{\sqrt{1+r^2} \cos \tau + r \cos \theta} \label{eq:t}\\
x &= \frac{\alpha r \sin \theta}{\sqrt{1+r^2} \cos \tau + r\cos \theta}\,, \label{eq:x} 
\end{align}
for arbitrary $\alpha > 0$, locally maps Poincar\'e AdS$_3$ of Eq.~\eqref{eq:Poincare} into global AdS$_3$ with metric\footnote{
To cover the region of global AdS$_3$ where $\sqrt{1+r^2} \cos \tau + r \cos \theta< 0$, one should consider the change of coordinates:
\begin{align}
z &= -\frac{\alpha}{\sqrt{1+r^2} \cos \tau + r \cos \theta} \\
t &= -\frac{\alpha \sqrt{1+r^2} \sin \tau}{\sqrt{1+r^2} \cos \tau + r \cos \theta}\\
x &= -\frac{\alpha r \sin \theta}{\sqrt{1+r^2} \cos \tau + r\cos \theta}\,. 
\end{align}
} 
\be \label{eq:global}
ds^2 = -(1+r^2) d \tau^2 + \frac{dr^2}{1+r^2} + r^2 d\theta^2\,,
\ee
and the particle trajectory in Eq.~\eqref{eq:trajectory} into $r=0$  \cite{Horowitz:1999gf,Nozaki:2013wia}.  
We choose $\tau, \theta \in [-\pi, \pi]$ with $\theta$ periodic, and notice that since $z>0$ and $r \ge 0$, for $t\ge 0$ ($t \le 0$), $\tau \in [0,\pi]$ ($\tau \in [-\pi, 0]$) and for $x \ge 0$ ($x \le 0$), $\theta \in [0,\pi]$ ($[-\pi, 0]$).

The metric outside a massive static object at $r=0$ in global AdS$_3$ is known to be
\be
\label{eq:BTZ}
ds^2 = -(r^2+1 -M) d\tau^2 + \frac{dr^2}{r^2+1 -M} + r^2 d\theta^2 \,,
\ee
where $M$ is related to the mass of the particle by \cite{Nozaki:2013wia}
\be\label{eq:M}
M = 8 \GN m \,.
\ee
For $M > 1$, the metric in Eq.~\eqref{eq:BTZ} is known as the BTZ black hole solution,
with black hole mass $M - 1$ \cite{Banados:1992wn}. 
The case of $M = 1$ is the zero mass BTZ black hole \cite{Banados:1992gq}.
Here, motivated by the observations made in Section \ref{sec:Holog_renorm}, we will focus on the case $0< M < 1$, in particular $M=3/4$, for which the solution has a naked conical singularity at the origin
$r=0$ \cite{Banados:1992gq, Deser:1983nh}. 
We will not be concerned about this singularity since 
we can, as in \cite{Nozaki:2013wia}, replace the metric with a non-singular 
metric, like that of a star, in an arbitrarily small neighborhood of the origin.
Since we will only be concerned with geodesics that do not intersect the origin, this 
has no effect on our results.

Given the geometry in Eq.~\eqref{eq:BTZ}, one can therefore apply the inverse change of coordinates of Eqs.~\eqref{eq:z}-\eqref{eq:x}: 
\begin{align}
r &= \frac{1}{z \alpha} \sqrt{x^2 \alpha^2 + \frac 1 4 \left[ -\alpha^2 +z^2 +x^2 -t^2 \right]^2} \label{eq:GtoP1}\\
\tan \tau & =  \frac{2 t \alpha}{\alpha^2 +z^2 +x^2 -t^2} \\
\tan \theta & = - \frac{2 x \alpha}{-\alpha^2 +z^2 +x^2 -t^2}\label{eq:GtoP4}
\end{align}
to work out the backreacted spacetime in Poincar\'e coordinates \cite{Horowitz:1999gf,Nozaki:2013wia}. Because of its length, we do not here give the expression of the geometry in Poincar\'e coordinates, but this can be readily obtained from Eq.~\eqref{eq:BTZ} through Eqs.~\eqref{eq:GtoP1}-\eqref{eq:GtoP4}. 

%%%%%%%%%%%%%%%%%%%%%%%%%%%%%%%%%%%%%%%%%%%%%%%%%%

\subsection{Entanglement entropy} \label{sec:EE}

To compute the entanglement entropy associated with a spatial region of 
the conformal boundary via the covariant generalization of the Ryu-Takayanagi formula \eqref{eq:RT}, one needs to solve for spacelike geodesics that extend between two boundary endpoints $(t_\infty, z_\infty, \ell_1)$, $(t_\infty, z_\infty, \ell_2)$ in the asymptotically Poincar\'e AdS space. 
As for the construction of the backreacted geometry, it is simpler to first solve for geodesics in the spacetime in Eq.~\eqref{eq:BTZ} connecting the points
\bea
\tau_\infty^{(i)} &= \tan^{-1} \left[ \frac{2 t_\infty \alpha}{\alpha^2 + \ell_i^2 -t_\infty^2} \right]\label{eq:tauinfs} \\
\theta_\infty^{(i)} &= \tan^{-1} \left[ -\frac{2 \ell_i \alpha}{\ell_i^2 -t_\infty^2 -\alpha^2}\right]  \label{eq:thetainfs} \\
r_\infty^{(i)} &= \frac{1}{z_\infty \alpha} \sqrt{ \ell_i^2 \alpha^2+ \frac 1 4 \left[ \ell_i^2 -t_\infty^2 - \alpha^2 \right]^2}
\end{align}
for $i =1,2$ and then use the coordinate transformation in Eqs.~\eqref{eq:z}-\eqref{eq:x} to map such geodesics into the asymptotically Poincar\'e AdS metric. Notice that, in principle, there are four possible solutions to 
Eqs.~\eqref{eq:tauinfs}-\eqref{eq:thetainfs} since $\tau, \theta \in [-\pi, \pi]$. However, the coordinate transformation in Eqs.~\eqref{eq:z}-\eqref{eq:x} only covers the region $\sqrt{1+r^2} \cos\tau + r\cos \theta >0$ with $r\ge 0$, so that $\tau_\infty^{(i)} \in [0,\pi]$ ($\tau_\infty^{(i)} \in [-\pi,0]$) for $t_\infty \ge 0$ ($t_\infty \le0$) and $\theta_\infty^{(i)} \in [0,\pi]$ ($\theta_\infty^{(i)} \in [-\pi,0]$) for $\ell_i \ge0$ ($\ell_i \le 0$), as we observed below Eq.~\eqref{eq:global}.  

The relevant geodesics have been worked out in \cite{Nozaki:2013wia} and we review their construction in Appendix \ref{sect:geodesics}. Here we just quote the result for the entanglement entropy \cite{Nozaki:2013wia}:
\be \label{eq:EE}
S = \frac{1}{4 \GN} \left[\ln \(r_\infty^{(1)}\ r_\infty^{(2)} \) + \ln \frac{2\( \cos\(\sqrt{1-M} \Delta \tau_\infty\) -\cos\(\sqrt{1-M} \Delta \theta_\infty\)\) }{1-M} \right]\,, 
\ee
where $\Delta \tau_\infty \equiv |\tau_\infty^{(2)} - \tau_\infty^{(1)}|$ and $\Delta \theta_\infty \equiv |\theta_\infty^{(2)} - \theta_\infty^{(1)}|$, and $\Delta \theta_\infty$ should be replaced by $2\pi -\Delta \theta_\infty$ if $\Delta \theta_\infty > \pi$. 

In the case of a symmetric interval with $-\ell_1 = \ell_2 \equiv \ell$, this reduces to 
\be \label{eq:EES}
S = \frac{1}{2 \GN}  \ln  \frac{2 r_\infty  \sin (\sqrt{1-M} \theta_\infty)}{\sqrt{1-M}} \,,
\ee
where $r_\infty \equiv r_\infty^{(1)} = r_\infty^{(2)}$, $\theta_\infty \equiv- \theta_\infty^{(1)} = \theta_\infty^{(2)}$, and $\theta_\infty \to \pi - \theta_\infty$ if $\theta_\infty > \frac \pi 2$.

The entanglement entropy diverges if we remove the IR cutoff $z_\infty$, making it convenient to define a renormalized entanglement entropy $\Delta S$, for instance by subtracting the value of the entanglement entropy in pure AdS ($M=0$) \cite{Nozaki:2013wia}:
\be \label{eq:EErengeneral}
\Delta S =\frac{1}{4 \GN} \ln \frac{\cos\(\sqrt{1-M} \Delta \tau_\infty\) -\cos\(\sqrt{1-M} \Delta \theta_\infty\)}{(1-M)\( \cos\( \Delta \tau_\infty\) -\cos\( \Delta \theta_\infty\)\)} \,,
\ee
which in the case of a symmetric interval becomes
\be \label{eq:REES}
\Delta S = \frac{1}{2 \GN} \ln \frac{\sin (\sqrt{1-M} \theta_\infty)}{\sqrt{1-M} \sin \theta_\infty} \,.
\ee
Looking ahead, we note that we will not need this kind of renormalization for the mutual information 
between non-overlapping intervals, which we compute in Sec.~\ref{sec:MI}, since it is always finite. 

\begin{figure}[h]
\begin{tabular}{ccc}
\includegraphics[width=0.4\textwidth]{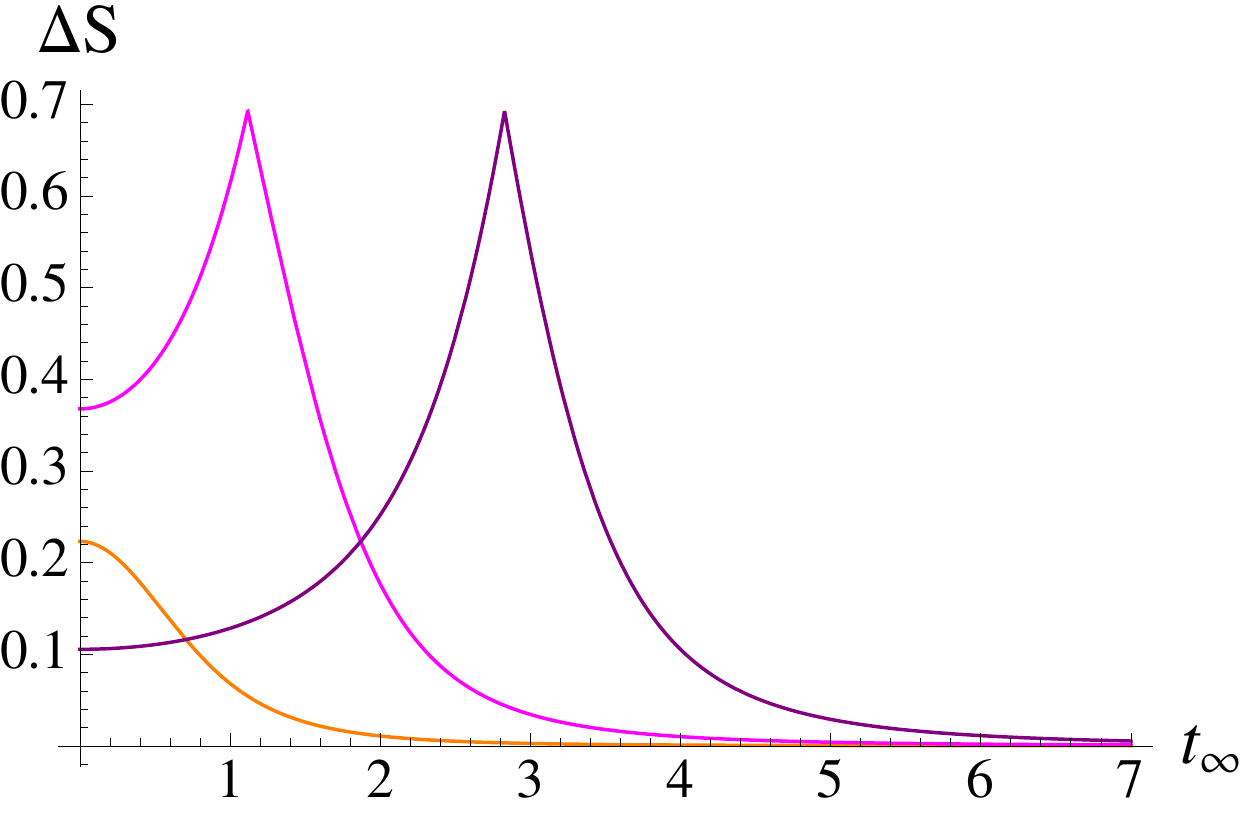} \hfill
&\qquad\qquad &
\includegraphics[width=0.4\textwidth]{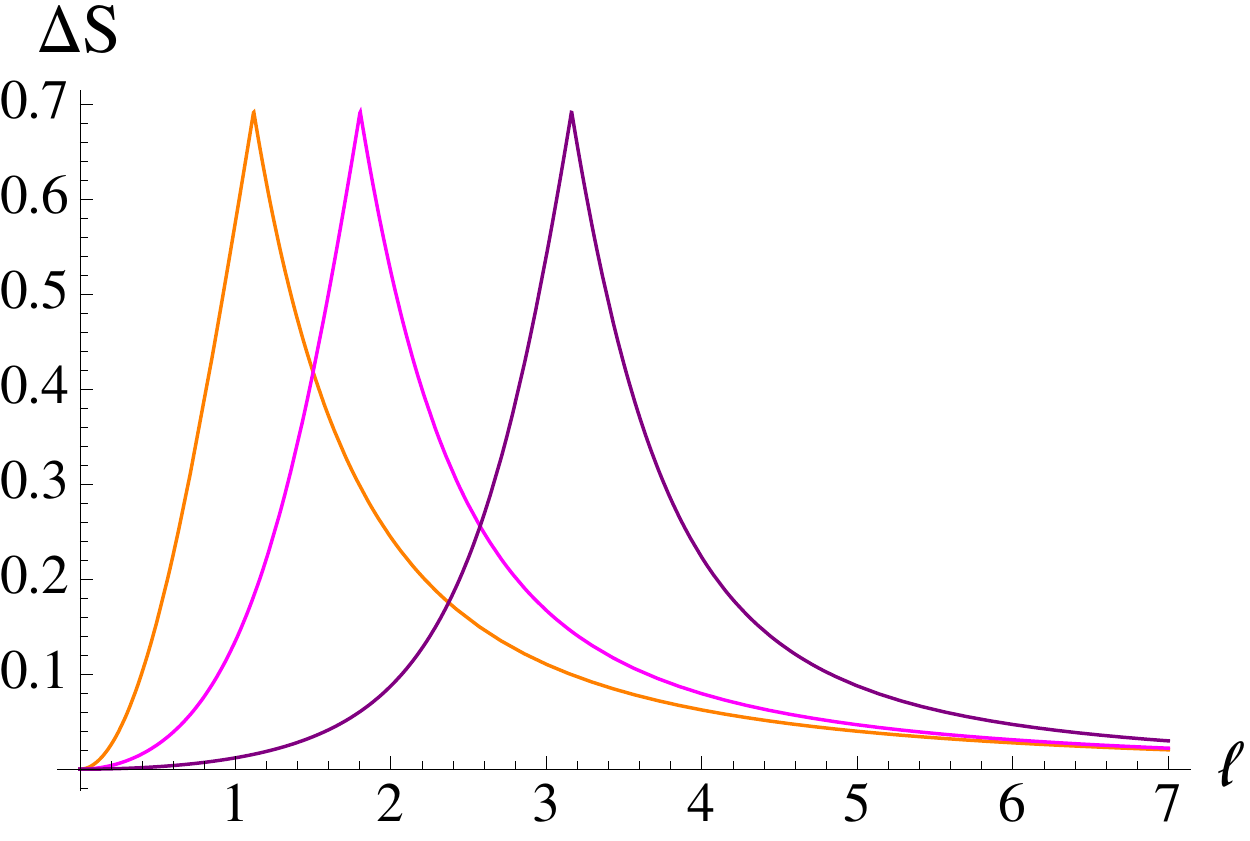}
\\
{\bf (A)} &\qquad\qquad & {\bf (B)}
\\
\includegraphics[width=0.4\textwidth]{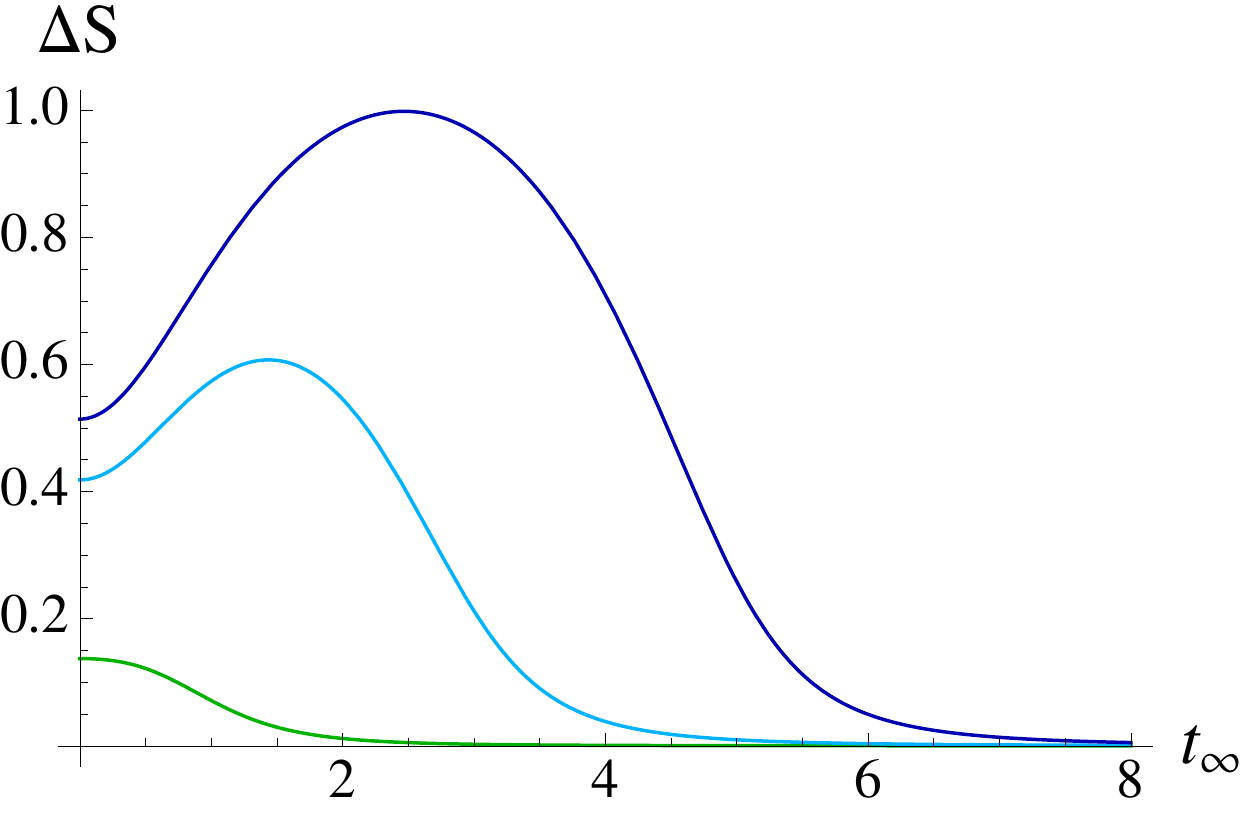} \hfill
&\qquad\qquad &
\includegraphics[width=0.4\textwidth]{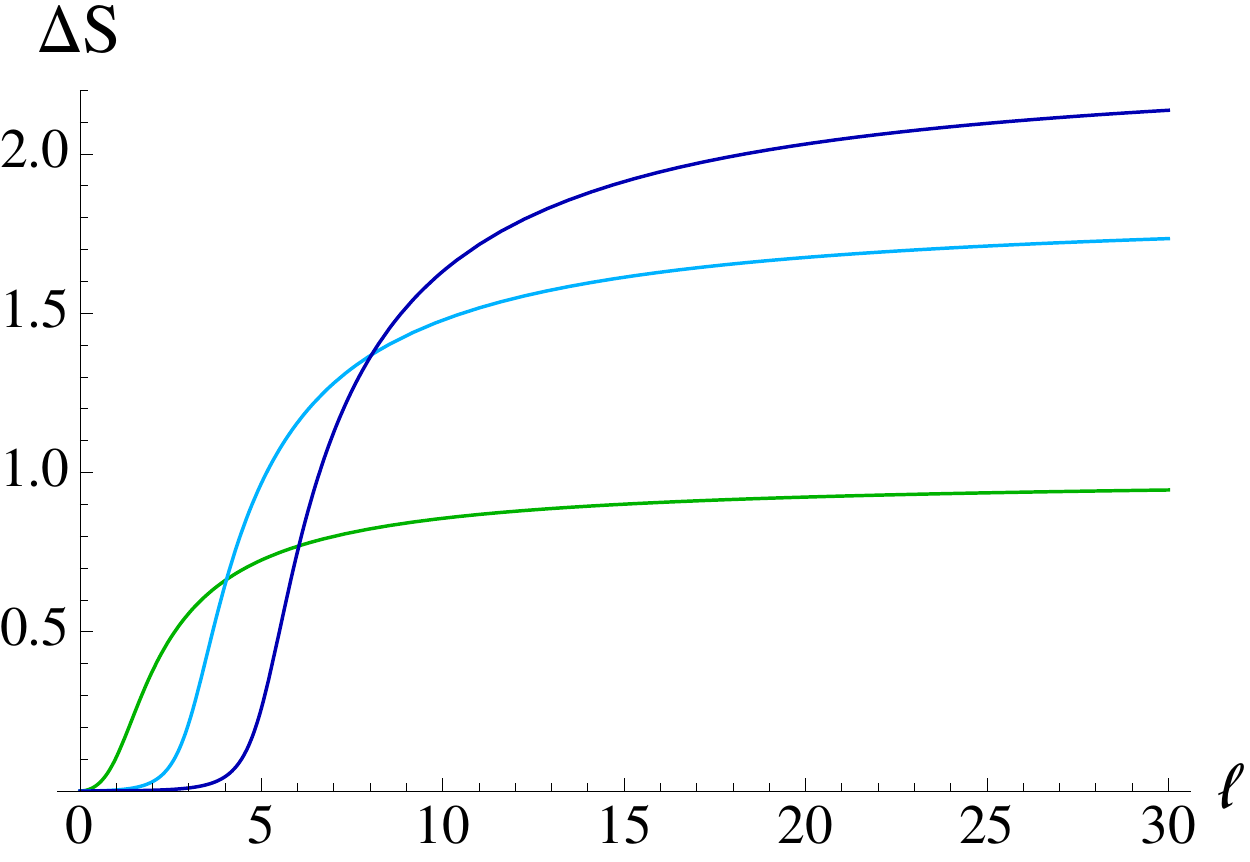}
\\
{\bf (C)} &\qquad\qquad & {\bf (D)}
\\
\includegraphics[width=0.4\textwidth]{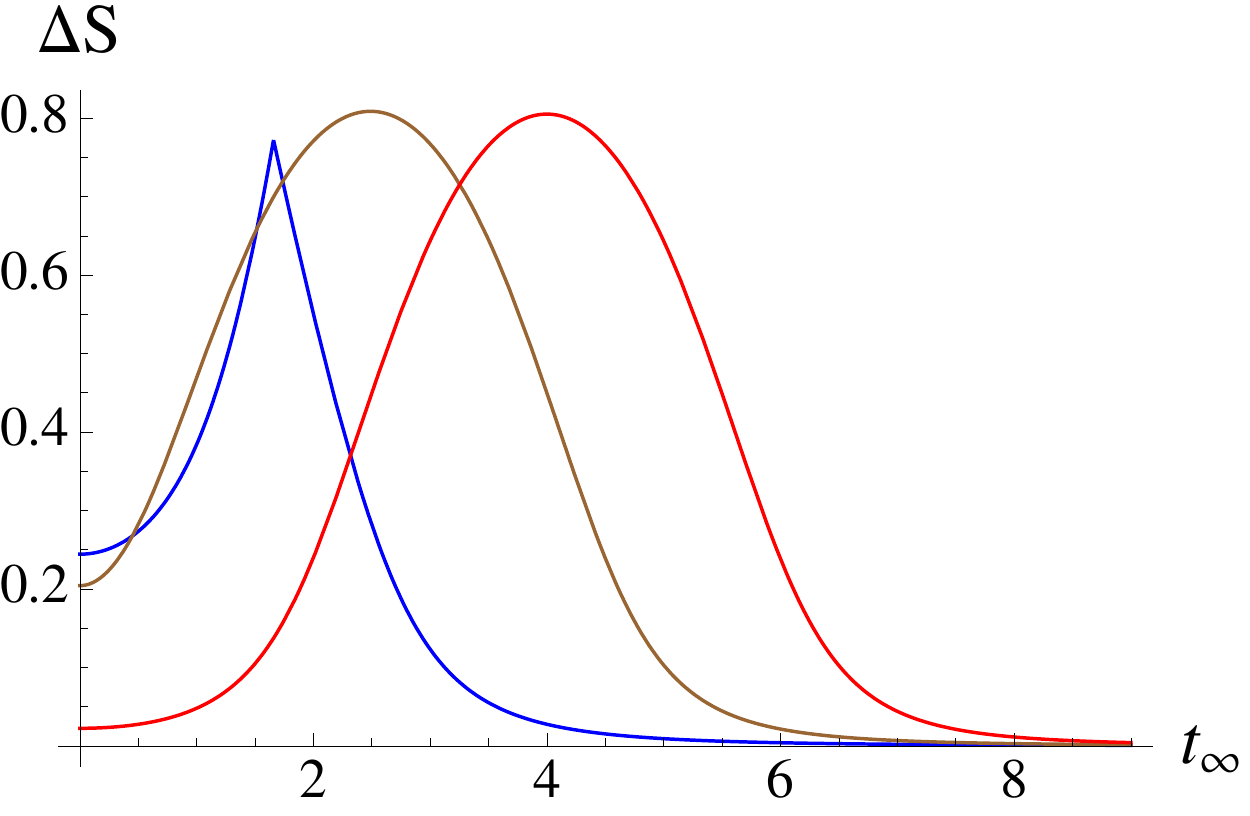} \hfill
&\qquad\qquad &
\includegraphics[width=0.4\textwidth]{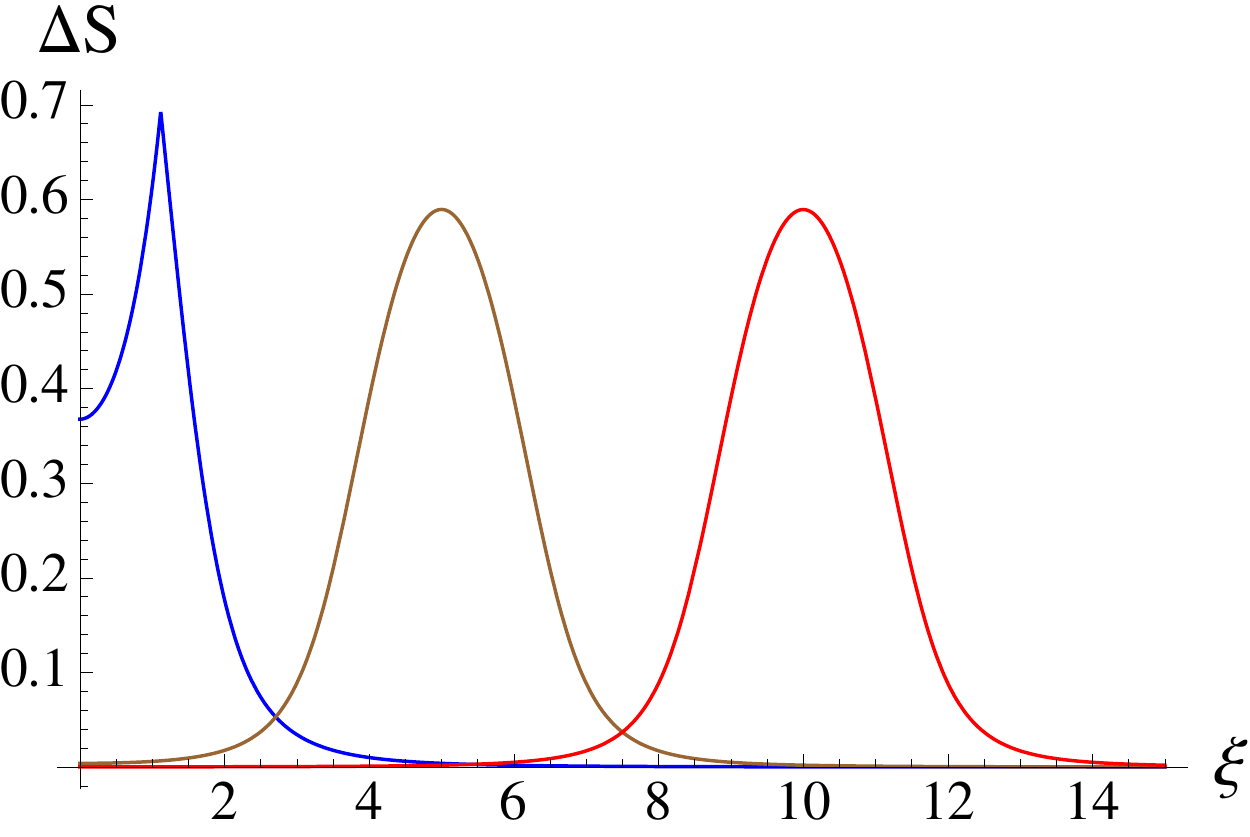}
\\
{\bf (E)} &\qquad\qquad & {\bf (F)}
\\
\end{tabular}
\caption{Renormalized entanglement entropy $\Delta S$ (rescaled by $4 \GN$) as a function of time $t_\infty$, interval length $\ell$ or interval midpoint position $\xi$ for:  a symmetric interval with {\bf (A)} $-\ell_1 = \ell_2 \equiv \ell = 0.5, 1.5, 3$ from left to right,  {\bf (B)} times $t_\infty = 0.5,1.5, 3$ from left to right; an interval with an excited endpoint $\ell_1 =0$ with {\bf (C)} $\ell_2 = 0.9,3,5$ from the bottom up, {\bf (D)}  $t_\infty =0.9,3,5$ from left to right; a generic interval with {\bf (E)} $\ell_1 = -1.5,0.5,2$, $\ell_2=\ell_1 +4$ from left to right, {\bf (F)} $\ell_1 = \xi - \frac 3 2$, $\ell_2 = \xi + \frac 3 2$ for $t=0,5,10$ from left to right. In the plots, we have set the parameters $\alpha=1$ and $M=3/4$.  
}
\label{fig:EEholo}
\end{figure}

We plot the renormalized entanglement entropy as a function of time, interval size or interval midpoint for various cases in Fig.~\ref{fig:EEholo}. 

%%%%%%%%%%%%%%%%%%%%%%%%%%%%%%%%%%%%%%%%%%%%%%%%%%

\subsection{$t, \ell \gg \alpha$ limit}

In this section, to get more analytic insight on the holographic results for the entanglement entropy and to compare with the CFT computations discussed in Section~\ref{sec:EE1}, we consider the regime in which the ratio of $\alpha$ (which in view of the discussion in Sec.~\ref{sec:Holog_renorm} is identified with the CFT regulator $\epsilon$) over all length and time scales goes to zero. 
We can keep the particle energy, Eq.~\eqref{eq:energy}, finite in this limit as long as the ratio $\alpha/M$ (in units of the AdS radius) is finite, since $M$ is related to the particle mass $m$ via Eq.~\eqref{eq:M}. 

For notational simplicity, in the following we denote the boundary time $t_\infty$ by $t$. 

%%%%%%%%%%%%%%%%%%%%%%%%%%%%%%%%%%%%%%%%%%%%%%%%%%

\subsubsection{Symmetric intervals about the origin}

For $\ell \gg \alpha$, the entanglement entropy, given by Eq.~\eqref{eq:EES}, for the symmetric interval of length $2 \ell$ reduces to
\be \label{eq:Ssalpha}
S = \frac{1}{2\GN} \ln \frac{2 \ell}{z_\infty}\,,
\ee
for times before and after the peak at $t = \ell$. For $\alpha \ll \ell, t$, it is natural to identify the IR bulk cutoff $z_\infty = \alpha$  $(=\epsilon)$. Using $c = \frac{3}{2 \GN}$, Eq.~\eqref{eq:Ssalpha} becomes 
\be
S = \frac{c}{3} \ln \frac{2 \ell}{\epsilon}\,,
\ee
i.e., the value for an interval of length $2\ell$ in a CFT on an infinite line in its ground state. 
On the other hand, we had seen the CFT computation led to Eq.~\eqref{eq:EEsinglesymm}, that is, the sum of the ground state entanglement entropies of two intervals $(0, \ell)$ in the half line. Indeed, while the local quench starts out without any entanglement between the two half lines in their separate ground states, giving rise to Eq.~\eqref{eq:EEsinglesymm}, the holographic setup of \cite{Nozaki:2013wia} has non-zero entanglement from the beginning and corresponds, at $t=0$, to a CFT on an infinite line in its ground state. 

%%%%%%%%%%%%%%%%%%%%%%%%%%%%%%%%%%%%%%%%%%%%%%%%%%

\subsubsection{General interval with $|\ell_1| \neq |\ell_2|$}

\begin{figure}[h]
\begin{center}
\includegraphics[width = 0.45 \textwidth]{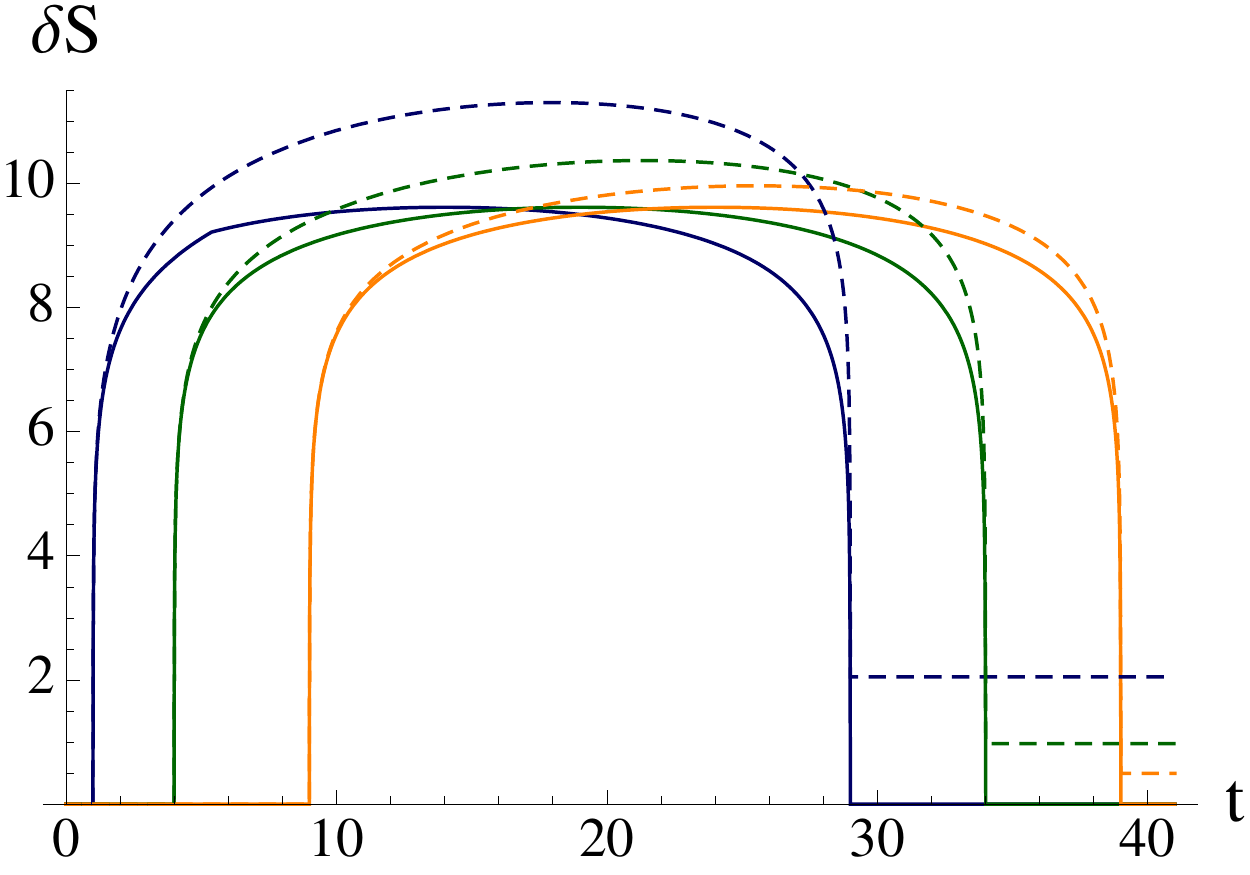} 
\caption{The time evolution of the entanglement entropy with the early time value subtracted (and rescaled by $4 \GN$) for intervals of length 30 and $\ell_1 = -1,4,9$. The universal part of the CFT results are shown with dashed lines for comparison. Here we have set $M=3/4$ and $\epsilon = \alpha  = 0.001$.}\label{fig:Stalpha}
\end{center}
\end{figure}

For $t, \ell \gg \alpha$, the entanglement entropy in Eq.~\eqref{eq:EE} for a generic interval with $\ell_2>0$ and $\ell_2 > |\ell_1|$ becomes: 
\be \label{eq:earlylatetime}
S(t < |\ell_1|) = S(t > \ell_2) = \frac{1}{2\GN} \ln \frac{(\ell_2 -\ell_1 )}{z_\infty} \,,
\ee
for $t < |\ell_1|$ and $t > \ell_2$, i.e., for early and late times. For intermediate 
times, $|\ell_1| < t < \ell_2$, we have
\be \label{eq:Saalpha}
S(0\le \ell_1 < t < \ell_2) = \frac{1}{4 \GN} \ln \frac{(\ell_2 -  \ell_1) (\ell_2 - t) (t -\ell_1) \tilde M }{\alpha z_\infty^2}
\ee
if $\ell_1  \ge 0$ and 
\be  \label{eq:Sbalpha}
S(\ell_1<0 < t < \ell_2) = \left\{\begin{array}{lc}
\frac{1}{4 \GN} \ln \frac{(\ell_2 -  \ell_1) (\ell_2 + t) (t + \ell_1) \tilde M}{\alpha z_\infty^2 } & |\ell_1| < t < \sqrt{-\ell_1\ell_2}\\ 
\frac{1}{4 \GN} \ln \frac{(\ell_2 -  \ell_1) (\ell_2 - t) (t -\ell_1)\tilde M}{\alpha z_\infty^2 } & \sqrt{-\ell_1\ell_2} \le t < \ell_2
\end{array}\right. 
\ee
if $\ell_1 < 0$. For compactness, here we have defined:
\be
\tilde M \equiv \frac{ \sin(\sqrt{1-M} \pi) }{\sqrt{1-M}}\,,
\ee
which is an increasing function of $M$ and goes from $0$ to $\pi$ as $M$ varies between zero and one. 
Although the holographic analytic expressions Eqs.~\eqref{eq:earlylatetime}-\eqref{eq:Sbalpha} do not match exactly the universal part of the CFT results in Eqs.~\eqref{eq:EEsingle1}-\eqref{eq:EEsingle3}, they do share a number of features. First of all, the two are time-dependent only for $t \in [\ell_1, \ell_2]$, in agreement with the quasiparticle picture. Moreover, they behave similarly in the regime where the non-universal contributions to Eqs.~\eqref{eq:EEsingle1}-\eqref{eq:EEsingle3} can be neglected. This can be seen in Fig.~\ref{fig:Stalpha}, where we plot $\delta S \equiv S -S(t <|\ell_1|)$ and compare it with the (universal part of the) CFT results. 

\begin{figure}[h]
\begin{tabular}{ccc}
\includegraphics[width = 0.45 \textwidth]{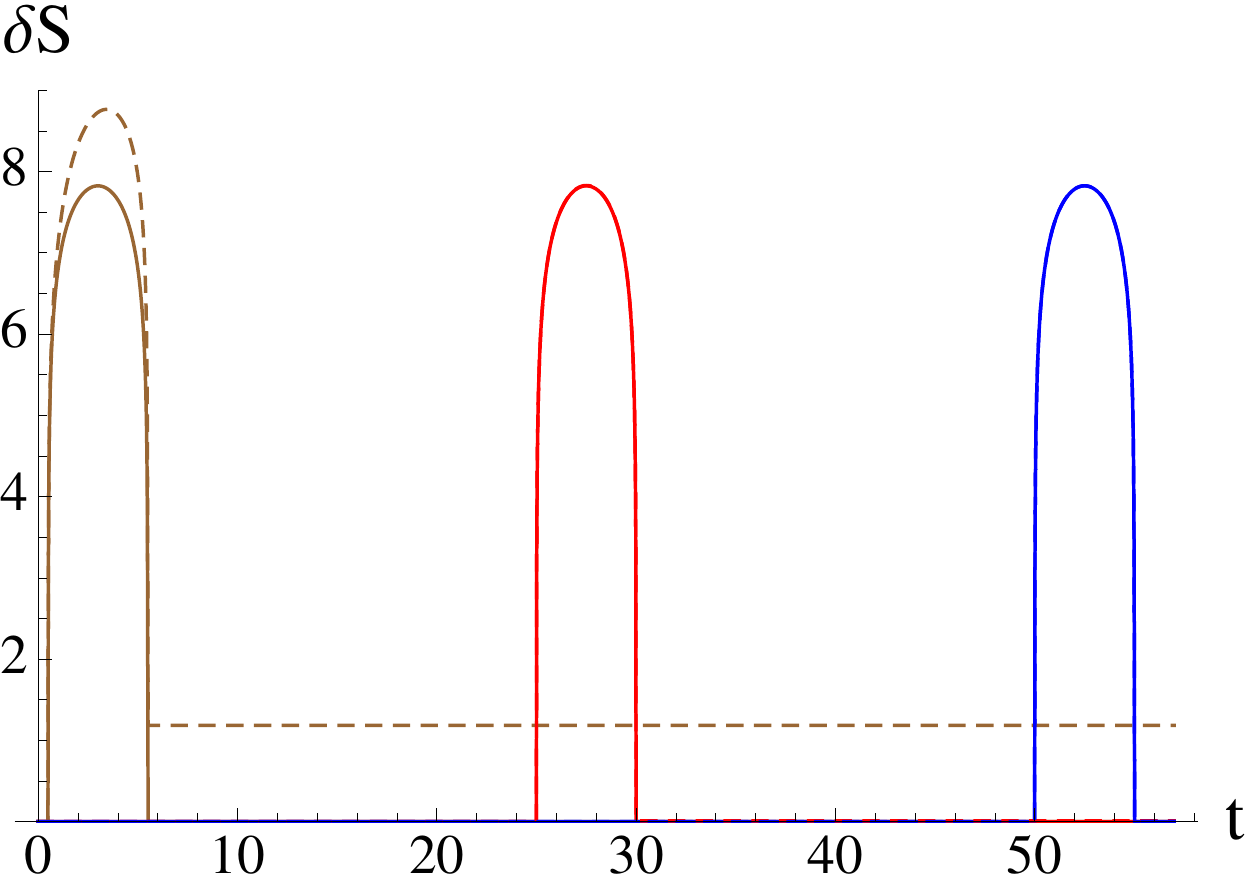} \hfill 
&\qquad\qquad &
\includegraphics[width = 0.45 \textwidth]{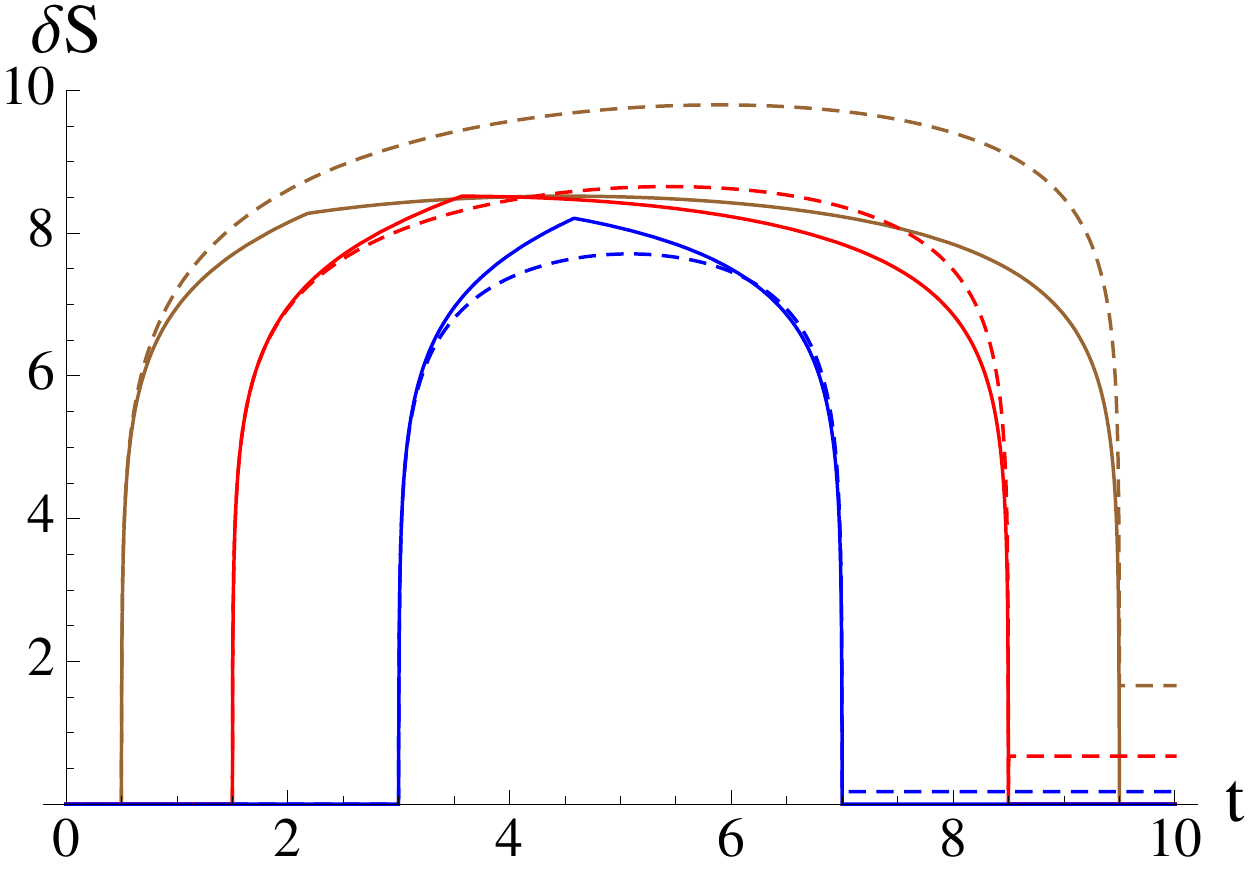}
\\
{\bf (A)} &\qquad\qquad & {\bf (B)}
\\
\end{tabular}
\caption{The entanglement entropy with the early time value subtracted (and rescaled by $4 \GN$) for intervals  {\bf(A)} of length 5 and $\ell_1 =0.5,25,50$ from left to right; {\bf (B)} of length 10 and $\ell_1 = -0.5,-1.5, -3$ from the outmost to the inmost. In dashed lines, the universal part of the CFT results, which in some cases overlap the holographic curves without appreciable difference. The parameters have been set to: $M=3/4$, $\epsilon = \alpha  = 0.001$.}\label{fig:Stalphaheight}
\end{figure}

Notice in particular that for $\ell_1 \ge 0$ the shape of the entanglement entropy curve is independent of the displacement of the interval from the origin respectively exactly and approximately in the holographic and CFT results (see panel {\bf(A)} of Fig.~\ref{fig:Stalphaheight}). As the defect at the origin penetrates deeper in the interval, the entanglement entropy instead decreases (Fig.~\ref{fig:Stalphaheight} {\bf(B)}). 

\begin{figure}[h]
\begin{center}
\includegraphics[width = 0.45 \textwidth]{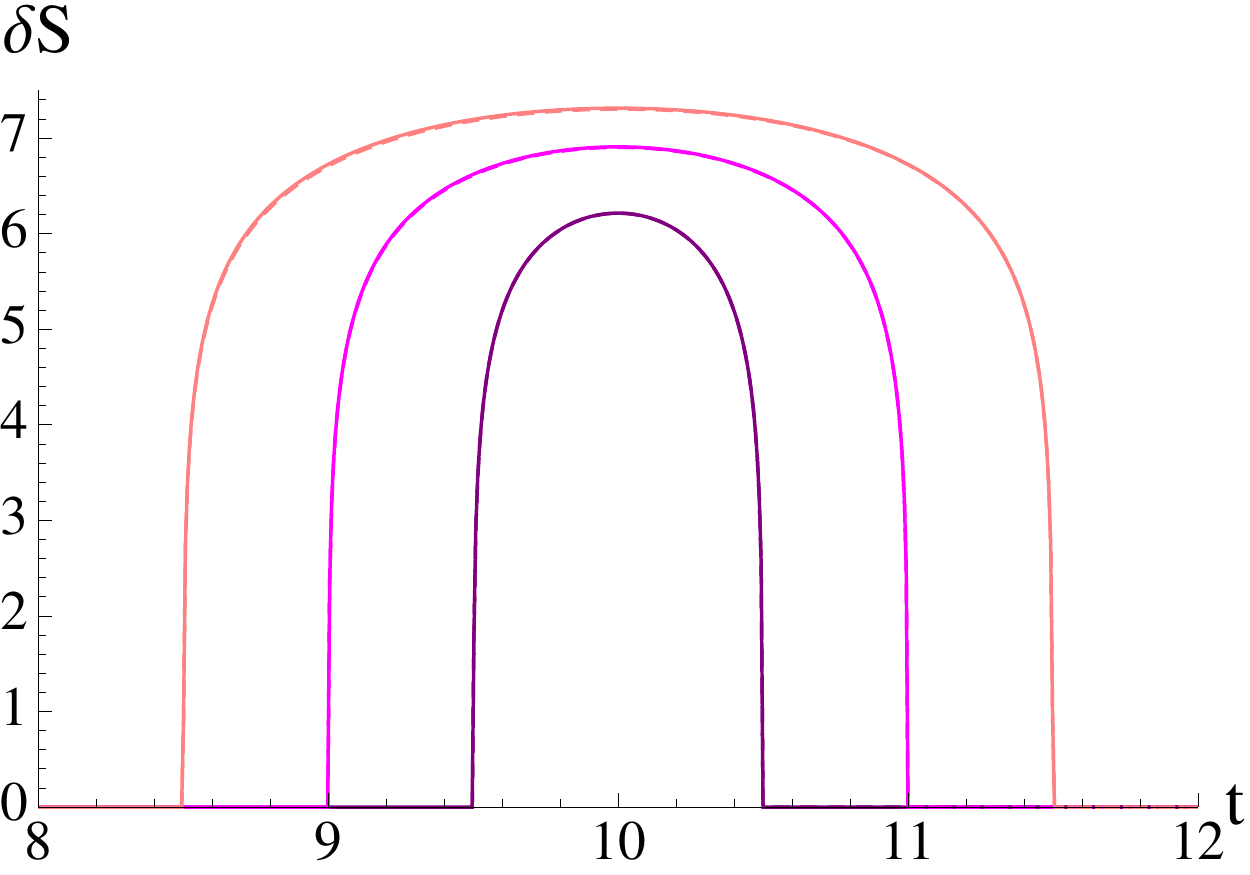} 
\caption{The entanglement entropy with the early time value subtracted (and rescaled by $4 \GN$) for intervals centered at 10 and of length $\ell_2-\ell_1 = 1,2,3$ from the interior to the exterior. In dashed lines, the universal contributions to the CFT result, which are essentially indistinguishable from the holographic ones. We have set: $M=3/4$, $\epsilon = \alpha  = 0.001$.}\label{fig:Stalphawidth}
\end{center}
\end{figure}

For $\ell_1 \ge 0$, the maximum value of the holographic entanglement entropy
\be
S_{\rm max} = \frac{1}{4 \GN} \ln \frac{(\ell_2-\ell_1)^3\tilde M}{4 \alpha z_\infty^2} 
\ee
is attained at $t = \frac{\ell_1 + \ell_2}{2}$, and it grows logarithmically in the length of the interval, in $1/\alpha$ and $1/z_\infty$. This is approximately also the case in the CFT expression in Eq.~\eqref{eq:EEsingle2}. 
We illustrate the dependence on the interval length in Fig.~\ref{fig:Stalphawidth}. 
%

%%%%%%%%%%%%%%%%%%%%%%%%%%%%%%%%%%%%%%%%%%%%%%%%%%
%%%%%%%%%%%%%%%%%%%%%%%%%%%%%%%%%%%%%%%%%%%%%%%%%%

\section{Holographic mutual information} \label{sec:MI}

%%%%%%%%%%%%%%%%%%%%%%%%%%%%%%%%%%%%%%%%%%%%%%%%%%

\begin{figure}[h]
\begin{center}
\includegraphics[width = 0.5 \textwidth]{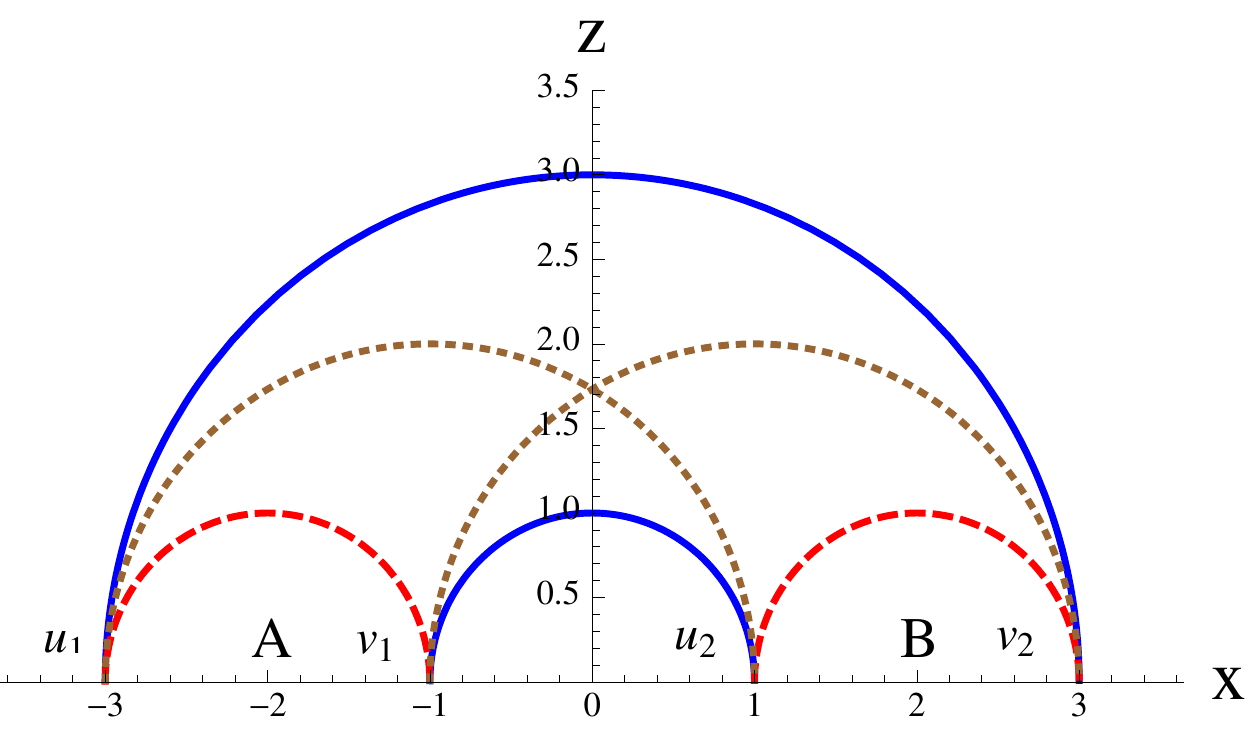} 
\caption{A schematic drawing of the three sets of geodesics competing to the holographic entanglement entropy of the union of two intervals $S_{A\cup B}$.
}\label{fig:configurations}
\end{center}
\end{figure}

To further explore the nature of the entanglement in this system and the relation between the CFT and holographic setups, we compute the mutual information, Eq.~\eqref{eq:MI},
\be \label{eq:MIhol}
I= S_A + S_B - S_{A\cup B}\,,
\ee
between two disjoint intervals in the setup we have been considering.
We have reviewed the entanglement entropy of a single interval in the previous Section. To compute the entanglement entropy of two disjoint intervals $S_{A\cup B}$ we assume the prescription that holds at equilibrium \cite{2010PhRvD..82l6010H} extends to the time-dependent setup of a falling mass in AdS. We therefore assume $S_{A\cup B}$ is given by the shortest collection of geodesics connecting the endpoints $A$ and $B$. 
For evidence in favor of this prescription in time-dependent cases, particularly 
strong subadditivity of the entanglement entropy and monogamy of the mutual information, see 
\cite{Callan:2012ip, Caceres:2013dma, Wall:2012uf}.
The three possible configurations are drawn schematically in Fig.~\ref{fig:configurations}. We will refer to them as the connected (solid, blue), the disconnected (red, dashed) and the ``intersecting'' configurations (brown, dotted), although in a time-dependent setup the latter geodesics do not generically intersect. 
The prescription then reads
\beq
\label{eq:Holog_SAB}
	S_{A\cup B} = \min\left\{S_\text{conn}, S_\text{disc}, S_\text{int} \right\} \,,
\eeq
where each $S$ is computed using the lengths of the curves described above, 
via the covariant holographic entanglement entropy proposal 
\cite{Hubeny:2007xt}.
The ``intersecting'' configuration never contributes at equilibrium \cite{2010PhRvD..82l6010H}, leaving only two competing configurations. However, there does not exist a proof of this statement in a generic time-dependent geometry so we will need to check this case by case.
In all cases we have checked explicitly the ``intersecting'' configuration turns out not to contribute, 
so that for our two intervals $A = [u_1, v_1]$ and $B = [u_2, v_2]$ we have
\beq
\label{eq:Holog_SAB_2}
	S_{A\cup B} = \min\left\{ S_{[u_1,v_1]} + S_{[u_2,v_2]}, S_{[u_1,v_2]} + S_{[v_1,u_2]} \right\} \,.
\eeq

In order to be able to obtain simple analytic expressions and to compare with the CFT results of Section~\ref{sec:form-MI} that were obtained for $\epsilon \ll \ell, t$, we here again work in the regime $\alpha \ll \ell, t$. 

%%%%%%%%%%%%%%%%%%%%%%%%%%%%%%%%%%%%%%%%%%%%%%%%%%

\subsection{Symmetric intervals of length $\ell$ separated a distance $d$}\label{sect:MISymm}

The mutual information for two symmetric intervals of length $\ell$ separated by a distance $d$ is obtained from Eq.~\eqref{eq:MIhol}, using Eq.~\eqref{eq:Ssalpha} and Eqs.~\eqref{eq:earlylatetime}-\eqref{eq:Sbalpha}. 

It is easy to check that the configuration with ``intersecting'' geodesics never contributes to the entanglement entropy, since $S_{\left[\frac d 2, \frac d 2 +\ell\right] }< S_{\left[-\frac d 2,\frac d 2+\ell \right]}$ for all times. By comparing the connected and disconnected configurations, one finds that the mutual information vanishes for $t < \frac d 2$ and $t > \frac d 2 +\ell$ whenever $d \ge (\sqrt 2 -1)\ell$, as in the vacuum \cite{2010PhRvD..82l6010H}. 
It is also zero in the time range $\frac d 2 < t < \frac d 2 +\ell$ if 
\be \label{eq:ineq}
\frac{\(\frac d 2 +\ell -t\)\(t - \frac d 2\)}{d(d+2\ell)}\le \frac{\alpha}{\tilde M \ell}\,,
\ee
for all $t \in (\frac d 2 , \frac d 2 +\ell)$, that is if 
\be
\frac d \ell \(\frac d \ell +2 \) \ge \frac{\tilde M \ell}{\alpha} \,. 
\ee
or
\be
d \ge \left[\sqrt{1 +\frac{\tilde M \ell}{4 \alpha} } -1 \right] \ell \equiv \bar d \,.
\ee
If this is not the case, since the two roots of Eq.~\eqref{eq:ineq} are $t_1 =\frac d 2 + O(\alpha)$ and $ t_2 = \frac d 2 +\ell + O(\alpha) $, the mutual information does not vanish in the whole time range $\frac d 2 < t  <  \frac d 2 +\ell $. 

The complete expression reads
\be 
I = \frac{1}{2\GN} \left\{\begin{array}{lc}
 \ln \frac{\ell^2}{d(d +2\ell)} & t < \frac d 2, t > \frac d 2 +\ell \\
 \ln \frac{\ell \( \frac d 2 +\ell  - t\) \(t -\frac d 2\) \tilde M}{\alpha \, d (d+2 \ell) } & \frac d 2 < t < \frac d 2 +\ell
\end{array}\right.
\ee
if $d< (\sqrt 2 -1) \ell$, or
\be 
I =\frac{1}{2\GN} \left\{\begin{array}{lc}
0 &  t< \frac d 2, t> \frac d 2 +\ell  \\
 \ln \frac{\ell \( \frac d 2 +\ell  - t\) \(t -\frac d 2\) \tilde M}{\alpha \, d (d+2 \ell)} & \frac d 2 < t < \frac d 2+ \ell 
\end{array}\right.
\ee
if $ (\sqrt 2 -1)\ell \le d < \bar d$, and $I  = 0$ for all times otherwise. 

\begin{figure}[h]
\begin{tabular}{ccc}
\includegraphics[width = 0.45 \textwidth]{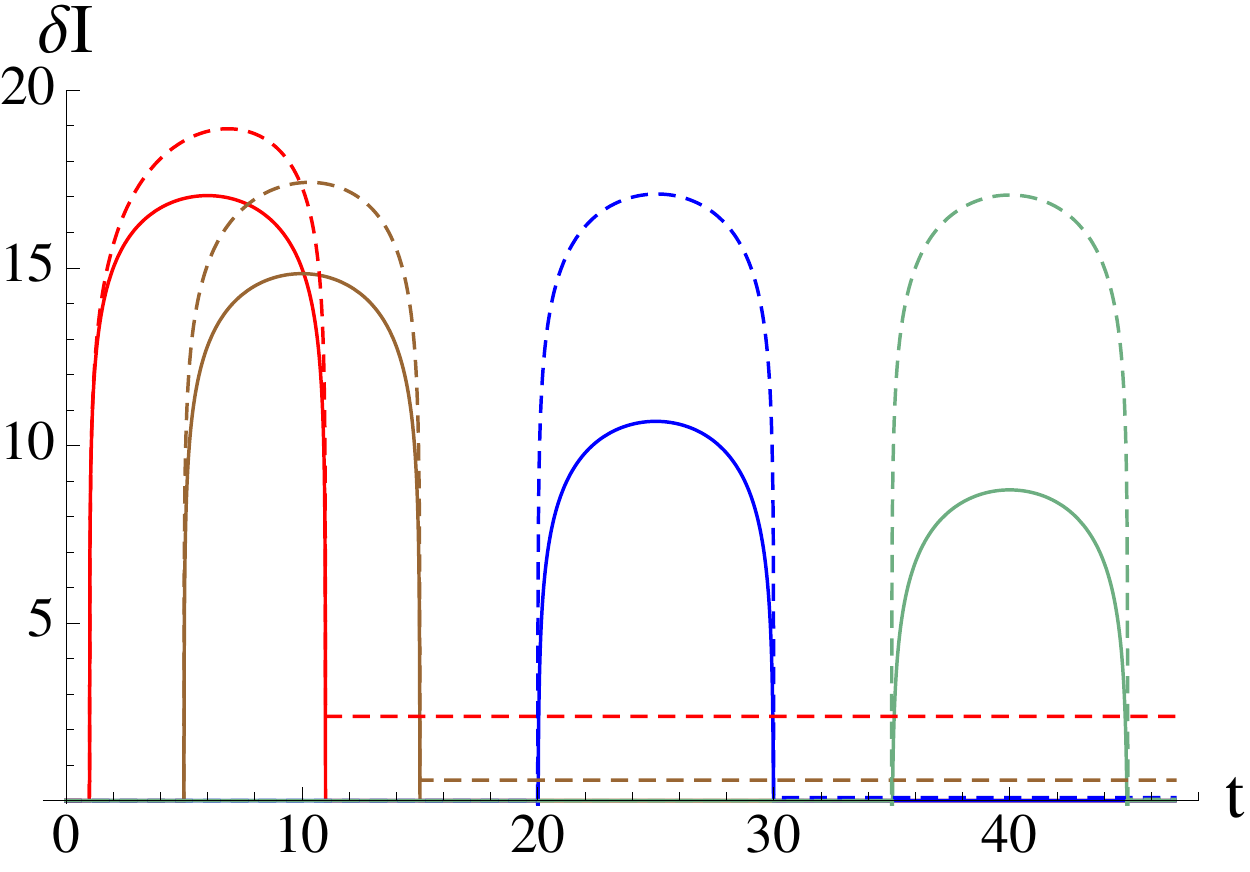} \hfill
&\qquad\qquad &
\includegraphics[width = 0.45 \textwidth]{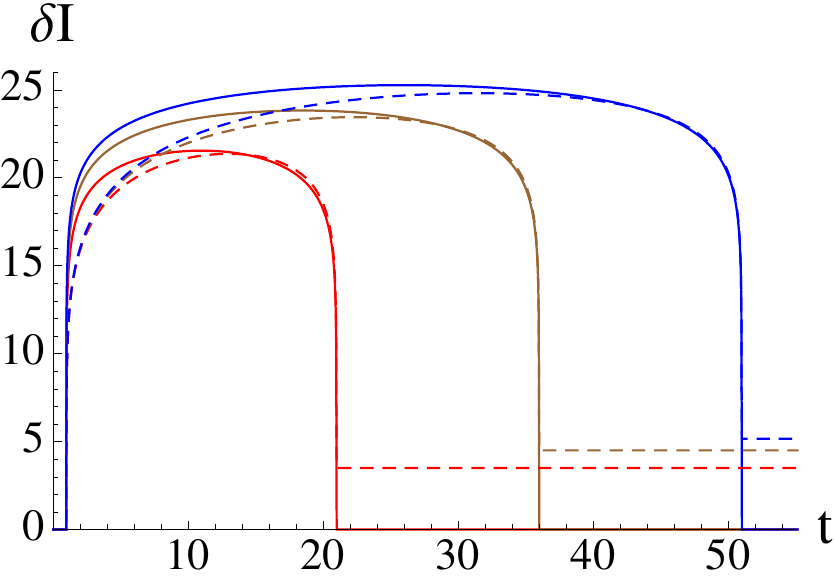} 
\\
{\bf (A)}&\qquad\qquad & {\bf (B)}
\\
\end{tabular}
\caption{Time evolution of the mutual information with the early time value subtracted (rescaled by $4 \GN$) for {\bf (A)} $\ell =10$, $d= 2,10,40,70$ from left to right and {\bf (B)} $d =2$, $\ell=20,35,50$ from left to right. The CFT results are shown with dashed lines and here we have set $M =3/4$ and $\alpha=\epsilon=0.001$.}\label{fig:Isymmtalpha}
\end{figure}

\begin{figure}[h]
\begin{center}
\includegraphics[width = 0.45 \textwidth]{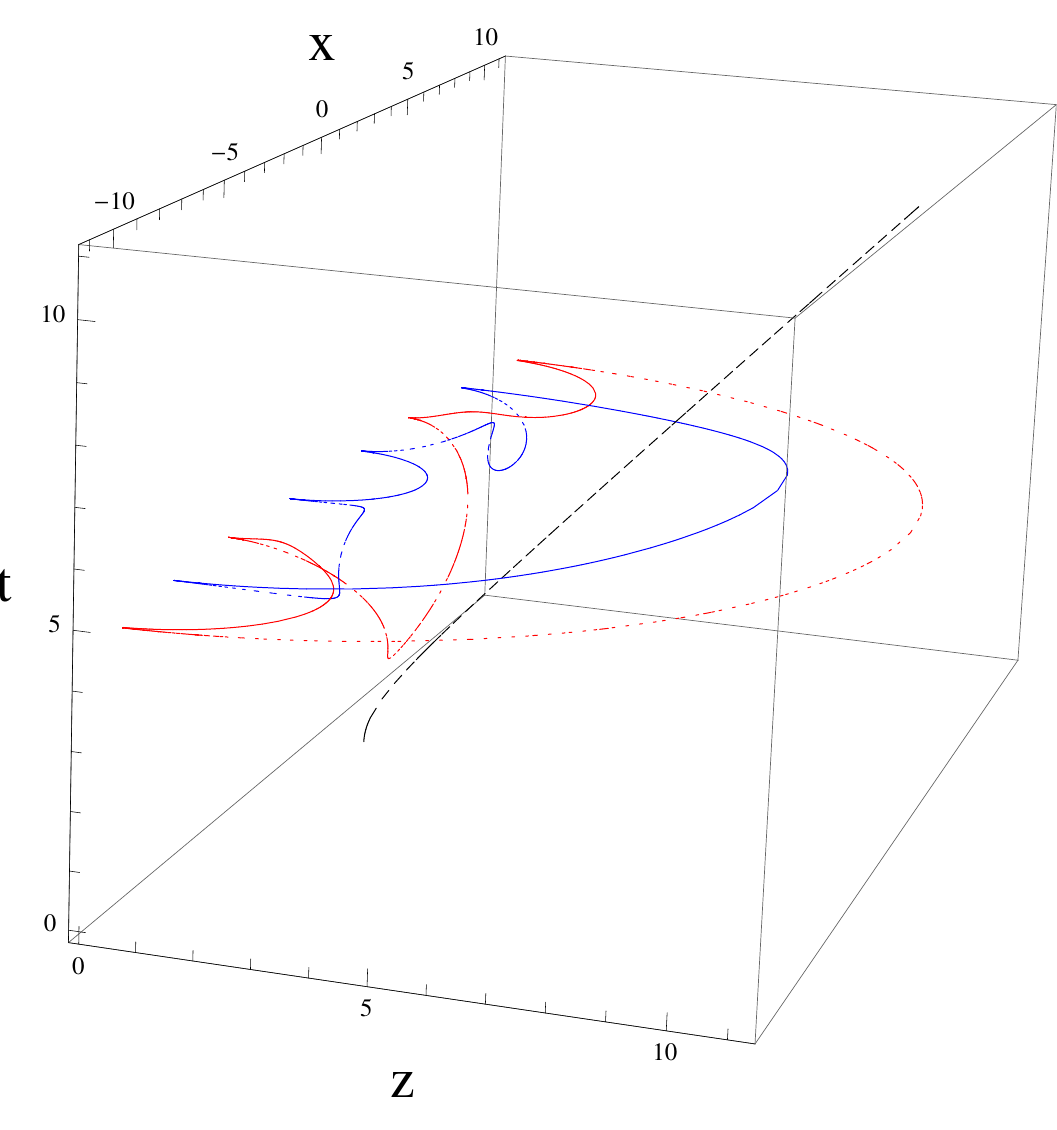} 
\caption{The two competing configurations of geodesics for the mutual information at $t=5$ for $\ell=6$ and $d=4$ (blue), $d=10$ (red) ($\alpha =1$, $M=3/4$). The configuration of minimal length is in solid lines. As we increase the separation between the two intervals, we observe a transition from the connected towards the disconnected configuration.}\label{fig:MIgeodesics}
\end{center}
\end{figure}

We plot in Fig.~\ref{fig:Isymmtalpha} the evolution of the mutual information with the early time value subtracted $\delta I \equiv I - I(t< \frac d 2)$ for two equal length intervals symmetric about the defect, for different separations and interval lengths, and compare with the (universal piece of the) CFT results \eqref{eq:MI_sym}.
Notice in particular that the maximum value of the mutual information is attained at $t= \frac{d+\ell}{2}$ and reads
\be
I_{\rm max} = \frac{1}{2\GN} \ln \frac{\ell^3 \tilde M}{4 \alpha \, d(d+2\ell) }\,,
\ee
which decreases logarithmically as we increase the separation $d$, while in the CFT calculation 
it is constant at large $d$ (see Eq.~\eqref{eq:smallr} and Figs.~\ref{fig:MI_trans} and \ref{fig:Isymmtalpha} {\bf (A)}). In both cases though, the maximal value increases as we increase the interval length $\ell$, while keeping the other quantities fixed (Fig.~\ref{fig:Isymmtalpha} {\bf (B)}). As we increase the separation $d$ between the two intervals, we observe a transition to a phase where the holographic mutual information is always vanishing, a feature that is not present in the CFT calculation. This was also found to be the case for the mutual information following a global quench when computed holographically in the AdS Vaidya geometry \cite{Balasubramanian:2011at,Allais:2011ys}. This holographic feature was again in contrast to the behavior of the mutual 
information after a CFT global quench \cite{Calabrese:2005in,Calabrese:2007rg}, which can be straightforwardly calculated using the same techniques we use in this paper. We will discuss this further in the Conclusion.
In Fig.~\ref{fig:MIgeodesics} we plot the two competing collections of extremal curves in the two phases, in which the connected or the disconnected configuration respectively minimize the length at intermediate times.
%

%%%%%%%%%%%%%%%%%%%%%%%%%%%%%%%%%%%%%%%%%%%%%%%%%%

\subsection{Equal length intervals on the same side of the quench} \label{sect:onesidealpha}

Let us consider two equal length intervals $A =[x, x+\ell]$, $B= [x + \ell +d, x +2\ell +d]$ on the same side of the quench ($x, \ell, d >0$) separated by a distance $d$, see Fig.~\ref{fig:intervals} $(iii)$. 
Using the formulae in Eqs.~\eqref{eq:earlylatetime}-\eqref{eq:Saalpha}, it follows that we have to look for the minimum among
\be \label{eq:onesideAB}
S_{\rm disc} = \frac{1}{4\GN} \left\{\begin{array}{lc}
\ln \frac{\ell^4}{z_\infty^4} &  t  \in \delta t_1, \delta t_3, \delta t_5\\
\ln \frac{\ell^3 (x+ \ell -t)(t -x) \tilde M}{\alpha z_\infty^4} & t \in \delta t_2 \\
\ln \frac{\ell^3 (x + 2\ell +d -t)(t -(x+\ell+d))\tilde M}{\alpha z_\infty^4} & t\in \delta t_4
\end{array}\right.
\ee
\be \label{eq:onesidedAdB}
S_{\rm conn} = \frac{1}{4\GN} \left\{\begin{array}{lc}
\ln \frac{d^2 (2\ell+d)^2}{z_\infty^4} &  t\in \delta t_1, \delta t_5 \\
\ln \frac{d^2(2\ell+d) (x+2\ell+d-t)(t -x)\tilde M}{\alpha z_\infty^4} & t\in \delta t_2, \delta t_4 \\
\ln \frac{d(2\ell+d) (x+\ell +d -t) (t -(x+\ell))(x+2\ell+d-t)(t -x) \tilde M^2}{\alpha^2 z_\infty^4} & t\in \delta t_3
\end{array}\right.
\ee
\be
S_{\rm int} = \frac{1}{4 \GN} 
\left\{\begin{array}{lc}
\ln \frac{(\ell+d)^4}{z_\infty^4} & t \in \delta t_1,  \delta t_5 \\
\ln \frac{(\ell+d)^3 (x+\ell+d-t)(t -x)\tilde M}{\alpha z_\infty^4} & t\in \delta t_2 \\
\ln \frac{(\ell+d)^2 (x+\ell+d-t)(t -x) (x+2\ell+d-t)(t -(x+\ell))\tilde M^2}{\alpha^2 z_\infty^4} & t\in\delta t_3 \\
\ln \frac{(\ell+d)^3 (x+2 \ell +d -t) (t- (\ell+x))\tilde M}{\alpha z_\infty^4} & t\in\delta t_4
\end{array}\right.
\ee
where $S_{\rm disc}$, $S_{\rm conn}$, $S_{\rm int}$ denote respectively the entanglement entropy associated to the disconnected, connected and ``intersecting'' configurations. As in Section~\ref{sec:MI_same_side}, we have defined the time intervals $\delta t_1 = (0,x)$, $\delta t_2 = (x,x+ \ell)$, $\delta t_3 = (x + \ell, x+\ell+d)$, $\delta t_4 = (x+\ell+d, x +2 \ell +d)$ and $\delta t_5 = (x +2 \ell + d, \infty)$. It is easy to see that $S_{\rm int} > S_{\rm disc}$ for $t < x +\ell +d$ and $t > x + 2\ell +d$, and $S_{\rm int} > S_{\rm conn}$ for $x+\ell + d < t< x +2\ell +d$, so that the ``intersecting'' configuration never contributes. 

Comparing Eq.~\eqref{eq:onesideAB} and Eq.~\eqref{eq:onesidedAdB} one finds the following result for the mutual information. 
If $d < \tilde d$, where
\be
\tilde d\equiv \left[ -1 + \sqrt{1 + \( \frac{4 \alpha }{\tilde M \ell}\)^{2/3}}\right] \ell\,,
\ee
we have
\be
I = \frac{1}{4\GN} \left\{\begin{array}{lc}
\ln \frac{\ell^4}{d^2(2\ell+d)^2} & t \in \delta t_1, \delta t_5\\
\ln \frac{\ell^3(x+ \ell  -t)}{d^2 (2\ell+d)(x+ 2\ell+d -t)} & t \in \delta t_2\\
\ln \frac{\ell^4 \alpha^2}{d(2\ell+d) (x+d+\ell-t) (t -(x+\ell))(x+2\ell+d-t)(t -x)\tilde M^2} &  t \in\delta t_3  \\
\ln \frac{\ell^3 (t - (x+\ell+d))}{d^2 (2\ell+d)(t-x)} & t \in \delta t_4 
\end{array}\right.
\ee
If instead $\tilde d \le d < (\sqrt{2}-1)\ell$
\be
I = \frac{1}{4\GN} \left\{\begin{array}{lc}
\ln \frac{\ell^4}{d^2(2\ell+d)^2} &  t < x, t > x +2\ell +d \\
\ln \frac{\ell^3(x+ \ell  -t)}{d^2 (2\ell+d)(x+ 2\ell+d -t)} & x < t <  \bar \ell_2 \\
0 & \bar \ell_2 < t < \bar \ell_3 \\
\ln \frac{\ell^3 (t - (x+\ell+d))}{d^2 (2\ell+d)(t -x)} & \bar \ell_3 < t < x +2\ell +d
\end{array}\right.
\ee
where
\bea
\bar \ell_2&= \ell +x +\frac{d^2(2\ell+d)}{d^2 + d\ell -\ell^2} \le \ell_2\\
\bar \ell_3 &= x - \frac{\ell^3}{d^2 + d\ell -\ell^2} \ge \ell_3 \,. 
\end{align}
Finally, the mutual information is identically vanishing if $d \ge (\sqrt{2}-1)\ell$. 

\begin{figure}[h]
\begin{tabular}{ccc}
\includegraphics[width = 0.45 \textwidth]{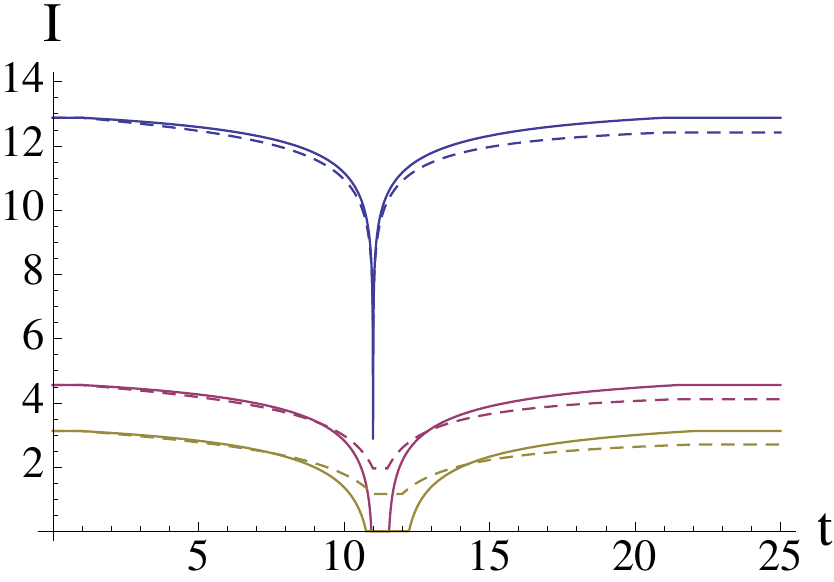} \hfill
&\qquad\qquad &
\includegraphics[width = 0.45 \textwidth]{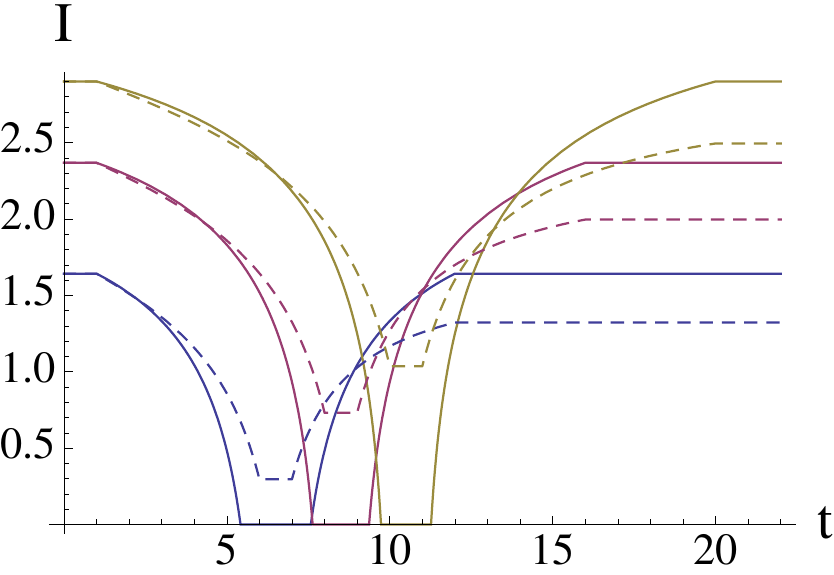} 
\\
{\bf (A)} &\qquad\qquad & {\bf (B)}
\\
\end{tabular}
\caption{Mutual information (rescaled by $4 \GN$) for {\bf (A)} $x=1$, $\ell=10$, $d=0.008, 0.5, 1$ from the top to the bottom, {\bf (B)} $x=1$, $d=1$, $\ell =5,7,9$ from the bottom to the top, as a function of time. The universal parts of the CFT results are shown in dashed lines. These have been individually shifted as to match the holographic results at $t=0$, to better visualize how the time dependence of the two results compares. Here $M =3/4$ and $\alpha=\epsilon=0.001$.}\label{fig:Ioneside}
\end{figure}

We plot in Fig.~\ref{fig:Ioneside} the mutual information for various separations and interval lengths, and compare with the field theory expressions Eqs.~\eqref{eq:SST1}-\eqref{eq:SST4}. To help visualize the comparison between the two results, the latter curves have been shifted so that at $t=0$ their values match the holographic ones. Notice however that the CFT results are universal for all times only if $d \gg \ell$ (and $x \gg d$ or $x\ll d$), in which case the holographic mutual information vanishes. The dashed curves in Fig.~\ref{fig:Ioneside} therefore do in general get non-universal corrections. 

\begin{figure}[h]
\begin{center}
\includegraphics[width = 0.45 \textwidth]{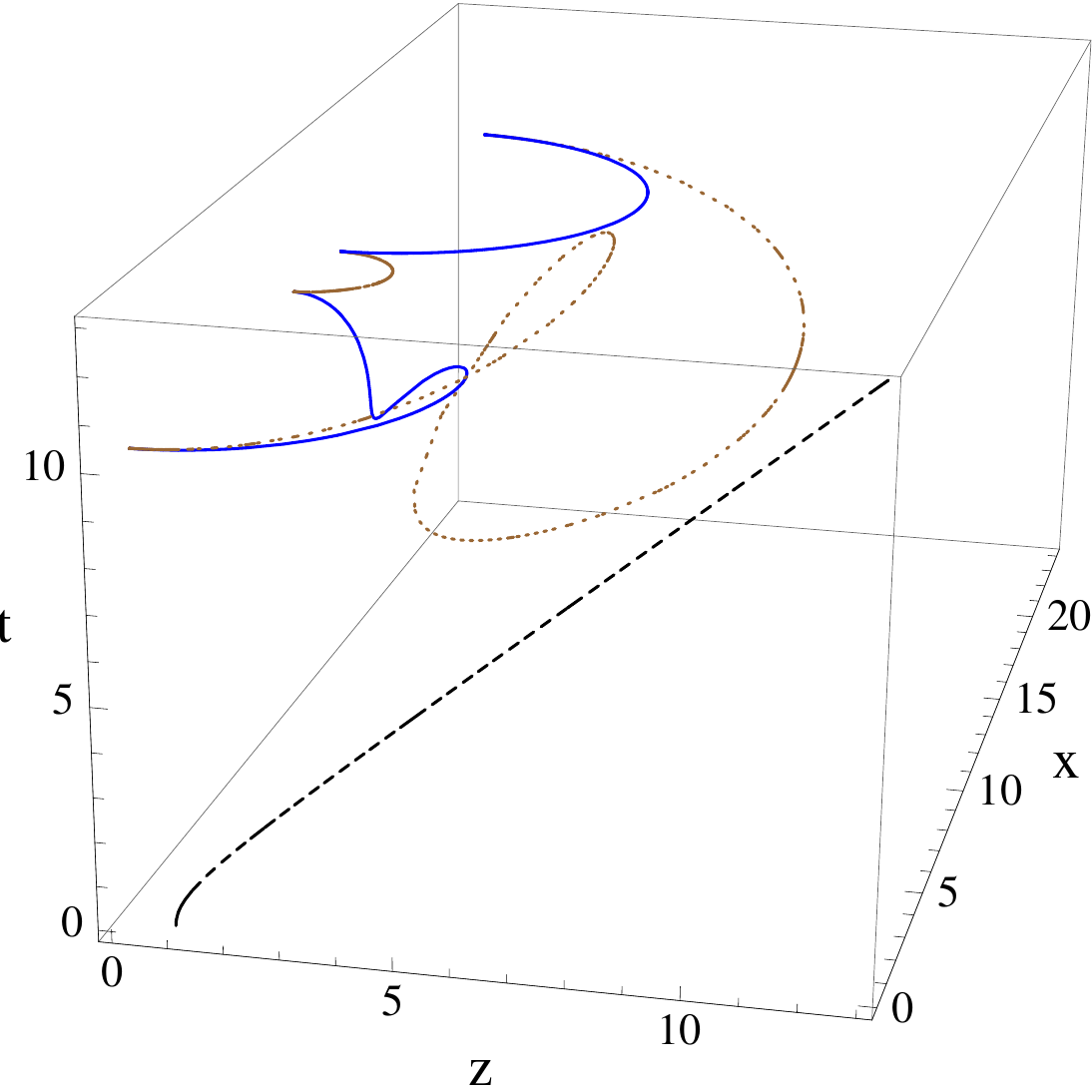} 
\caption{The two competing configurations of geodesics for the mutual information at $t =10 \in [\bar \ell_2,\bar \ell_3]$ for $x=1$, $\ell=10$ and $d=3$ ($\alpha =1$, $M=3/4$). In this time range, the outmost geodesic develops a long curl and the disconnected configuration (in solid lines) becomes the minimal one.}\label{fig:MIgeodesicsoneside}
\end{center}
\end{figure}

The most striking feature of the mutual information is that for $d < (\sqrt{2}-1)\ell$ it decreases at intermediate times, eventually reaching zero. This feature is not captured by the quasiparticle picture. 
However, it qualitatively matches the CFT computations, although there the minimal value attained by the mutual information is always positive. 
In terms of geodesics, this decrease is due to the fact that the outmost curve in the connected configuration develops a long curl (see Fig.~\ref{fig:MIgeodesicsoneside}). 
For large enough separations, the holographic mutual information vanishes identically, as for symmetric intervals. Again, this is not the case in the CFT results. 

%%%%%%%%%%%%%%%%%%%%%%%%%%%%%%%%%%%%%%%%%%%%%%%%%%

\subsection{Asymmetric intervals and overlapping the quench} \label{sect:generalalpha}

Finally, we consider the case of (equal length) asymmetric intervals about the origin 
$A = [ -(d-x+\ell),-(d-x)]$, $B=[x,x+\ell]$ 
(see Fig.~\ref{fig:intervals} $(ii)$) and the case of one interval overlapping the quench $A = [-(\ell -y), y] $, $B=[y+d, y+d+\ell]$. 
Because the analytic formulae for these cases are long and not particularly illuminating, 
we content ourselves with a qualitative discussion and some plots of a few cases, 
generated by direct calculation using Eqs.~\eqref{eq:MIhol}, \eqref{eq:Holog_SAB_2} and \eqref{eq:earlylatetime}-\eqref{eq:Sbalpha}.

\begin{figure}[h]
\begin{tabular}{ccc}
\includegraphics[width = 0.45 \textwidth]{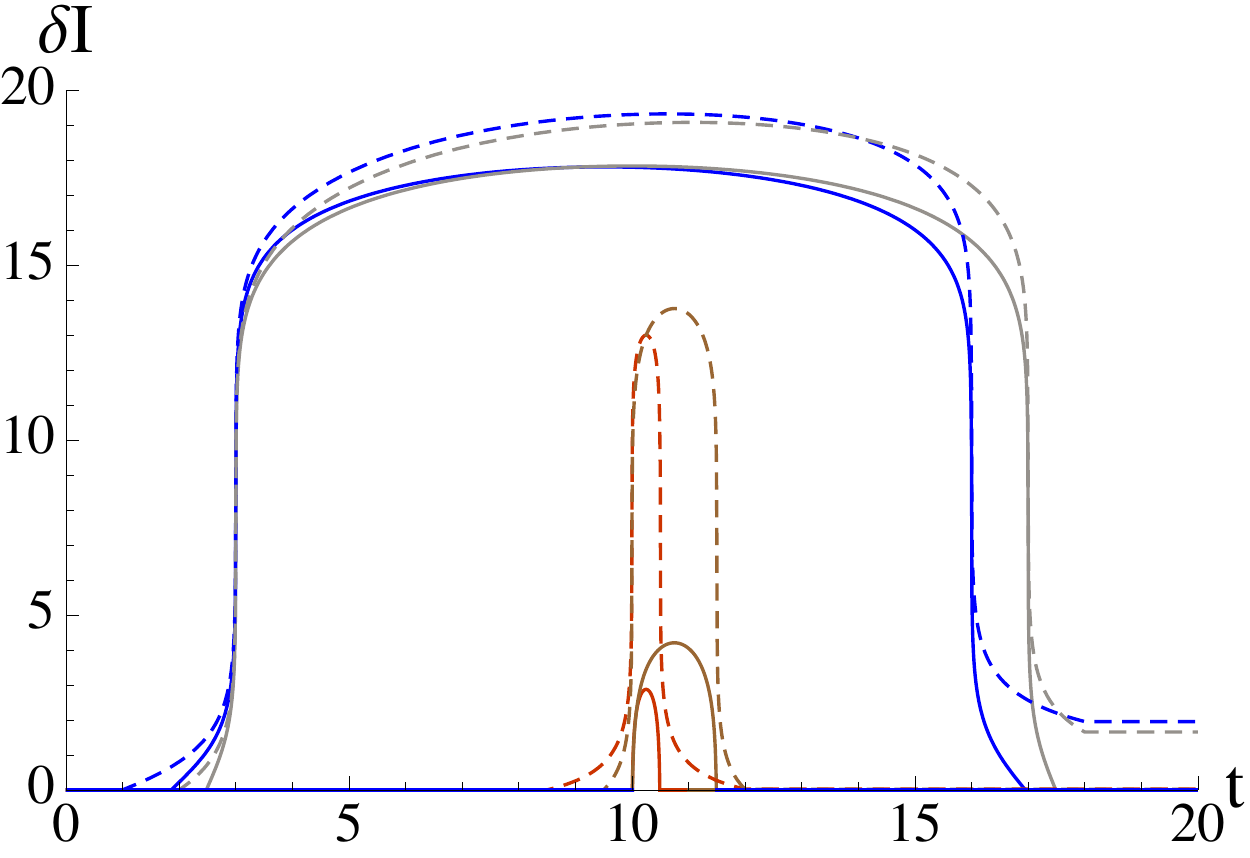} \hfill
&\qquad\qquad &
\includegraphics[width = 0.45 \textwidth]{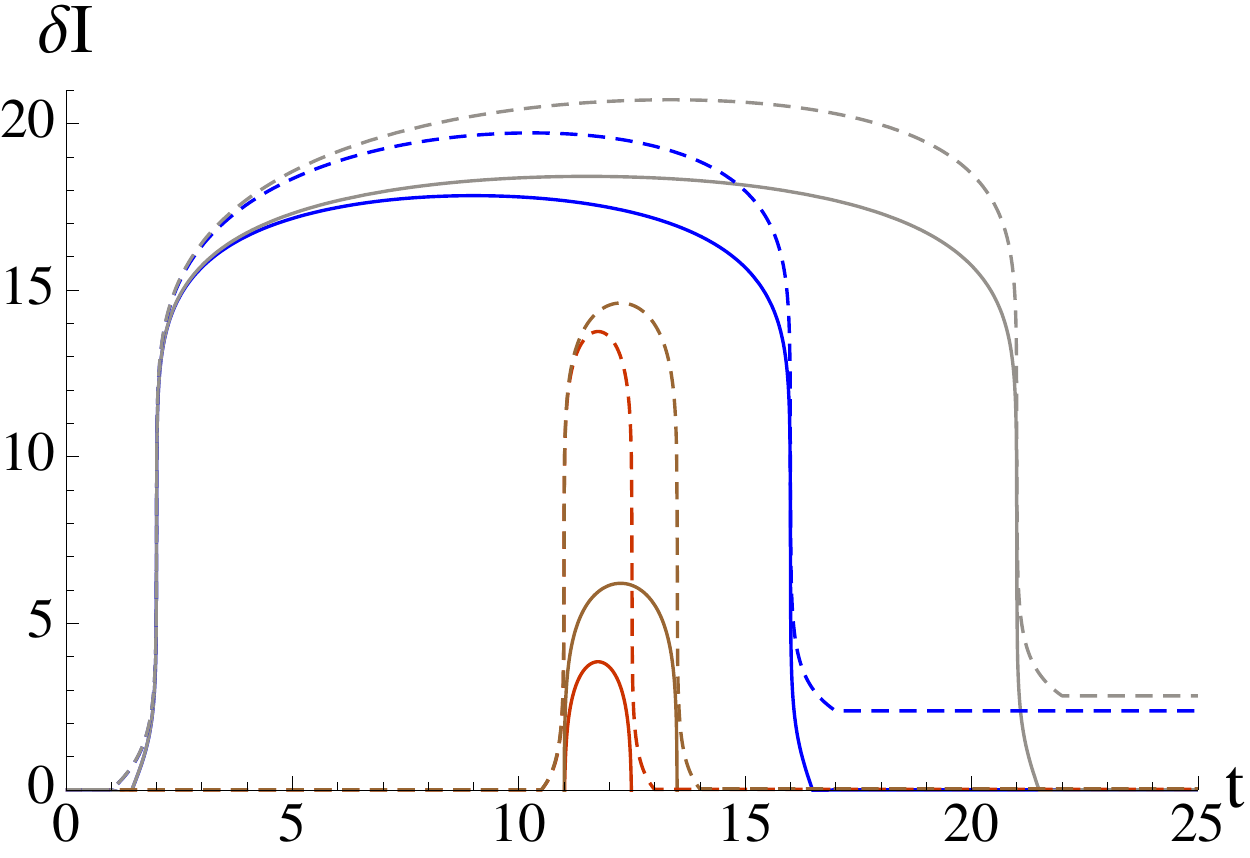} 
\\
{\bf (A)} &\qquad\qquad & {\bf (B)}
\\
\end{tabular}
\caption{Mutual information (rescaled by $4 \GN$) for asymmetric intervals about the quench with {\bf (A)} $x=10$, $\ell=2$, $d=18.5,19,19.5$ and $x=3$, $\ell=15$, $d=4,5$ {\bf (B)} $x=11$, $d=21.5$, $\ell =2, 3$ and $x=2$, $d=3$, $\ell=15,20$. We have set $M =3/4$ and $\alpha=\epsilon =0.001$ and plotted the CFT results in dashed lines.}\label{fig:Iasymm}
\end{figure}

The first situation is similar to that described in Section~\ref{sect:MISymm}. The mutual information is not 
just a constant only if both intervals are intersected by the lightcone of the quench 
at the same time, i.e.,
only if $d-x$ or $d-x+\ell$ are in the interval $(x, x+\ell)$ respectively for times $(d-x, x+\ell)$ or $(x,d-x+\ell)$, in agreement with the quasiparticle picture. The height of the bump also increases 
logarithmically if we increase the lengths of the intervals $(d-x, x+\ell)$ or $(x,d-x+\ell)$, as can be seen in Fig.~\ref{fig:Iasymm}. We compare with the universal CFT results of Section~\ref{sec:form-MI} in the universal regimes, that is for $x/\ell, (d-x)/\ell \gg1$ and $x/\ell, (d-x)/\ell \ll1$. The mismatch between the amplitudes of the two results is smaller in the second case, as anticipated by the analysis of the symmetric intervals. 

The case with one interval overlapping the defect qualitatively resembles that of two intervals on the same side of quench, analyzed in Section~\ref{sect:onesidealpha}. Typically, the mutual information starts and ends at the vacuum value, but drops at intermediate times. Its height decreases as the defects penetrates deeper in the interval, eventually vanishing everywhere, and increases as the length of the intervals increases. Contrary to the case of the two intervals on the same side of the quench, the decrease of the mutual information towards its minimal value is not monotonic and shows a spike. 

\begin{figure}[h]
\begin{tabular}{ccc}
\includegraphics[width = 0.45 \textwidth]{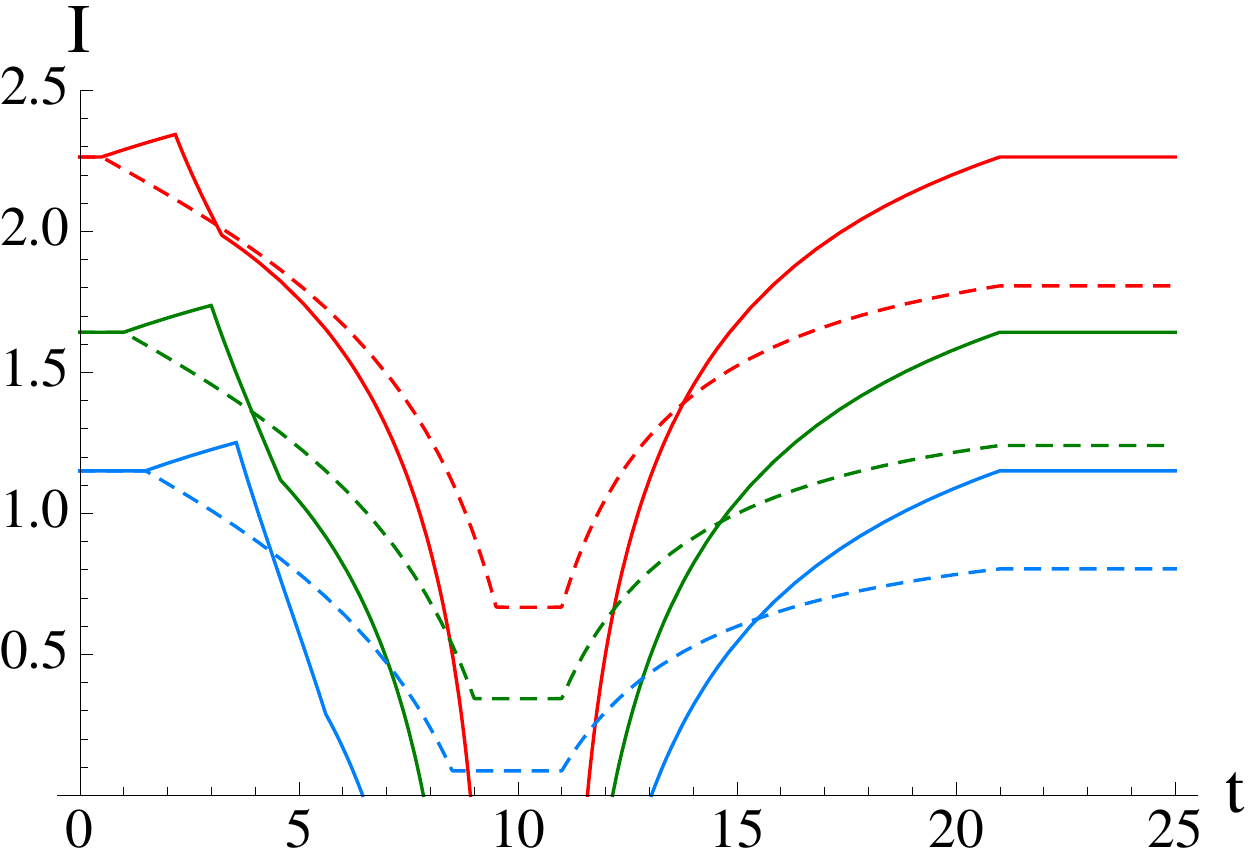} \hfill
&\qquad\qquad &
\includegraphics[width = 0.45 \textwidth]{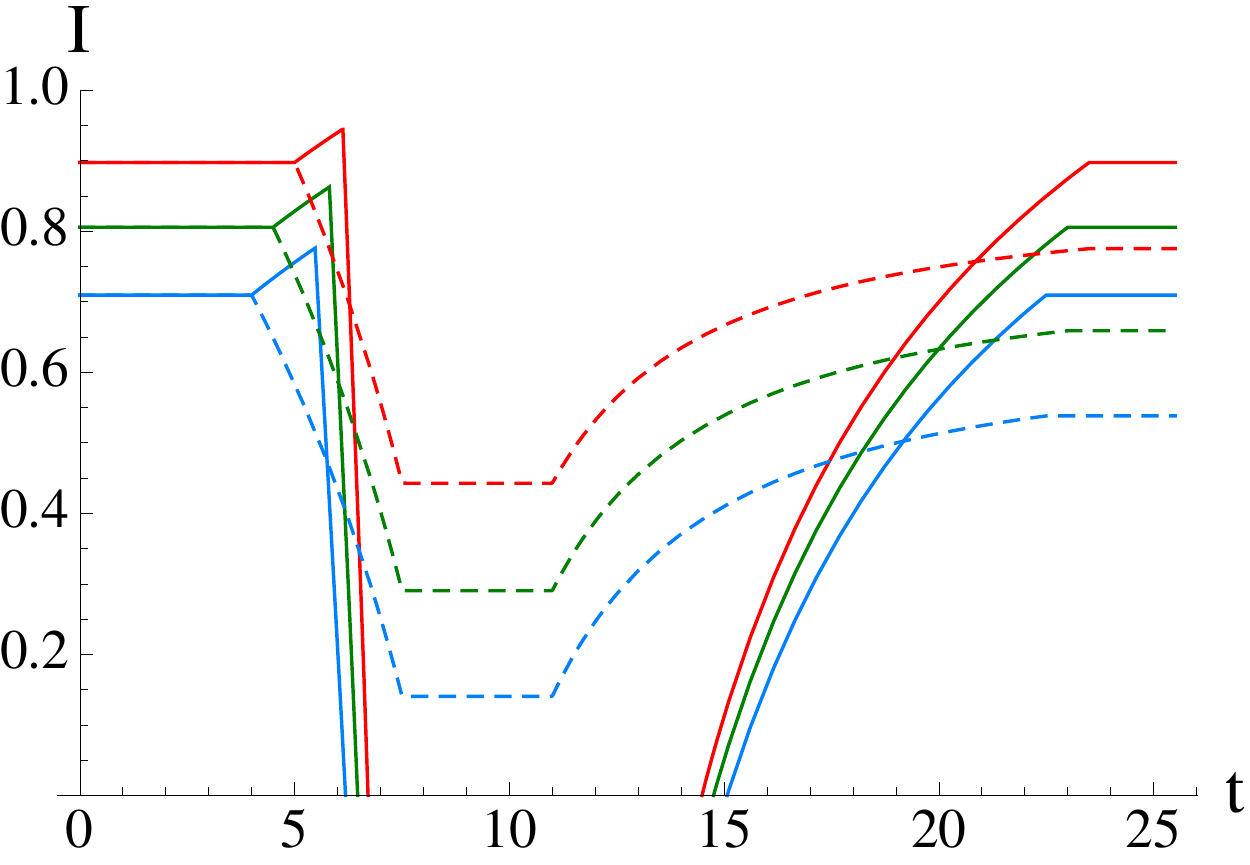} 
\\
{\bf (A)} &\qquad\qquad & {\bf (B)}
\\
\end{tabular}
\caption{Mutual information (rescaled by $4 \GN$) for non zero overlap with the defect and {\bf (A)} $\ell=10$, $d=1.5,2,2.5$, $y = 9.5,9,8.5$ from the top to the bottom, {\bf (B)} $y=7.5$, $d=3.5$, $\ell =11.5,12,12.5$ from the bottom to the top, as a function of time. The solid lines are the holographic results and the dashed lines the universal part of the CFT ones (the latter have been shifted so as to match the holographic result at $t=0$, to facilitate the comparison between the two). Here $M =3/4$ and $\alpha=\epsilon =0.001$.}\label{fig:Ioverlap}
\end{figure}

We plot some examples in Fig.~\ref{fig:Ioverlap}, together with the universal contributions to the CFT curves. Notice however that the CFT results get non-universal corrections in the regimes that are plotted. In the universal regime $d \gg \ell$, the holographic results identically vanish. 
%

%%%%%%%%%%%%%%%%%%%%%%%%%%%%%%%%%%%%
%%%%%%%%%%%%%%%%%%%%%%%%%%%%%%%%%%%%

\section{Alternative method for computing the entanglement entropy}
\label{sec:alt_method}

We saw in Sec.~\ref{sec:Holog_renorm} that the bulk geometry we have been 
studying can be obtained from a diffeomorphism of pure $\AdS_3$, in Eq.~\eqref{eq:diffeo}-\eqref{eq:diffeo-u}, 
specified at the conformal boundary by the maps 
\beq \label{eq:null_soln2}
	f_{\pm}(x_\pm) = x_\pm + \sqrt{x_\pm^2 + \epsilon^2}\,. 
\eeq 
given in Eq.~\eqref{eq:null_soln}.
In the last two sections we have used and extended the results of \cite{Nozaki:2013wia} to 
compute the holographic entanglement entropy and mutual information, but 
in fact these quantities may also be computed directly in terms of the $f_\pm$ \cite{Roberts:2012aq}. 
This leads to the same results, but we present this method here as it is often more efficient to use for explicit computations and applies to more general situations 
as well.

The bulk diffeomorphism in Eqs.~\eqref{eq:diffeo}-\eqref{eq:diffeo-u} specified by $f_\pm$ allows one to compute how 
various geometric quantities transform. We are particularly interested in 
lengths of geodesics anchored on the conformal boundary, 
which we can use to compute the holographic entanglement 
entropy via the Ryu-Takayanagi proposal \cite{Ryu:2006bv,Hubeny:2007xt}. 
We can, for example, use the formula for the lengths of geodesics 
in pure $\AdS_3$ and then pull back the result. 
This is explicitly done for one class of geodesics in \cite{Roberts:2012aq} 
and we review and extend those results here.

Let us use coordinates $(t,y,u)$ on the Poincar\'e patch of pure $\AdS_3$,
with metric  
\beq
	ds^2 = \frac{-dt^2 + dy^2 + du^2}{u^2}\,.
\eeq
For an interval that lies between two 
space-like separated boundary points $(t_\infty^{(1)},y_\infty^{(1)}, u_\infty^{(1)})$ and $(t_\infty^{(2)}, y_\infty^{(2)},u_\infty^{(2)})$,  
the Ryu-Takayanagi formula gives the entanglement entropy in terms of the 
length of the space-like geodesic $\gamma$ that goes between them, 
a semi-circle whose diameter has  proper length 
\be
s = \sqrt{\(y_\infty^{(2)} - y_\infty^{(1)}\)^2 - \(t_\infty^{(2)} - t_\infty^{(1)}\)^2}\,.
\ee
As in \cite{Nozaki:2013wia}, we need to introduce two radial cutoffs $u_\infty^{(i)}$, one for each endpoint, in the $(t,y,u)$ coordinates in order to obtain a result in terms of the transformed coordinates $(x_\pm, z)$ that depends on a single regulator $z_\infty$.

\begin{figure}[h]
\begin{center}
\def\svgwidth{12.0cm}
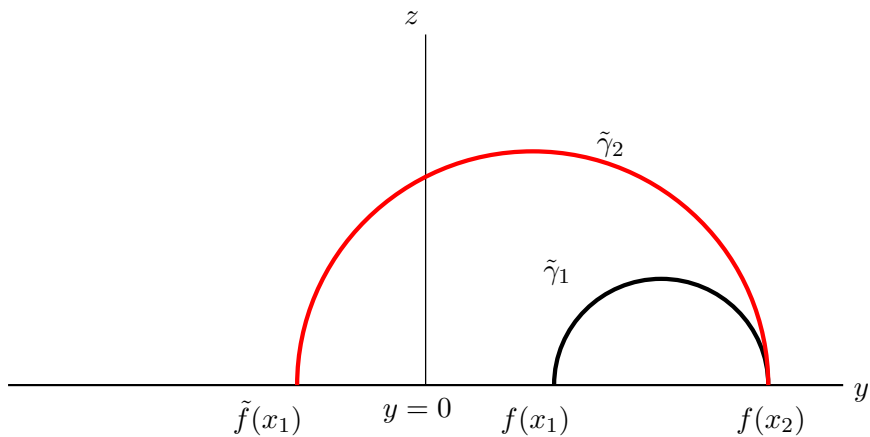
\end{center}
\caption{Schematic drawing of the geodesics contributing to the entropy 
formula in Eq.~\eqref{eq:L_ent_form}. The semicircles $\tilde\gamma_1$ and $\tilde\gamma_2$ 
are mapped to the geodesics $\gamma_1$ and $\gamma_2$, respectively, 
under the map $g$ defined on the conformal boundary by Eq.~\eqref{eq:null_inv}. 
}
 \label{fig:geodesics_y}
\end{figure}

The length ${\cal L}(\gamma)$ of these pure $\AdS_3$ geodesics is
\beq
\label{eq:geo1}
	{\cal L}(\gamma) =  \ln  \frac{\( y_\infty^{(2)} - y_\infty^{(1)}\)^2 - \(t_\infty^{(2)} - t_\infty^{(1)}\)^2}{u_\infty^{(1)} u_\infty^{(2)}} \,.
\eeq
In terms of null coordinates $y_\pm = y \pm t$, it becomes
\beq
\label{eq:null_entropy1}
	{\cal L}(\gamma) =  \ln 
		\frac{\(y_{\infty +}^{(2)} -  y_{\infty +}^{(1)}\) \(y_{\infty - }^{(2)}- y_{\infty-}^{(1)}\)}{u_\infty^{(1)} u_\infty^{(2)}} \,.
\eeq
Just as with the energy-momentum tensor, this result can be pulled back under 
the action of $f_\pm$ in Eq.~\eqref{eq:null_soln2}. As in \cite{Roberts:2012aq}, we can start with the case of a geodesic $\gamma_1$ 
whose image $\tilde{\gamma}_1$, under the diffeomorphism in Eqs.~\eqref{eq:diffeo} and 
\eqref{eq:diffeo-u}, is a semi-circle with endpoints satisfying $y_{\infty\pm}^{(1)} >0$ and $y_{\infty\pm}^{(2)}>0$ (see Fig.~\ref{fig:geodesics_y}). 
In that case, the length in the transformed system with $(x_\pm,z)$ coordinates can be written 
as \cite{Roberts:2012aq}
\beq
\label{eq:Matt_ent1}
	{\cal L}(\gamma_1) = \frac{1}{2} \ln \frac{\( f_+(x_{2+}) - f_+(x_{1+})\)^2 
	\( f_-(x_{2-}) - f_-(x_{1-})\)^2 }{f'_+(x_{2+}) f'_+(x_{1+}) f'_-(x_{2-}) f'_-(x_{1-}) 
	z_\infty^4} \,,
\eeq
where we have used the asymptotic behavior of the diffeomorphism in Eqs.~\eqref{eq:diffeo}-\eqref{eq:diffeo-u}
\begin{align}
y_{\infty\pm} &=f_\pm (x_{\pm}) \\
u_{\infty} &= z_\infty \sqrt{f_+' (x_+) f_-'(x_-)}
\end{align}
and identified the $z$ cutoffs $z_\infty^{(1)} =z_\infty^{(2)}\equiv z_\infty$.

In general we will need to go beyond this formula, as already noted in \cite{Roberts:2012aq},
since it misses some of the other possible geodesics.
This is similar to the situation we would encounter if we were interested 
in the BTZ black hole metric, which can be written in the form of 
Eq.~\eqref{eq:metric_gen} with constant $L_\pm$ (proportional to the 
black hole mass) and which would correspond to maps 
$f_\pm^{\text{BTZ}}(x_\pm) = \exp(2 \sqrt{L_\pm} x_\pm)$ \cite{Carlip:1994gc, 1999AIPC..484..147B}.
As reviewed in Appendix~\ref{sect:geodesics}, it is clear from writing the BTZ metric in global coordinates that there are 
two geodesics that anchor at a given pair of boundary points, one that wraps 
around the horizon and one that does not (the latter is homotopic to the boundary 
interval stretching between the points). The formula in Eq.~\eqref{eq:Matt_ent1} 
only yields the length of the non-wrapping geodesic, and so we need to go further, 
in general, to find the geodesic of minimal length as required by the Ryu-Takayanagi 
proposal. 
We have the same issue of multiple geodesics in the model we 
consider here; 
from our analysis in the previous two sections, we can interpret these as 
geodesics that either 
wrap around the trajectory of the massive particle, or not.

This issue can be fixed in a way that requires only slight modifications of Eq.~\eqref{eq:null_soln2}.
First of all, notice that what we are missing above are geodesics (semi-circles) 
$\tilde\gamma_2$ that stretch from boundary points with $y_{\infty \pm} > 0$ to boundary points with $y_{\infty \pm} < 0$ (see Fig.~\ref{fig:geodesics_y}). To include this case, it suffices to  
add another branch to the boundary maps, whose ranges cover $y_\pm <0$:
\beq
\label{eq:null_soln3}
	\tilde{f}_{\pm}(x_\pm) = x_\pm - \sqrt{x_\pm^2 + \epsilon^2}\,.
\eeq
This is of course still a solution of the differential equation \eqref{eq:Sch_null}. 

To explain and justify this choice, it is useful to go the other direction, finding maps 
$g_\pm(y_{\infty \pm})$ whose restrictions to $y_{\infty \pm} >0 $ are invertible and with inverses given by 
$f_\pm$ in Eq.~\eqref{eq:null_soln2}. In practice this just means inverting 
$f_\pm$:
\beq
	\label{eq:null_inv}
	g_\pm(y_{\infty \pm}) = \frac{y_{\infty \pm}^2 - \epsilon^2}{2 y_{\infty \pm}}\,.
\eeq
These functions are defined for all $y_{\infty \pm} \neq 0$
and form a double cover of the $x_\pm$ plane.
They send $0$ to $\pm\infty$ but we can 
ignore this since we are only interested in boundary points with $x_\pm \neq \pm\infty$.
To promote 
this to a map on the full geometry we would need to extend $g_\pm$ into the 
bulk, but for just computing anchored geodesic lengths we do not need this.
Inverting $g_\pm$ leads to a double valued map with the two branches given by 
the maps in Eqs.~\eqref{eq:null_soln2} and \eqref{eq:null_soln3}. 

It turns out, then, that one can capture the other possible geodesic $\gamma_2$
by applying $\tilde{f}_\pm(x_\pm)$ to one of the boundary anchor points, say 
the one with boundary coordinates $x_{1\pm}$, see Fig.~\ref{fig:geodesics_y}.
The result is the formula 
\beq
\label{eq:Matt_ent2}
	{\cal L}(\gamma_2) = \frac{1}{2} \ln \frac{( f_+(x_{2+}) - \tilde{f}_+(x_{1+}))^2 
	( f_-(x_{2-}) - \tilde{f}_-(x_{1-}))^2 }{f'_+(x_{2+}) 
		\tilde{f}'_+(x_{1+}) f'_-(x_{2-}) \tilde{f}'_-(x_{1-}) 
	z_\infty^4} \,.
\eeq
The formula for the entropy is then
\beq
\label{eq:L_ent_form}
	S = \frac{1}{4 \GN} \min\left\{ {\cal L}(\gamma_1),{\cal  L}(\gamma_2) \right\}.
\eeq
To get rid of the dependence on the cutoff $z_\infty$ we can subtract off, as we did in Sec.~\ref{sec:EE}, 
the entropy in pure $\AdS_3$ in $x_\pm$ coordinates,
\beq
	S_{\text{vac}} = \frac{1}{8 \GN} \ln 
		\frac{(x_{2+} - x_{1+})^2  (x_{2-}- x_{1-})^2}{z_\infty^4} \,,
\eeq
i.e., compute $\Delta S \equiv S - S_{\text{vac}}$. The cutoff dependence also disappears 
when computing the mutual information. 

\begin{figure}[h]
\includegraphics[width = 0.5 \textwidth]{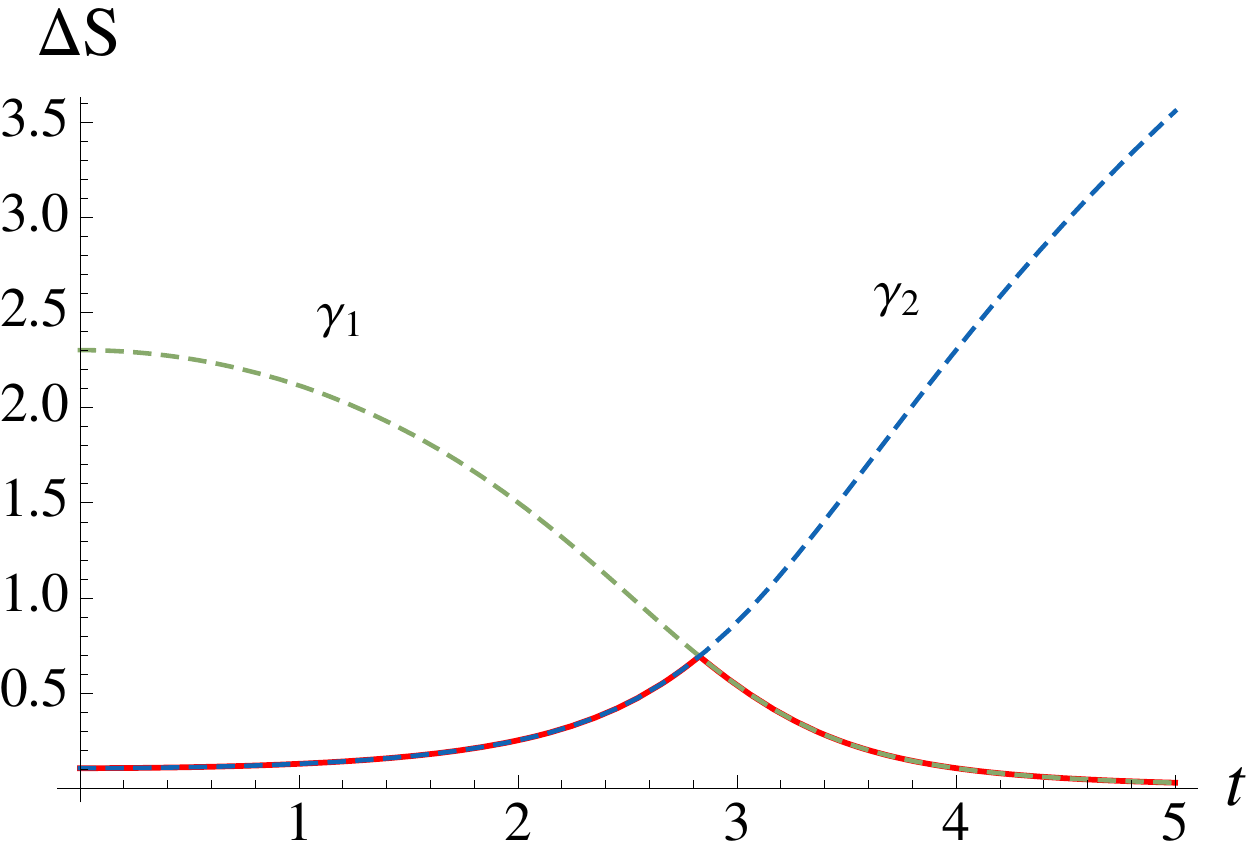} 
\caption{Comparison of the renormalized entanglement entropy (rescaled by $4 \GN$) of a symmetric interval of length 6 computed with the two methods we have described. The dashed lines show the contributions coming from the two competing geodesics $\gamma_1$ and $\gamma_2$ and the solid red line 
shows the result computed in Sec.~\ref{sec:EE}. 
}\label{fig:EE_alternative}
\end{figure}

As we already noted, these formulas give equivalent results to those obtained in Sec.~\ref{sec:EE} (for $M=3/4 R^2$), and are often more efficient for explicit calculations. In Fig.~\ref{fig:EE_alternative}, we give an example of an explicit check between the two results. 
On the other hand, starting in global coordinates does make some features of the geometry more 
transparent, such as the role of wrapping and non-wrapping geodesics.
This is because in that case we do not have to deal with double-valued maps like $g_\pm^{-1}$. 

When this work was almost completed,
material closely related to that in this section, as well as in Sec.~\ref{sec:Holog_renorm}, 
appeared in the paper \cite{Ugajin:2013xxa}. 
Following the AdS/BCFT 
proposal \cite{Takayanagi:2011zk,Fujita:2011fp}, 
\cite{Ugajin:2013xxa} introduces a boundary 
 in the bulk spacetime (different from the conformal boundary)
and relates it to the boundary 
appearing in the BCFT formalism used in CFT analyses of quenches. 
Using this proposal, \cite{Ugajin:2013xxa} gives a new formula for the 
holographic entanglement entropy in these quench setups, including local quenches.
We do not enter into a detailed analysis of this work, but note that, as 
also noted in \cite{Ugajin:2013xxa}, this new model does not appear to alter the behavior of 
the entanglement entropy after a local quench for late times or for intervals 
far from the quench. In particular, we have explicitly verified that the mutual 
information for symmetric intervals also decays logarithmically in the separation 
distance, and the analogous plot of the holographic mutual information shown in 
Fig.~\ref{fig:Isymmtalpha} {\bf (A)} is unchanged except at very early times.

%%%%%%%%%%%%%%%%%%%%%%%%%%%%%%
%%%%%%%%%%%%%%%%%%%%%%%%%%%%%%

\section{Conclusions}
\label{sec:Concl}

In the first part of this paper we have used CFT techniques to compute the 
universal parts of the mutual information $I_{A,B}$ after a local quench in a general
(1+1)-dimensional CFT. 
One significant finding is long range entanglement, i.e.,
that the leading contribution (in the small $\epsilon$ limit) 
to $I_{A,B}$ is 
of order the central charge and independent of the separation distance
for symmetric intervals (equal time, well-separated intervals of fixed length located on opposite sides of 
the quench and intersected by the lightcone of 
the quench event), see Eq.~\eqref{eq:smallr} and Fig.~\ref{fig:MI_trans}. 
This long range entanglement is consistent with the 
quasiparticle picture of a local quench, 
from which we would expect non-decaying entanglement between regions 
intersected by the trajectories of the entangled quasiparticles. 
This has also been observed in studies of spin chains 
\cite{2010PhRvB..81j0412S, 2010arXiv1006.1422B}, although these studies 
use a measure of entanglement different than the mutual information.

This contrasts with the behavior of the mutual information in the 
ground state of (non-holographic) CFTs, 
which at large separation decays as a power law in the separation distance
\cite{2011JSMTE..01..021C}.
Mutual information in the ground state of holographic CFTs was studied in detail in 
\cite{2010PhRvD..82l6010H}, 
where it was shown that in this state there is a phase transition to 
$I_{A,B} = 0$ for intervals at sufficiently 
large separation. 
More specifically, from conformal symmetry it is known that $I_{A,B}$ 
in the ground state is a 
function of the cross-ratio of the interval endpoints
\beq
	\eta = \frac{(v_1 -u_1)(v_2 - u_2)}{(u_2 - u_1)(v_2 - v_1)}.
\end{equation}
In the case of symmetric intervals, 
$A = [-(\ell+d/2),-d/2], B= [d/2, \ell+d/2]$, we have $\eta = \ell^2/(d+\ell)^2$, and small 
$\eta$ corresponds to small $\ell/d$.
It was shown that $I_{A,B}$ is of order the central charge for small 
separations ($\eta > 1/2$)
while it is order $1$ for large separations ($\eta \leq 1/2$). 
This implies that 
the holographic mutual information, as computed by the Ryu-Takayanagi formula, 
should go to zero for large separations, which it does when computed in pure $\AdS$ 
\cite{2010PhRvD..82l6010H}.  

In the recent paper \cite{Nozaki:2013wia} the authors proposed a model 
for a holographic local quench.
We have shown in Sec.~\ref{sec:Holog_renorm} that for some specific 
parameter values this model precisely matches the bulk geometry one would 
obtain from applying holographic renormalization to the energy-momentum tensor 
expectation value corresponding to the CFT local quench.
In the second part of this paper we 
have studied the holographic mutual information in this model. 
We found that the mutual information is not 
independent of the separation distance 
for equal length intervals intersected by the two 
pieces of the quench lightcone, i.e.,
intervals centered on $x = \pm \sqrt{\alpha^2 + t^2}$.
Instead, we found a logarithmic decay and then a transition to zero mutual 
information above a certain separation distance, see Sec.~\ref{sect:MISymm}. 
This is the same qualitative behavior as seen in the ground state 
of holographic CFTs that we mentioned above. The transition distance grows 
as $\alpha$ decreases but for fixed energy and interval length 
the mutual information always 
transitions to zero at a finite separation, see Eq.~\eqref{eq:ineq}.
In the ``shockwave limit," where one takes $\alpha$ and 
$M$ to zero such that the energy $E = M/8\GN \alpha$ is fixed, 
the transition to zero mutual information for symmetric intervals of length 
$\ell$ occurs at a separation distance 
$\bar{d} = \ell (\sqrt{1 + \pi\GN\ell E} -1)$. 

The authors of \cite{Nozaki:2013wia}
seem to observe long range entanglement and entangled pairs of excitations 
that travel away from each other and the quench event, while in our holographic analysis  
we find a vanishing mutual information between the propagating excitations
at large enough separation. The origin of this discrepancy might lie in the interpretation of entanglement 
entropy in terms of entanglement density of \cite{Nozaki:2013wia}, which seems to 
be a reliable measure only in regimes where the entanglement entropy behaves additively, i.e., extensively,
as in the case of thermal states. We expect the same cautionary remark to extend to higher 
dimensional versions of entanglement density \cite{Nozaki:2013vta}. 

The differences in the behavior of the CFT calculation of the mutual 
information and the holographic mutual information, 
exemplified by the plot in Fig.~\ref{fig:Isymmtalpha} {\bf (A)},
indicate significant differences between the state produced by the 
CFT local quench and the state described by the bulk geometry we 
studied, even though they have exactly the same $\bra T_{\mu\nu}(x,t) \ket$ 
for all $x$ and $t$.\footnote{We are thinking here of extending all 
of our CFT results to $t<0$ by unitary evolution. 
In other words, even though we analyzed the state by implementing a sudden 
quench at $t=0$, we can equivalently omit the quench event and simply take the 
state just after the quench as an initial condition that we may evolve forward 
and backward in time (with the connected CFT Hamiltonian).}
We note that there are no other primary fields with non-zero expectation 
values.
This is because for this kind of local quench the state just after the quench is
the fusion of the two ground states of the two half-lines.
The operator associated with the ground state under the operator-state 
correspondence is the identity operator so this fusion is
expected to only excite fields in the conformal family of the identity, i.e.,
the energy-momentum tensor and associated Virasoro algebra. 
More details of this argument are found in \cite{2011JSMTE..08..019S}. 
This justifies studying the process holographically using pure gravity.
However, local quenches with different boundary conditions may excite other 
primary fields, and it would be interesting to study these further.

In general, one expects the classical geometry to describe, to a good 
approximation in the appropriate large $N$, large coupling limit, 
typical CFT states 
with the matching $\bra T_{\mu\nu}(x,t) \ket$ \cite{Martinec:1998wm,Balasubramanian:2008da},
so one way to interpret our results is that the state created by a CFT local 
quench is an atypical state with the given $\bra T_{\mu\nu}(x,t) \ket$. 
This is consistent with the arguments in \cite{Asplund:2012rg} that the correct
holographic dual (away from the classical limit in the bulk) 
to a local quench state would require significant entanglement in the 
bulk. Thus it may be fruitful to apply recent generalizations of the 
holographic entanglement entropy proposal \cite{Barrella:2013wja, Faulkner:2013ana}, 
which include quantum effects in the bulk, 
to this kind of situation.

We can, however, provide an interpretation for the CFT state dual to 
the bulk geometry of \cite{Nozaki:2013wia}. Namely, as the result of transforming 
the ground state of a CFT in Minkowski space under the Lorentzian transformation 
given in Eq.~\eqref{eq:null_soln}. We saw in Sec.~\ref{sec:Holog_renorm} that 
this Lorentzian transformation can be extended to a (bulk) diffeomorphism 
of $\AdS_3$, given by Eqs. \eqref{eq:diffeo} and \eqref{eq:diffeo-u}, 
and that this leads to the same geometry as \cite{Nozaki:2013wia}, 
and hence to equivalent results for entanglement entropy, mutual 
information, etc.

It is instructive to compare this situation to that of a global quench. 
A review of the CFT analysis of global quenches is contained in \cite{2009JPhA...42X4005C}. 
The technique for computing entanglement entropy after global quench in 
$\CFT_2$ is the same as that described in this paper in Sec.~\ref{sec:CFT_one_two_EE}, but with a different conformal map: 
$z^\text{gq} = \exp(\pi w/2\tau_0)$, where $\tau_0$ is inversely proportional to the initial mass gap. Thus, one may also compute the mutual 
information for a pair of intervals using Eq.~\eqref{eq:MIz}. 
There is again a universal regime for equal length intervals that 
are sufficiently well separated.  
In that case, one 
can show that there is again long range entanglement in the sense of 
non-decaying mutual information between intervals separated by arbitrarily 
large distances.\footnote{The formula for entanglement entropy of multiple intervals in this regime was already given 
in \cite{Calabrese:2005in} (App. C), up to an unspecified asymptotic value $S_A(\infty)$.}
The asymptotically $\AdS_3$ Vaidya geometry, corresponding to a shell of 
matter falling from the conformal boundary and collapsing to form a 
black hole, has been proposed 
as the holographic dual to a $\CFT_2$ global quench 
\cite{AbajoArrastia:2010yt,Balasubramanian:2010ce,Balasubramanian:2011ur}.
In \cite{Balasubramanian:2011at,Allais:2011ys} the authors studied holographic mutual 
information and found that the mutual information between two intervals decayed logarithmically 
in the separation distance until transitioning to zero, just as we found 
in this paper and just as in pure $\AdS_3$. The situation is thus closely 
analogous to our results for a local quench, indicating that the CFT global quench state 
has stronger entanglement properties than the state dual to the 
$\AdS_3$ Vaidya geometry. We note that there have been several studies of 
quantum corrections to such holographic thermalization setups
\cite{Ebrahim:2010ra,Chesler:2011ds,Baron:2013cya}.
In addition, the recent paper \cite{Hartman:2013qma} has proposed 
a different holographic dual to a global quench, which seems to match the 
CFT global quench state more closely in certain respects.
It would be very interesting to study mutual information in these setups and 
to investigate whether such ideas can be applied to local quenches. 
We plan to pursue these questions in future work. 

Our results indicate that
the state produced by a local quench provides an example of an 
Einstein-Podolksy-Rosen (EPR) pair in CFT. 
More precisely, the state contains a separated pair of 
excitations in an entangled state with mutual information of order the 
central charge $c$ (a traditional EPR pair 
has mutual information of order $\log 2$). 
There have been several recent papers discussing the holographic dual of an 
EPR pair in the boundary CFT 
\cite{Maldacena:2013xja, Jensen:2013ora, Chernicoff:2013iga,Sonner:2013mba,
Faulkner:2013ana,Marolf:2013dba}. In particular, in \cite{Maldacena:2013xja} the authors 
propose a conjecture that they summarize as ``EPR = ER," roughly, that an EPR pair 
in a holographic CFT is dual or equivalent to an Einstein-Rosen (ER) bridge in the bulk. 
Our results indicate that the EPR type state created by a local quench is not 
dual to the source-free classical bulk geometry 
with the matching holographic energy-momentum tensor. 
It would be interesting to investigate further the relation of our results with the conjecture proposed in \cite{Maldacena:2013xja}.  

For applications to more general holographic dualities and on general grounds 
it is natural to consider local 
quenches in higher dimensions. Unfortunately, many of the CFT results we used 
do not hold in higher dimensions, so a direct CFT calculation would be difficult.
However, see \cite{Cardy:2013nua} for recent results on the mutual information in 
higher-dimensional CFTs (in the ground state).
We expect certain qualitative aspects, which follow from the localized creation 
of entangled particles,
would also hold in higher dimensions. Numerical simulations on 
higher dimensional critical 
spin chains should be possible and would provide valuable data on such 
processes. Another direction to pursue would be to use measures of 
entanglement other than the mutual information, such as 
entanglement negativity, which has been recently analyzed in detail in 
two dimensional CFTs \cite{2012PhRvL.109m0502C, 2013JSMTE..02..008C}. 
Finally, in the ultimate pursuit of the holographic dual of a CFT local quench 
of the kind considered here, it may be useful to use the AdS/BCFT proposals 
in \cite{Takayanagi:2011zk, Fujita:2011fp, Nozaki:2012qd, Jensen:2013lxa}, 
as in \cite{Ugajin:2013xxa} (see the note at the end of 
Sec.~\ref{sec:alt_method}),
or to modify the bulk in various other ways, such as by adding more 
complicated sources, as in, e.g., \cite{Jensen:2013ora,Chernicoff:2013iga}, attaching 
additional boundary regions, possibly 
with different asymptotics or different signatures, as in, e.g., \cite{Hartman:2013qma}, or including 
bulk quantum effects, of the kind proposed in \cite{Asplund:2012rg} or 
in the recent studies \cite{Barrella:2013wja, Faulkner:2013ana}. 
At the practical level, the alternative method for computing the holographic entanglement entropy discussed in Sec.~\ref{sec:alt_method} could be useful 
for performing calculations in some of these more complicated setups. 

%%%%%%%%%%%%%%%%%%%%%%%%%%%%%%%%%%%%
%%%%%%%%%%%%%%%%%%%%%%%%%%%%%%%%%%%%

\begin{acknowledgments}

We thank S.~Avery, F.~Galli, M.~Headrick, J.~Maldacena, R.~Monten, R.~Myers and 
M.~Roberts for discussions on this work, and especially B.~Craps and F.~Denef for comments on a draft of the paper. 
This work is supported by the FWO-Vlaanderen, Project No. G.0651.11, by the Belgian Federal Science Policy Office through the Interuniversity Attraction Pole P7/37, by the European Science Foundation Holograv Network, by COST Action MP1210 The String Theory Universe, by the John Templeton Foundation and by the Odysseus program of the FWO. AB is a Postdoctoral Researcher FWO-Vlaanderen. The opinions expressed in this publication are those of the authors and do not necessarily reflect the views of the John Templeton Foundation.

\end{acknowledgments}

%%%%%%%%%%%%%%%%%%%%%%%%%%%%%%%%%%%%%%%%%%%%%%%%%%%%%%%%%%%%%%%%%%%%%%%%%%
%%%%%%%%%%%%%%%%%%%%%%%%%%%%%%%%%%%%%%%%%%%%%%%%%%%%%%%%%%%%%%%%%%%%%%%%%%

\appendix

%%%%%%%%%%%%%%%%%%%%%%%%%%%%%%%%%%%%%%%%%%%%%%%%%%%%%%%%%%%%%%%%%%%%%%%%%

\section{Cross-ratios and non-universal factors}\label{sec:crossratios}

In this Appendix we discuss in detail the cross-ratios and non-universal factors 
that were introduced in Sec.~\ref{sec:EEcomputation}.
In particular, we are interested in finding the regimes where these cross-ratios go to $0$, $1$ or $\infty$
so that the functions $\tilde{\mathcal{F}}_{n,2N} \to 1$. 
In this regime the results for the entanglement entropy in Eqs.~\eqref{eq:SAsingle} and \eqref{eq:Sdouble} (and subsequent) are universal, up to an 
additive constant, and the result for the mutual information in Eq.~\eqref{eq:MIz} is also universal.

For any four points $z_a, z_b, z_c$ and $z_i$ in the complex plane the cross-ratio 
is defined as
\footnote{Here we use the definition in \cite{2007JSMTE..10....4C,2009JPhA...42X4005C}.}
\beq
\label{eq:cr}
	\eta_i \equiv \eta_{abci} \equiv  
		\frac{(z_a - z_b)(z_c - z_i)}{(z_c - z_b)(z_a - z_i)}
\eeq
and is invariant under global conformal (M\"obius) transformations 
acting on $z_a, z_b, z_c$ and $z_i$. 
From Eq.~\eqref{eq:cr} we see $\eta_a = \infty$, $\eta_b =1$, and $\eta_ c = 0$,
while for $i\neq a,b,c$, $\eta_i$ depends nontrivially on $z_a, z_b, z_c$ and  $z_i$. 
For a general collection of $2N$ points we have, then, 
$2N-3$ independent, nontrivial, conformally invariant quantities. 
Note that the choice of which points to call 
$z_a, z_b$ and $z_c$ is arbitrary, and any choice results in $2N-3$ 
independent, nontrivial cross-ratios. 

The correlators we need to compute, 
in Eqs.~\eqref{eq:4twist}
and \eqref{eq:wcorr1int}, correspond to an eight-point and a four-point function in the $z$ plane, 
respectively. Conformal symmetry can be used to show that a  
general $2N$-point function does not depend on the locations of all $2N$ points
separately. Instead, they can be written as a function of 
the $2N-3$ cross-ratios $\{\eta_i\}$ described above
 \cite{2009JPhA...42X4005C, 1970ZhPmR..12..538P}.

Let us begin by considering the entanglement entropy of one interval, which, in the notation of Fig.~\ref{fig:map}, corresponds to the insertion of twist operators in the $W$ plane at $w_3 = \ell_1 + i \tau$, $w_4 = \ell_2 + i \tau$, where we assume without loss of generality $0< |\ell_1| < \ell_2$. Under the mapping in Eq.~\eqref{eq:UHPmap}, these endpoints correspond to $\epsilon z_p = \ell_p + i \tau + \rho_p e^{i \theta_p}$ where 
\be
\rho_p^2 =\sqrt{\(\epsilon^2 +\ell_p^2 -\tau^2\)^2 + 4 \ell_p^2 \tau^2}\,, \qquad \theta_p = \frac 1 2 \tan^{-1} \frac{2 \ell_p \tau}{\epsilon^2 +\ell_p^2 -\tau^2}\,,
\ee
and $p=1,2$. The associated cross ratio has been computed in \cite{2007JSMTE..10....4C,2009JPhA...42X4005C} and in the notation of Fig.~\ref{fig:map} reads:
\begin{align} \label{eq:crossratiosingle}
	\eta &\equiv \eta_{3645} =
	\frac{(z_3 - z_6)(z_4 - z_5)}{(z_4 - z_6)(z_3 - z_5)}= \frac{(z_3 +\bar z_3)(z_4+\bar z_4)}{(z_4 +\bar z_3)(z_3+\bar z_4)}\,. 
\end{align}
The expression simplifies in the limit $\ell_1,\ell_2, t \gg \epsilon$ and the real time evolution is obtained by analytically transforming $\tau \to i t$. The computation is simplified by the observation that in this limit we have:
\begin{align}
\rho_p \cos \theta_p  &\to  \max[\ell_p,t] \( 1 +  \frac{\epsilon^2}{2 \left[(\max[\ell_p,t])^2 -(\min[\ell_p,t])^2\right]} \)+O(\epsilon^4) \\
\rho_p \sin \theta_p  &\to  i \min[\ell_p,t] \( 1 +  \frac{\epsilon^2}{2 \left[(\min[\ell_p,t])^2 -(\max[\ell_p,t])^2\right]} \)+O(\epsilon^4) 
\end{align}
and leads, to first nontrivial order in $\epsilon$, to \cite{2007JSMTE..10....4C,2009JPhA...42X4005C}
\be \label{eq:crosssingle1}
\eta(t< |\ell_1|)  = \left\{\begin{array}{lc}
\frac{\epsilon^2 |\ell_1| \ell_2 }{(\ell_1^2 -t^2)(\ell_2^2 -t^2)} & \ell_1 <0  \\
4 \frac{\ell_1 \ell_2}{(\ell_1 + \ell_2)^2} & \ell_1 >0 \,,
\end{array}\right.
\ee
\be\label{eq:crosssingle2}
\eta(|\ell_1| < t < \ell_2) = \frac{2 \ell_2 (\ell_1 +t )}{(\ell_1 + \ell_2)(\ell_2 +t)}
\ee
\be\label{eq:crosssingle3}
\eta(t>\ell_2) = 1\,.
\ee
When $\eta \simeq 0, 1$ or $\infty$, the function $\tilde{\cal F}_{1,2} (\eta)$ goes to one. This happens, for instance, for all times when $\ell_2 \gg |\ell_1|$ or for $|\ell_1| \gg \ell_2 -\ell_1( >0)$, and in these cases the results for the single interval entanglement entropy in Eqs.~\eqref{eq:EEsingle1}-\eqref{eq:EEsingle2} are universal. 
 
Notice that the cross-ratio associated with an interval to the left of the quench with endpoints at (in the notation of Fig.~\ref{fig:map})  $w_1 = -\ell_2 + i \tau$, $w_2 =- \ell_1 + i \tau$, with $0 < \ell_1 < \ell_2 $ is also given by Eq.~\eqref{eq:crossratiosingle} and Eqs.~\eqref{eq:crosssingle1}-\eqref{eq:crosssingle3}, as expected. However, the singular behavior of the map \eqref{eq:UHPmap} for points in the left half plane in the $\ell_1,\ell_2, t \gg \epsilon$ limit does not allow one to recover the explicit expressions  \eqref{eq:crosssingle1}-\eqref{eq:crosssingle3} as a limit of $\eta_{1827}$.
		
For the entanglement entropy of two intervals there are five distinct cross-ratios $\eta_i \equiv \eta_{364i}$, with $i=5,7,8,1,2$. We already computed $\eta_5$ in Eq.~\eqref{eq:crossratiosingle} and it is instructive to look at the explicit expressions of the remaining cross ratios for the cases of symmetric intervals about the quench and of two intervals on the same side of the quench, which are discussed in detail in the main text. 

For two symmetric intervals of equal length $\ell$ separated a distance $d$ about the quench, to leading nontrivial order in $\epsilon$ the independent cross ratios are $\eta_5$ with $\ell_1 \to \frac d 2$, $\ell_2 \to \frac d 2 +\ell$ and:
\begin{align} 
\left. \begin{array}{ccc}
\eta_7 = \left\{\begin{array}{lc}
\frac{d(d+2\ell-2t)}{(d+\ell)(d-2 t)} & t < \frac d 2 \\
\frac{\ell(d+2t)}{(d+\ell)(2t-d)} & \frac d 2 < t< \frac d 2 +\ell \\
1 & t> \frac d 2 +\ell
\end{array}\right.
& \quad &
\eta_8 = \left\{\begin{array}{lc}
\frac{d(d+2\ell-2t)}{(d+\ell)(d-2 t)} & t < \frac d 2 \\
\frac{(d+2\ell-2t)^2 (4 t^2-d^2)}{4 \epsilon^2 (d+\ell)(d+\ell -2t)} & \frac d 2 < t< \frac d 2 +\ell \\
1 & t> \frac d 2 +\ell
\end{array}\right.
\end{array}\right.
\end{align}
\begin{align}
\left. \begin{array}{ccc}
\eta_1 = \left\{\begin{array}{lc}
\frac{d(d+2\ell-2t)}{(d+\ell)(d-2 t)} & t < \frac d 2 \\
\frac{((d+2\ell)^2-4t^2) (4 t^2-d^2)}{4 \epsilon^2 (d+\ell)^2} & \frac d 2 < t< \frac d 2 +\ell \\
\frac{(d+\ell)(d-2t)}{d(d+2\ell-2t)} & t> \frac d 2 +\ell
\end{array}\right.
& \quad &
\eta_2 = \left\{\begin{array}{lc}
\frac{d(d+2\ell-2t)}{(d+\ell)(d-2 t)} & t < \frac d 2 \\
\frac{((2t -d)(d+2t)^2(d+2\ell-2t)}{4 \epsilon^2 d (d+\ell)} & \frac d 2 < t< \frac d 2 +\ell \\
\frac{(d+2\ell)(d-2t)}{(d+\ell)(d+2\ell-2t)}  & t> \frac d 2 +\ell
\end{array}\right.
\end{array}\right.
\end{align}
Notice they all go as $0,1$ or $\infty$ for all times if the scales set by the intervals length and their distance are well separated, that is if $\ell/d \ll 1$ or $d/\ell \ll1$. 

In Section~\ref{sec:MI_same_side} we consider two non-overlapping intervals 
on the same side of the quench: $A = [\ell_1, \ell_2]=[x, x + \ell]$, $B =[\ell_3, \ell_4]= [x+\ell+d,x+2 \ell+d]$ with $x,\ell,d>0$. 
The cross ratio $\eta_5$ is given by Eqs.~\eqref{eq:crosssingle1}-\eqref{eq:crosssingle3} with $\ell_1 \to x+\ell +d$ and $\ell_2 \to x+2\ell+d$, and the others read: 
\begin{align} 
\left. \begin{array}{ccc}
\eta_7 = \left\{\begin{array}{lc}
\frac{2 (x+\ell+d)(2x +3\ell+d)}{(2x +2 \ell+d)(2x +3 \ell+ 2 d)}& t < \ell_3 \\
\frac{(2x +3\ell+d)(x+\ell+d+t)}{(2x+3\ell+2d)(x+\ell+t)} &\ell_3 < t< \ell_4 \\
1 & t> \ell_4
\end{array}\right.
& \quad &
\eta_8 = \left\{\begin{array}{lc}
\frac{2 (x+\ell+d)(2x+2\ell+d)}{(2x +\ell+d)(2x +3 \ell+2 d)}& t < \ell_3 \\
\frac{(2x+2\ell+d)(x+\ell+d+t)}{(2x +3\ell+2 d)(x+t)} &\ell_3 < t< \ell_4 \\
1 & t> \ell_4
\end{array}\right.
\end{array}\right.
\end{align}
\begin{align}
\eta_1 = \left\{\begin{array}{lc}
\frac{2(x+\ell+d)(2\ell+d)}{(\ell+d)(2x +3\ell+2d)} & t <\ell_1 \\
\frac{2(x+\ell+d)(x+2\ell+d-t)}{(2x+3\ell+2d)(x+\ell+d -t)} & \ell_1 < t< \ell_3 \\
\frac{4(t-x)(x+2\ell+d-t)\(t^2 -(x+\ell+d)^2\)}{\epsilon^2 d(2x+3\ell+2d)} & \ell_3 < t< \ell_4\\
\frac{(2\ell+d)(x+\ell+d -t)}{(\ell+d)(x+2\ell+d-t)}& t> \ell_4
\end{array}\right.
\end{align}
\begin{align}
\eta_2 = \left\{\begin{array}{lc}
\frac{2(x+\ell+d)(\ell+d)}{d(2x+3\ell+2d)} & t <\ell_2 \\
\frac{2(x+\ell+d)(x+2\ell+d-t)}{(2x+3\ell+2d)(x+\ell+d -t)} & \ell_2 < t< \ell_3 \\
\frac{4\(t-(x+\ell)\)(x+2\ell+d-t)\(t^2 - (x+\ell+d)^2\)}{\epsilon^2 d(2x +3\ell+2d)} & \ell_3 < t< \ell_4\\
\frac{(\ell+d)(x+\ell+d -t)}{d(x+2\ell+d-t)}& t> \ell_4
\end{array}\right.
\end{align}
and one can verify that once again $\eta_i \approx 1$ or $\eta_i \to \infty$, as 
the length to separation ratio $\ell/d$ tends to zero and the distance from the defect to separation ratio $x/d$ tends to zero or diverges. 

For equal length intervals situated asymmetrically about the quench $A = [-(\ell_1+\ell),-\ell_1]$, $B= [\ell_2, \ell_2+\ell]$ with $\ell_1, \ell_2 >0$, the cross-ratios have more complicated expressions. However, they still approach $0,1$ or $\infty$ for $\ell_1, \ell_2 \gg \ell$ and $\ell_1,\ell_2\ll \ell$. 

More generally, whenever there is a large separation of scales one expects 
universal behavior. We do not perform a complete analysis of the cross ratios here, but it is straightforward to calculate their values in any given configuration and to determine the universal regimes. 

%%%%%%%%%%%%%%%%%%%%%%%%%%%%%%%%%%%%%%%%%%%%%%%%%%%%%%%%%%%%%%%%%%%%%%%%%%

\section{Geodesics in the falling mass geometry}\label{sect:geodesics}

In this Appendix we review the geodesic solutions of  \cite{Nozaki:2013wia} in the falling mass geometry. To construct the geodesics joining the boundary endpoints $(t_\infty, z_\infty, \ell_1)$, $(t_\infty, z_\infty, \ell_2)$ in the asymptotically Poincar\'e AdS geometry, the strategy is to first construct the geodesics in the global metric, Eq.~\eqref{eq:BTZ},
\be
\label{eq:BTZApp}
ds^2 = -(r^2+1 -M) d\tau^2 + \frac{dr^2}{r^2+1 -M} + r^2 d\theta^2 \,,
\ee
connecting the endpoints
\bea
\tau_\infty^{(i)} &= \tan^{-1} \left[ \frac{2 t_\infty \alpha}{\alpha^2 + \ell_i^2 -t_\infty^2} \right] \label{eq:tauinf} \\
\theta_\infty^{(i)} &= \tan^{-1} \left[ -\frac{2 \ell_i \alpha}{\ell_i^2 -t_\infty^2 -\alpha^2}\right] \label{eq:thetainf} \\
r_\infty^{(i)} &= \frac{1}{z_\infty \alpha} \sqrt{ \ell_i^2 \alpha^2+ \frac 1 4 \left[ \ell_i^2 -t_\infty^2 - \alpha^2 \right]^2}
\end{align}
for $i=1,2$, and then apply the coordinates transformation in Eq.~\eqref{eq:z}-\eqref{eq:x}. Recall that $\tau_\infty^{(i)} \in [0,\pi]$ ($\tau_\infty^{(i)} \in [-\pi,0]$) for $t_\infty \ge 0$ ($t_\infty \le0$) and $\theta_\infty^{(i)} \in [0,\pi]$ ($\theta_\infty^{(i)} \in [-\pi,0]$) for $\ell_i \ge0$ ($\ell_i \le 0$). 

Parametrizing the geodesics by the geodesic length $\lambda$ and using the symmetries of the metric \eqref{eq:BTZApp}, the geodesic equations can be written as 
\begin{align}
(r^2+1 -M) \frac{d \tau}{d\lambda} &= \frac{A}{B} \\
r^2 \frac{d \theta}{d \lambda} &= \frac 1 B \\
- ( r^2+1 -M) \left(\frac{d \tau}{d\lambda} \right)^2 + r^2 \left(\frac{d \theta}{d \lambda} \right)^2 &+ \left(\frac{d r}{d \lambda}\right)^2 \frac{1}{r^2+1 -M} =1 
\end{align}
where $A$ and $B$ are integration constants. It follows that
\be
\(\frac{d r}{d \lambda} \)^2=  \frac{ A^2 r^2 +(B^2 r^2 -1)(r^2+1-M)}{B^2 r^2}\,.
\ee
We can trade the $\lambda$ dependence for $r$ to obtain the $\pm$ branches:
\begin{align}
\frac{d \tau_{\pm}}{dr} &= \pm \frac{A r}{(r^2 +1 -M) \sqrt{A^2 r^2 +(B^2 r^2 -1)(r^2+1-M)}}
 \label{eq:taur}\\
\frac{d \theta_{\pm}}{dr} &= \pm \frac{1}{r \sqrt{A^2 r^2 +(B^2 r^2 -1)(r^2+1-M)}} \label{eq:thetar}
\end{align}
and the geodesic length 
\be \label{eq:geodesiclength1}
{\cal L} = \sum_{i=1}^2 \int_{r_*}^{r_\infty^{(i)}} dr \frac{B r }{\sqrt{A^2 r^2 + (B^2 r^2 -1)(r^2 +1-M)}}\,.
\ee
Here
\be
r_* = \frac{\sqrt{1-A^2 -B^2(1-M) +\sqrt{(1-A^2-B^2(1-M))^2 +4 B^2(1-M)}}}{\sqrt 2 B} 
\ee
denotes the minimal value of the radial coordinate attained along the geodesic and we have assumed $B>0$ without loss of generality. 
Notice that depending on the sign of $A$, the derivatives of the $\tau_{\pm}$ and $\theta_{\pm}$ branches have the same or opposite sign. Alternatively, we can assume $A \ge 0$, but remember that each branch $\tau_+$, $\tau_-$ can be combined with both branches $\theta_{\pm}$.  In the following we first deal with the case $A>0$ and discuss the specific case $A = 0$ at the end. 

The geodesic equations \eqref{eq:taur} and \eqref{eq:thetar} can be integrated using
\be \label{eq:integral}
\int \frac{dx}{x \sqrt{R}} = \frac{1}{\sqrt{-a}} \sin^{-1} \frac{2a + bx}{x\sqrt{b^2 - 4 ac}}\,,
\ee
where $R = a + b x +c x^2$, $a<0$ and $\Delta \equiv 4 a c -b^2 <0$, respectively with $x= r^2$ and $x =r^2 +1-M$, to obtain
\begin{align}
\tau_{\pm}(r) &= \tau^0_{\pm} \pm \frac{1}{2\sqrt{1-M}} \sin^{-1} \frac{-2 A^2 (1-M) +(A^2 -B^2 (1-M)-1)(r^2 +1-M)}{(r^2+1-M)\sqrt{(A^2 -B^2(1-M)-1)^2 +4 A^2 B^2 (1-M)}}\\
\theta_{\pm}(r)& = \theta^0_{\pm} \pm \frac{1}{2\sqrt{1-M}} \sin^{-1} \frac{-2(1-M) +(A^2 +B^2(1-M)-1)r^2}{r^2\sqrt{(A^2 +B^2(1-M)-1)^2 + 4B^2(1-M)}} \,. \label{eq:thetasol}
\end{align}
To compute the entanglement entropy associated to an interval $\ell_1 \le x \le \ell_2$ at time $t_\infty$ on the spacetime boundary in Poincar\'e coordinates, we fix $A$, $B$ and the four integration constants $\tau^0_{\pm}$, $ \theta^0_{\pm}$ with the joining conditions
\begin{align}
\tau_+(r_*) &= \tau_-(r_*) \label{eq:joiningtau}\\
\theta_+(r_*) &= \theta_-(r_*)  \,
\end{align}
at the geodesic turning point, and assuming $\tau_\infty^{(1)} < \tau_\infty^{(2)}$, the boundary conditions
\begin{align}
\lim_{r \to \infty} \tau_+(r) &= \tau_\infty^{(2)} \qquad \lim_{r \to \infty} \tau_-(r) = \tau_\infty^{(1)}  \label{eq:BC1} \\
\lim_{r \to \infty} \theta_+(r)& = \theta_\infty^{(2)}   \qquad  \lim_{r \to \infty} \theta_-(r) = \theta_\infty^{(1)} \,, \label{eq:BC4}
\end{align}
or 
\begin{align}
\lim_{r \to \infty} \tau_+(r) &= \tau_\infty^{(2)} \qquad \lim_{r \to \infty} \tau_-(r) = \tau_\infty^{(1)}  \label{eq:BC1a} \\
\lim_{r \to \infty} \theta_+(r)& = \theta_\infty^{(1)}   \qquad  \lim_{r \to \infty} \theta_-(r) = \theta_\infty^{(2)} \,. \label{eq:BC4a}
\end{align}
Thus for each pair of points $(\tau_\infty^{(1)}, r_\infty^{(1)}, \theta_\infty^{(1)})$, $(\tau_\infty^{(2)}, r_\infty^{(2)}, \theta_\infty^{(2)})$ there exist two geodesics of boundary separation, in the $\theta$ coordinate, respectively  $ |\theta_\infty^{(2)} -\theta_\infty^{(1)}|$ and $2\pi - |\theta_\infty^{(2)} -\theta_\infty^{(1)}|$.    

\begin{figure}[h]
\begin{center}
\includegraphics[width = 0.4 \textwidth]{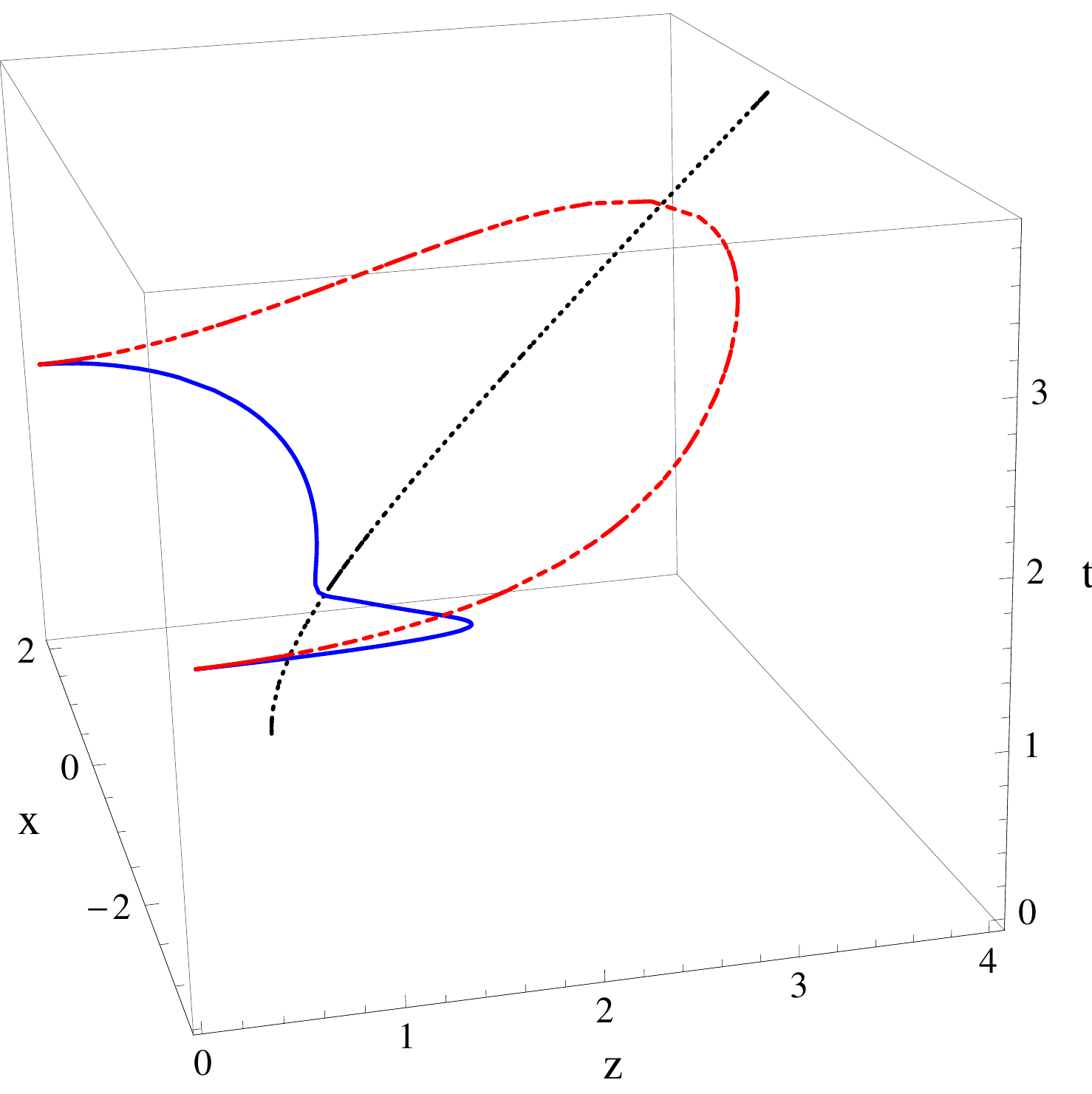} 
\caption{The two solutions for an interval $\ell_1 = -3.44$, $\ell_2 = 2.06$ at $t_\infty=1.93$. The geodesic of minimal length is the red dashed curve. The particle trajectory is the black dotted curve and here we have set $\alpha =1$, $M=0.4$.}\label{fig:multiple}
\end{center}
\end{figure}

Defining:
\bea
\Delta \tau_\infty \equiv |\tau_\infty^{(2)} -\tau_\infty^{(1)}| &= & \frac{1}{\sqrt{1-M}} \left[ \frac \pi 2 + \sin^{-1} \frac{-B^2(1-M) +A^2 -1}{\sqrt{(B^2(1-M) -A^2 +1)^2 +4 A^2B^2 (1-M)}}\right] \nonumber \\
&&\\
\Delta \theta_\infty \equiv  |\theta_\infty^{(2)} -\theta_\infty^{(1)}| &= & \frac{1}{\sqrt{1-M}} \left[ \frac \pi 2 + \sin^{-1} \frac{B^2(1-M) +A^2 -1}{\sqrt{(B^2(1-M) + A^2 - 1)^2 +4 B^2 (1-M)}}\right] \,,\nonumber \\
&& \label{eq:Deltathetainf}
\end{align}
we have
\bea
A &=  \left|  \frac{\sin (\sqrt{1-M} \Delta \tau_\infty)}{\sin(\sqrt{1-M} \Delta \theta_\infty)} \right| \label{eq:A} \\
B &= \frac{1}{\sqrt{1-M}} \left| \frac{\cos ( \sqrt{1-M} \Delta \tau_\infty) - \cos (\sqrt{1-M} \Delta \theta_\infty)}{\sin(\sqrt{1-M} \Delta \theta_\infty)} \right|\,, \label{eq:B}
\end{align}
while the geodesics satisfying the boundary conditions with boundary separation $2\pi -\Delta \theta_\infty$ instead are obtained by replacing $\Delta \theta_\infty$ by $2\pi -\Delta \theta_\infty$
 in Eq.~\eqref{eq:A}-\eqref{eq:B}.  Applying the coordinate transformation in Eq.~\eqref{eq:z}-\eqref{eq:x}, we finally obtain the geodesics in the falling mass geometry in Poincar\'e coordinates. In Fig.~\ref{fig:multiple}, we plot an example of the two solutions for a given boundary separation and time.  
In presence of multiple extremal solutions in a time-dependent background, the prescription of \cite{Hubeny:2007xt} to compute holographically the entanglement entropy amounts to taking the minimal length one. 

To decide which one has minimal length, we compute the geodesic length,
Eq.~\eqref{eq:geodesiclength1},
\bea
{\cal L} &= \sum_{i=1}^2 \int_{r_*}^{r_\infty^{(i)}} dr \frac{B r }{\sqrt{A^2 r^2 + (B^2 r^2 -1)(r^2 +1-M)}} \nonumber \\
&= \ln \(r_\infty^{(1)}\ r_\infty^{(2)} \) + \ln \frac{4 B^2}{\sqrt{(B^2(1-M) +A^2 -1)^2 +4 B^2 (1-M)}} \,, \label{eq:lengthpartial}
\end{align}
where we have used 
\be
\int \frac{dx}{\sqrt R} = \frac{1}{\sqrt c} \ln \( 2 \sqrt{c R} +2 c x +b\) \,,
\ee
for $R = a +bx+cx^2$, with $c>0$, $x =r^2$ and taken the limit $r_\infty^{(i)} \to \infty$. Using Eqs.~\eqref{eq:A}-\eqref{eq:B}, Eq.~\eqref{eq:lengthpartial} becomes
\be \label{eq:length1}
{\cal L} = \ln \(r_\infty^{(1)}\ r_\infty^{(2)} \) + \ln \frac{2|\cos\(\sqrt{1-M}  \Delta \tau_\infty \) -\cos\(\sqrt{1-M} \Delta \theta_\infty \)|}{1-M}\,, 
\ee
where $\Delta \theta_\infty$ should be replaced by $2\pi - \Delta \theta_\infty$ for the second solution. Since 
\be
\cos\(\sqrt{1-M}  \Delta \tau_\infty \) -\cos\(\sqrt{1-M} \Delta \theta_\infty \) = \frac{2 B^2 (1-M)}{\sqrt{\( B^2(1-M)+ A^2 -1 \)^2 + 4 B^2(1-M)}} >0
\ee
and $\cos(\sqrt{1-M} \Delta \theta_\infty) \ge \cos(\sqrt{1-M}(2\pi- \Delta \theta_\infty))$ for $\Delta \theta_\infty \le \pi$ (and $M \le1$), the length, Eq.~\eqref{eq:length1}, of the geodesic of minimal length becomes
\be \label{eq:lengthmin}
{\cal L} = \ln \(r_\infty^{(1)}\ r_\infty^{(2)} \) + \ln \frac{2 \( \cos\(\sqrt{1-M}  \Delta \tau_\infty \) -\cos\(\sqrt{1-M} \Delta \theta_\infty \) \)}{1-M}\,, 
\ee
if $\Delta \theta_\infty \le \pi$, and with $\Delta \theta_\infty$ replaced by $2\pi - \Delta \theta_\infty$ otherwise. 

If $A =0$, it follows from Eq.~\eqref{eq:taur} and the joining condition at the geodesic endpoint, Eq.~\eqref{eq:joiningtau}, that $\tau_+(r) = \tau_-(r) = const$. Notice the minimal radial value attained along the geodesic is now $ r_* = \frac{1}{|B|}$. From the boundary conditions \eqref{eq:tauinf}-\eqref{eq:thetainf}, it follows that this case describes a symmetric interval with $- \ell_1 = \ell_2 \equiv \ell$ and $-\theta_\infty^{(1)} =\theta_\infty^{(2)} \equiv \theta_\infty$ for $0<  \theta_\infty < \pi$. 

For $A=0$ the solution \eqref{eq:thetasol} reduces to 
\be
\theta_{\pm}(r) = \theta^0_{\pm} \pm \frac{1}{2\sqrt{1-M}} \sin^{-1} \frac{-2 r_*^2 (1-M) + (1-M -r_*^2)r^2}{r^2 (1-M +r_*^2)}\,.
\ee
Imposing 
\bea
\theta_{\pm}(r_*) &=0 \\
\lim_{r\to \infty}\theta_{\pm}(r) &= \pm \theta_{\infty}\,,
\end{align}
we obtain 
\bea
\theta^0_{\pm} &=  \pm \frac{1}{2\sqrt{1-M}} \frac \pi 2 \\
\theta_\infty &= \frac{1}{2\sqrt{1-M}} \left( \frac \pi 2 + \sin^{-1} \frac{1-M-r_*^2}{1-M+r_*^2}\right) \,,
\end{align}
so that explicilty
\be \label{eq:thetasymm}
\theta_{\pm}(r) = \pm \frac{1}{2\sqrt{1-M}} \left(\frac \pi 2 + \sin^{-1}   \frac{-2 r_*^2 (1-M) + (1-M -r_*^2)r^2}{r^2 (1-M +r_*^2)}\right)\,.
\ee
The length, Eq.~\eqref{eq:length1}, of the the symmetric geodesic is given by
\be \label{eq:lengthsymm}
{\cal L} = 2  \ln  \frac{2 r_\infty \sin (\sqrt{1-M} \theta_\infty)}{\sqrt{1-M}} \,, 
\ee
where $r_\infty \equiv r_\infty^{(1)} = r_\infty^{(2)}$. 
The second solution, satisfying the boundary condition 
\be
\lim_{r\to \infty}\theta_{\pm}(r) = \mp \theta_{\infty}\,,
\ee
has length given by Eq.~\eqref{eq:lengthsymm} with $ \theta_\infty$ replaced by $\pi - \theta_\infty$ and is the one of minimal length for $ \frac \pi 2 < \theta_\infty  < \pi$. 

\begin{figure}[h]
\begin{tabular}{ccc}
\includegraphics[width=5cm,height =5cm]{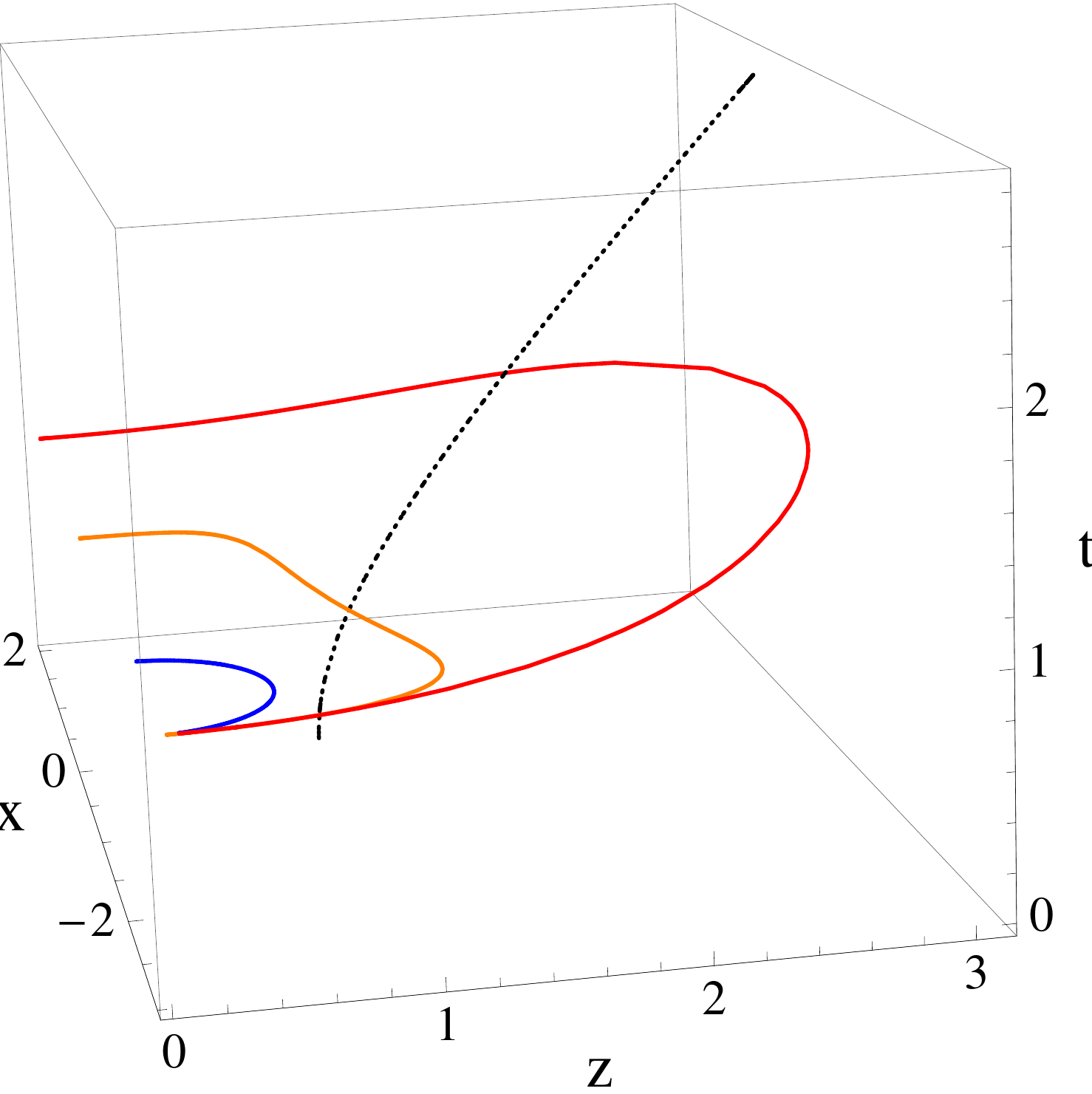}
&
\includegraphics[width=5.5cm,height =5.5cm]{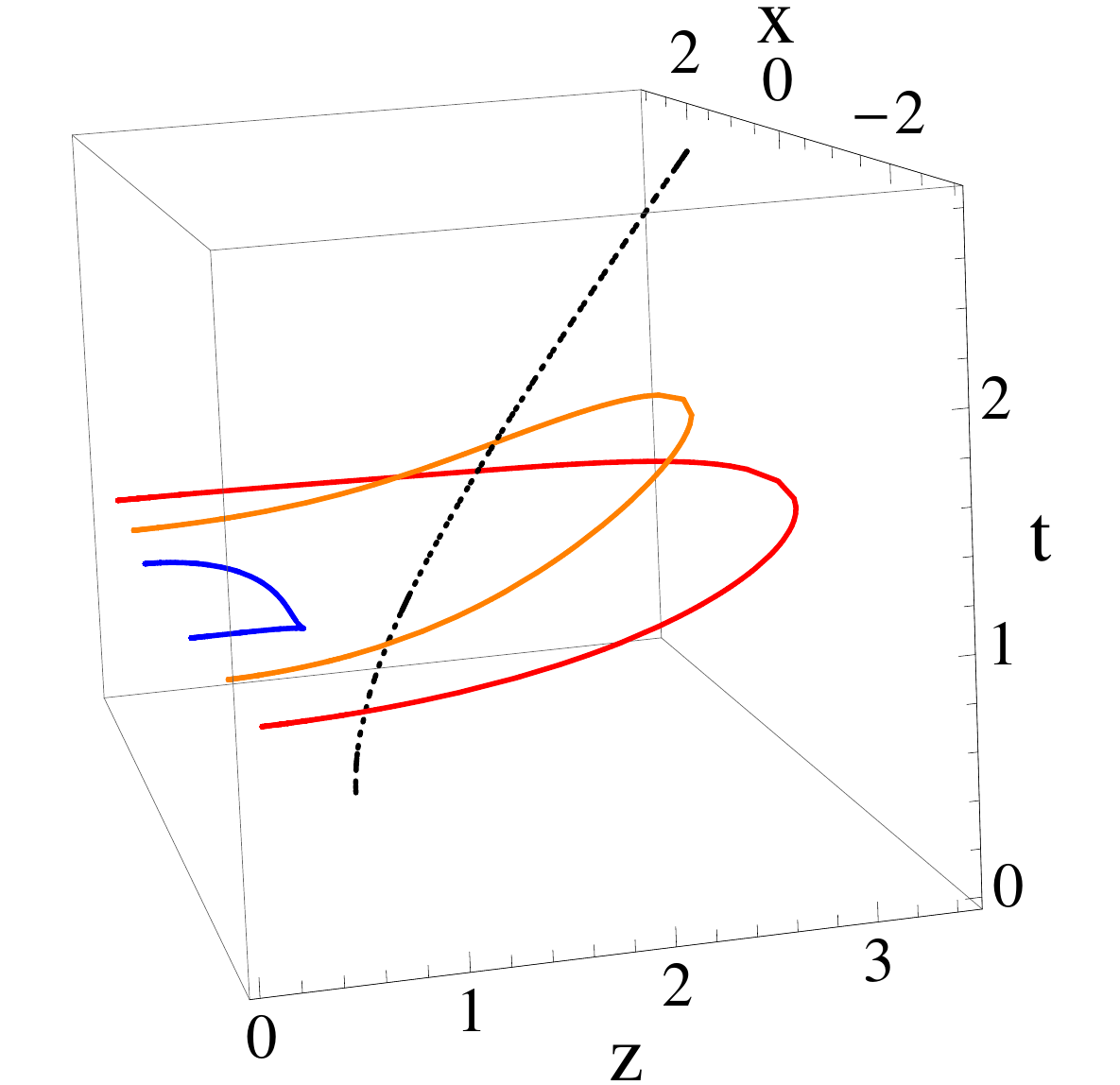}
&
\includegraphics[width=5cm,height =5cm]{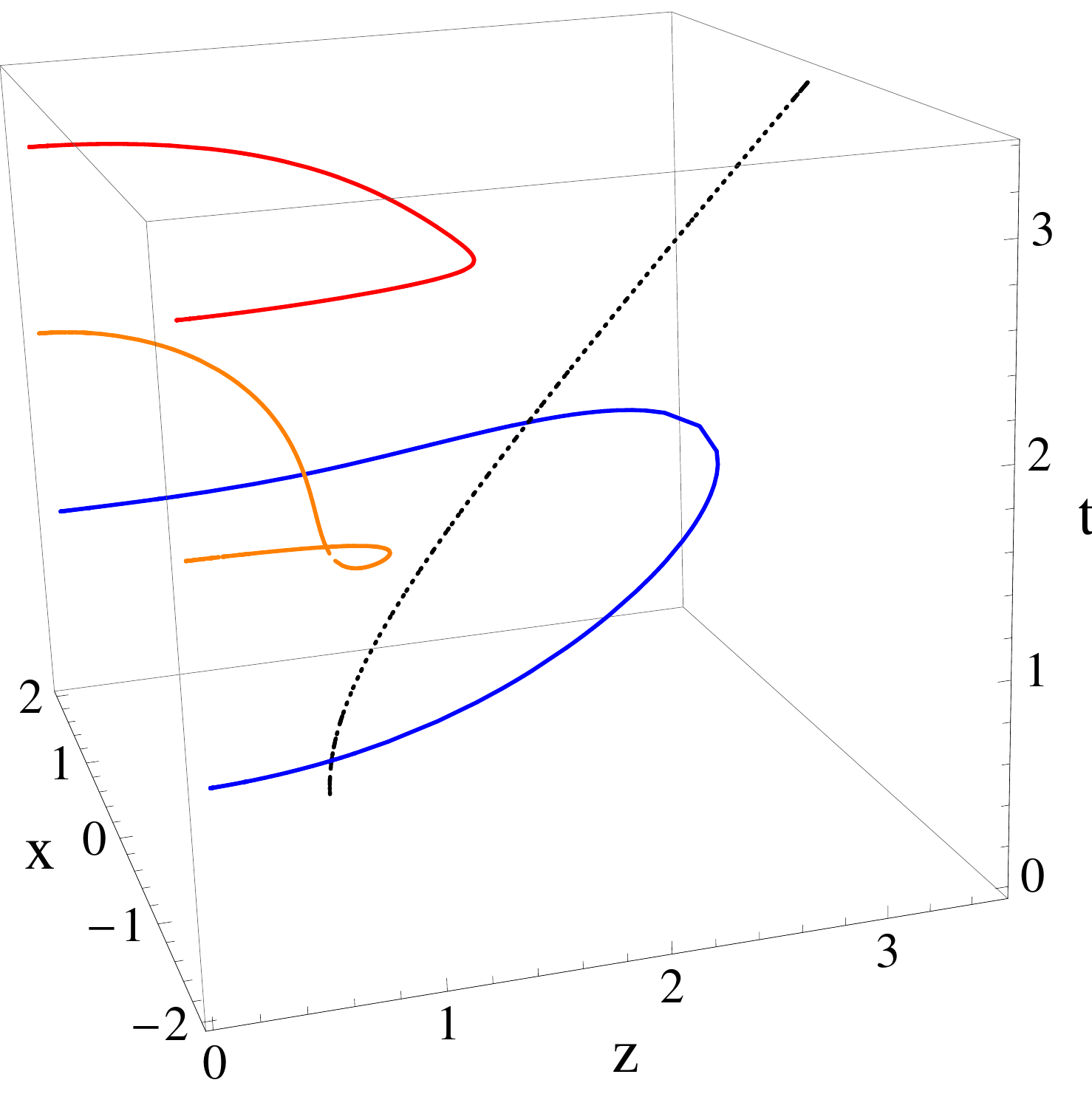}
\\
{\bf (A)} & {\bf (B)} & {\bf (C)} 
\\
\end{tabular}
\caption{Geodesics of minimal length for: {\bf (A)} $t_\infty = 1$, $\ell_1 =-3$, $\ell_2 = -2,0,2$, {\bf (B)} $t_\infty = 1$, $-\ell_1 = \ell_2 = 1,2,3$,  {\bf (C)} $t_\infty =1,2,3$,  $-\ell_1 = \ell_2 = 2$. The particle trajectory is the black dotted curve. In the plot, we have set the parameters $\alpha=1$ and $M=3/4$.  
}
\label{fig:geodesics}
\end{figure}

In Fig.~\ref{fig:geodesics}, we plot the minimal length geodesics that compute the entanglement entropy in the falling mass geometry for various values of the boundary time and interval choices. 

%%%%%%%%%%%%%%%%%%%%%%%%%%%%%%%%%%%%%%%%%%%%%%%%%%%%%%%%%%%%%%%%%%%%%%%%%%%%%%%%%%%
%%%%%%%%%%%%%%%%%%%%%%%%%%%%%%%%%%%%%%%%%%%%%%%%%%%%%%%%%%%%%%%%%%%%%%%%%%%%%%%%%%%

%%%%%%%%%%%%%%%%%%%%%%%%%%%%%%%%%%%%
%%%%%%%%%%%%%%%%%%%%%%%%%%%%%%%%%%%%

\input{MI-loc-quench_12_14.bbl}

%%%%%%%%%%%%%%%%%%%%%%%%%%%%%%%%%%%%
%%%%%%%%%%%%%%%%%%%%%%%%%%%%%%%%%%%%

\end{document}